\documentclass[journal,12pt,onecolumn,draftclsnofoot,]{IEEEtran}
\usepackage[a4paper, total={8.5in, 11in}, margin = 1in]{geometry}
\usepackage{amsmath}
\usepackage{graphicx}
\usepackage{caption2}
\usepackage{amsthm}
\usepackage{float}
\usepackage{mathrsfs}
\usepackage{verbatim}
\usepackage{epstopdf}
\usepackage{amssymb}
\usepackage{amsfonts}
\usepackage{subfigure}
\usepackage{color}
\usepackage{cite}
\usepackage{cancel}
\usepackage[shortlabels]{enumitem}
\usepackage[breaklinks=true,letterpaper=true,colorlinks=false,bookmarks=false]{hyperref}
\usepackage{algorithm}
\usepackage{algpseudocode}
\usepackage{arydshln}
\usepackage[official]{eurosym}
\usepackage{comment}
\usepackage{amsmath,amssymb,amsthm,mathrsfs,amsfonts,dsfont}
\usepackage[shortlabels]{enumitem}

\DeclareMathOperator*{\argmax}{arg\,max}
\newcommand\independent{\protect\mathpalette{\protect\independenT}{\perp}}
\def\independenT#1#2{\mathrel{\rlap{$#1#2$}\mkern2mu{#1#2}}}

\newtheorem{theorem}{Theorem}
\newtheorem{proposition}{Proposition}
\newtheorem{remark}{Remark}
\newtheorem{corollary}{Corollary}[theorem]
\newtheorem{example}{Example}
\newtheorem{lemma}{Lemma}
\newtheorem{conjecture}{Conjecture}
\newtheorem{definition}{Definition}

\begin{document}
\title{On Perfect Privacy}
\author{Borzoo Rassouli$^1$, and Deniz G\"und\"uz$^2$\\
\small{$^1$ School of Computer Science and Electronic Engineering, University of Essex, Colchester CO4 3SQ, UK}\\
\small{$^2$ Department of Electrical and Electronic Engineering, Imperial College London, , London SW7 2AZ, UK}\\
{\tt\small b.rassouli@essex.ac.uk}, {\tt\small  d.gunduz@imperial.ac.uk}
\thanks{
A conference version of this paper is provided in \cite{RG18}.
}
}

\maketitle
\begin{abstract}
The problem of private data disclosure is studied from an information theoretic perspective. Considering a pair of dependent random variables $(X,Y)$, where $X$ and $Y$ denote the private and useful data, respectively, the following problem is addressed: What is the maximum information that can be revealed about $Y$ (measured by mutual information $I(Y;U)$, in which $U$ is the revealed data), while disclosing no information about $X$ (captured by the condition of statistical independence, i.e., $X\independent U$, and henceforth, called \textit{perfect privacy})? 
We analyze the supremization of \textit{utility}, i.e., $I(Y;U)$ under the condition of perfect privacy for two scenarios: \textit{output perturbation} and \textit{full data observation} models, which correspond to the cases where a Markov kernel, called \textit{privacy-preserving mapping}, applies to $Y$ and the pair $(X,Y)$, respectively.  When both $X$ and $Y$ have a finite alphabet, the linear algebraic analysis involved in the solution provides some interesting results, such as upper/lower bounds on the size of the released alphabet and the maximum utility.
Afterwards, it is shown that for the jointly Gaussian $(X,Y)$, perfect privacy is not possible in the output perturbation model in contrast to the full data observation model. Finally, an asymptotic analysis is provided to obtain the rate of released information when a sufficiently small leakage is allowed. In particular, in the context of output perturbation model, it is shown that this rate is always finite when perfect privacy is not feasible, and two lower bounds are provided for it; When perfect privacy is feasible, it is shown that under mild conditions, this rate becomes unbounded.


\end{abstract}
\section{Introduction}
With the explosion of machine learning algorithms, and their applications in many areas of science, technology, and governance, data is becoming an extremely valuable asset. However, with the growing power of machine learning algorithms in learning individual behavioral 
patterns from diverse data sources, privacy is becoming a major concern, calling for strict regulations on data ownership and distribution. On the other hand, many recent examples of de-anonymization attacks on publicly available anonymized data ({\color{black}e.g.,}\!\!\cite{Nar},\cite{Ding}) show that regulation alone will not be sufficient to limit access to private data. An alternative approach, also considered in this paper, is to process the data at the time of release, such that no private information is leaked, called \textit{perfect privacy}. Assuming that the joint distribution of the observed data, useful data and the private data that should be kept private is known, an information-theoretic study is carried out in this paper to characterize the fundamental limits on perfect privacy. 

Consider a situation in which Alice wants to release some  \textit{useful} information about herself to Bob, represented by random variable $Y$, and she receives some utility from this disclosure of information. This may represent data measured and recorded by a health monitoring system \cite{mhealth}, her smart meter measurements \cite{Gomez}, or the sequence of a portion of her DNA to detect potential illnesses \cite{Abol}. At the same time, she wishes to conceal from Bob some \textit{private} information which depends on $Y$, represented by $X$. To this end, instead of letting Bob have a direct access to $Y$, a \textit{privacy-preserving mapping} is applied, whereby a distorted version of $Y$, denoted by $U$, is revealed to Bob. In this context, privacy and utility are competing goals that result in the \textit{utility-privacy trade-off}: The more $Y$ is distorted by the privacy-preserving mapping, the less information can Bob infer about $X$, but also the less the utility that can be obtained. This trade-off is the very result of the dependencies between $X$ and $Y$. An extreme point of this trade-off is the scenario termed as \textit{perfect privacy}, which refers to the situation where nothing is allowed to be inferred about $X$ by Bob through the disclosure of $U$. This condition is modelled by the statistical independence of $X$ and $U$. 

The concern of privacy and the design of privacy-preserving mappings have been the focus of a broad area of research in recent years, e.g., \cite{Dwork, Sweeney, LD, LL}, while the information-theoretic view of privacy has gained increasing attention more recently \cite{SG19}. 
In \cite{Makhdoumi}, the utility-privacy trade-off under the \textit{log-loss} cost function is considered, called as the \textit{privacy funnel}, which is closely related to the \textit{information bottleneck} introduced in \cite{Tishby}. In \cite{Calmon2} and \cite{info7010015}, the utility-privacy trade-off is investigated from an information theoretic perspective, and bounds on the optimal trade-off are derived. Measuring both the privacy and the utility in terms of mutual information, perfect privacy is fully characterized in \cite{Shahab1} for the binary case. Furthermore, a new quantity is introduced to capture the amount of private information about the latent variable $X$ carried by the useful data $Y$. In \cite{HuangK,Tripathy, advers}, the authors address this trade-off in a data-driven approach by setting an adversarial game between the competing neural networks.

We study the information theoretic perfect privacy in this paper, and {\color{black}our main contributions can be briefly summarized as follows:
\subsection{Non-asymptotic analysis - Sections \ref{S-Output perturbation}, \ref{S-Non-private} and \ref{S-Full data} }
\subsubsection{Output perturbation model (sections \ref{S-Output perturbation} and \ref{S-Non-private})}
\begin{itemize}
     \item Denoting the supremum of $I(Y;U)$ under perfect privacy by $g_0(X,Y)$, we analyze its solution through a linear programming (LP) for finite alphabets to obtain upper and lower bounds on the cardinality of the released data, where the former is a sufficient condition, and the latter is necessary. 
     \item From the LP solution, upper and lower bounds on $g_0(X,Y)$ are derived, which are tighter than any other known bounds in the literature in certain scenarios.
     \item For a jointly Gaussian $(X,Y)$, we obtain $g_\epsilon(X,Y)$ for the whole permissible range $\epsilon\in[0,I(X;Y)]$. Furthermore, we generalize this result for $\epsilon>0$ to any joint distribution that satisfies smoothness, and for $\epsilon=0$ to the additive case, i.e., $X=Y+N$, with $N(\independent Y)$ being Gaussian. 
     \item In the same setting, in the case of finite release alphabet, say $M$,
      we show that the utility reaches its maximum of $\log M$ for a vanishingly small leakage. This is shown by using two types of practical filters: equiprobable and uniform quantizers.
      
     \item We show that in the case of finite release alphabet, the supremum in the definition of $g_\epsilon(X,Y)$ is actually a maximum, which is in spite of the non-compactness of the search space.
     \item We establish the relationship between $g_0(X,Y)$ and \textit{non-private information about $X$ carried by $Y$}, $D_X(Y)$, as defined in \cite{Shahab1}, and provide the necessary and sufficient conditions when the two aforementioned quantities are equal.
\end{itemize}
\subsubsection{Full data observation model (section \ref{S-Full data})}
\begin{itemize}
    \item We provide the necessary and sufficient condition for the feasibility of perfect privacy. In this context, the maximum utility is denoted by $G_0(X,Y)$. 
     \item We provide a lower bound on $G_0(X,Y)$, which can become relevant to the \textit{maximal leakage} defined in \cite{Issa}.
     \item We show that for a jointly Gaussian $(X,Y)$, we have $G_0(X,Y)=\infty$, which is the direct opposite of $g_0(X,Y)=0$. We actually state this result for the broader class of additive noise, i.e., $Y=X+N$, in which, $N$ does not need to be independent of $X$, but it needs to admit a density for each realization $x$ of $X$. 
\end{itemize}
\subsection{Asymptotic analysis in the context of output pertaurbation model - Section \ref{asymp}}
\begin{itemize}
    \item We show that when perfect privacy is not feasible, the slope of $g_\epsilon(X,Y)$ at origin, i.e., $\epsilon=0$, is always finite, and provide two lower bounds on this slope, which are tighter than the previously known bounds in the literature.
     \item We show that when perfect privacy is feasible, for a broad range of cases, this slope at origin is infinite.
     \item We provide a general lower bound on this slope when perfect privacy is feasible.
\end{itemize}
}

\textbf{Notations.} Random variables are denoted by capital letters, their realizations by lower case letters, and their alphabets by capital letters in calligraphic font. Matrices and vectors are denoted by bold capital and bold lower case letters, respectively. For a matrix $\mathbf{A}_{m\times k}$, the null space, rank, and nullity are denoted by $\textnormal{Null}(\mathbf{A})$, $\textnormal{rank}(\mathbf{A})$, and $\textnormal{nul}(\mathbf{A})$, respectively, with $\textnormal{rank}(\mathbf{A})+\textnormal{nul}(\mathbf{A})=k$.
For integers $m\leq n$, we have the discrete interval $[m:n]\triangleq\{m, m+1,\ldots,n\}$, and the tuple $(a_m,a_{m+1},\ldots,a_n)$ is written in short as $a_{[m:n]}$. The set $[n]$ is written in short as $[n]$.
For an integer $n\geq 1$, the notations $\mathbf{1}_n$, and $\mathbf{0}_n$ denote the $n$-dimensional all-one, and all-zero column vectors, respectively. For a random variable $X\in\mathcal{X}$, with finite $|\mathcal{X}|$, the probability simplex $\mathcal{P}(\mathcal{X})$ is the standard $(|\mathcal{X}|-1)$-simplex given by
\begin{equation*}
\mathcal{P}(\mathcal{X})=\bigg\{\mathbf{v}\in\mathbb{R}^{|\mathcal{X}|}\bigg|\mathbf{1}_{|\mathcal{X}|}^T\cdot\mathbf{v}=1,\ v_i\geq 0,\ \forall i\in [|\mathcal{X}|]\bigg\},
\end{equation*}
whose interior is denoted by $\textnormal{int}(\mathcal{P}(\mathcal{X}))$. Furthermore, to each probability mass function (pmf) on $\mathcal{X}$, denoted by $p_{X}(\cdot)$, corresponds a matrix $\mathbf{P}_X=\textnormal{diag}(\mathbf{p}_X)$, where $\mathbf{p}_X$ is a probability vector  in $\mathcal{P}(\mathcal{X})$, whose  $i$-th element is $p_X(x_i)$ ($i\in[|\mathcal{X}|]$). For a pair of random variables $(X,Y)$ with joint pmf $p_{X,Y}$, $\mathbf{P}_{X,Y}$ is an $|\mathcal{X}|\times|\mathcal{Y}|$ matrix with $(i,j)$-th entry equal to $p_{X,Y}(i,j)$. Likewise, $\mathbf{P}_{X|Y}$ is an $|\mathcal{X}|\times|\mathcal{Y}|$ matrix with $(i,j)$-th entry equal to $p_{X|Y}(i|j)$.
$F_{Y}(\cdot)$ denotes the cumulative distribution function (CDF) of random variable $Y$, and if it admits a density, its probability density function (pdf) is denoted by $f_Y(\cdot)$. 
Throughout the paper, for a random variable $Y$ with the corresponding probability vector $\mathbf{p}_Y$, $H(Y)$ and $H(\mathbf{p}_Y)$ are written interchangeably, and so are the quantities $D(p_Y(\cdot)||q_Y(\cdot))$ and $D(\mathbf{p}_Y||\mathbf{q}_Y)$. {\color{black}All the logarithms in this paper are to the base of 2.} Given two positive integers $a,b$, $a$ modulo $b$ is abbreviated as $a\textnormal{ mod }b$. Finally, $d_{\textnormal{TV}}$, $\lfloor\cdot\rfloor$, and $\lceil\cdot\rceil$ denote the total variation distance, the floor, and the ceiling operators, respectively.
 
\section{System model and preliminaries}
Consider a triplet of random variables $(X,Y,W)\in\mathcal{X}\times\mathcal{Y}\times\mathcal{W}$, distributed according to the joint distribution $p_{X,Y,W}$. Let $X$ denote the \textit{private/sensitive data} that the user/curator wants to conceal, $Y$ denote the \textit{useful data} the user wishes to disclose, and $W$ denote the \textit{observable data} that the curator observes, which can be regarded as a noisy version of $(X,Y)$. Assume that the \textit{privacy-preserving mapping}/\textit{data release mechanism} takes $W$ as input, and maps it to the \textit{released data}, denoted by $U$. In this scenario, $(X,Y)-W-U$ form a Markov chain, and the privacy-preserving mapping is captured by the conditional distribution $p_{U|W}$.
{\color{black}\begin{definition}
\textit{Perfect privacy} is feasible if there exists a privacy-preserving mapping $p_{U|W}$ whose output ($U$) is statistically dependent on the useful data ($Y$), while  being  statistically independent  of  the  private  data ($X$);  that  is, $Y\not\independent U$ and $X\independent U$.
\end{definition}}
Unless otherwise stated explicitly, we assume that all the alphabets/supports $\mathcal{X,Y,W}$ are finite. In this context, we assume that $p_X(x),p_Y(y),p_W(w)>0,\forall (x,y,w)\in\mathcal{X}\times\mathcal{Y}\times\mathcal{W}$, since otherwise the alphabets could have been modified accordingly. This equivalently means that the corresponding probability vectors $\mathbf{p}_X,\mathbf{p}_Y, \mathbf{p}_W$ are in the interior of their corresponding probability simplices, i.e., $\mathcal{P}(\mathcal{X}),\mathcal{P}(\mathcal{Y}),\mathcal{P}(\mathcal{W})$, respectively.

The following proposition states the necessary and sufficient condition for the feasibility of perfect privacy.
\begin{proposition}
Perfect privacy is feasible for $(X,Y,W)\in\mathcal{X}\times\mathcal{Y}\times\mathcal{W}$ if and only if
\begin{equation}\label{Fullshart}
\textnormal{dim}\bigg(\textnormal{Null}(\mathbf{P}_{X|W})\backslash\textnormal{Null}(\mathbf{P}_{Y|W})\bigg)\neq 0.
\end{equation}
\end{proposition}
\begin{proof}
The proof is a simple generalization of \cite[Theorem 4]{Berger}, by noting that both $X-W-U$ and $Y-W-U$ form Markov chains. In other words, we have $X\independent U$ if and only if for all $u\in\mathcal{U}$, $\mathbf{p}_X=\mathbf{P}_{X|W}\mathbf{p}_{W|u}$. On the other hand, we have $Y\not\independent U$ if and only if there exists a $u'\in\mathcal{U}$, such that $\mathbf{p}_Y\neq\mathbf{P}_{Y|W}\mathbf{p}_{W|u'}$.
Equivalently, there exists a vector $\mathbf{v}'\triangleq\mathbf{p}_{W}-\mathbf{p}_{W|u'}$ in $\textnormal{Null}(\mathbf{P}_{X|W})$ ($\mathbf{v}'\neq\mathbf{0}$) that does not belong to $\textnormal{Null}(\mathbf{P}_{Y|W})$, which is equivalent to (\ref{Fullshart}).
\end{proof}
The special cases of \textit{full data observation} and \textit{output perturbation} (\!\!\cite{Ishwar}) refer to the scenarios in which the privacy-preserving mapping has direct access to both the private and useful data (i.e., $W=(X,Y)$) and only to the useful data (i.e., $W=Y$), respectively. The whole paper is devoted to these two models.

By adopting mutual information as the measure of both \textit{utility} and \textit{privacy} (i.e., $I(Y;U)$, and $I(X;U)$, respectively), the optimal utility-privacy trade-off in the output perturbation model is defined as\footnote{This is the same notation as in \cite{Shahab1}.}
\begin{equation}\label{def}
g_{\epsilon}(X,Y)\triangleq\sup_{\substack{p_{U|Y}:\\X-Y-U\\I(X;U)\leq\epsilon}}I(Y;U),
\end{equation}
and in the full data observation model, the trade-off can be formulated as
\begin{equation}\label{Geps}
   G_\epsilon(X,Y)\triangleq \sup_{\substack{p_{U|X,Y}:\\I(X;U)\leq\epsilon}} I(Y;U),
\end{equation}
where the effective range of $\epsilon$ is $[0,I(X;Y)]$.

Finally, we can say that perfect privacy being feasible in the output perturbation and full data observation models is equivalent to having $g_0(X,Y)>0$ and $G_0(X,Y)>0$, respectively. 
\section{Output perturbation model}\label{S-Output perturbation}
In this model, we have $X-Y-U$ form a Markov chain, and in order to derive $g_0(X,Y)$, we proceed as follows. From the singular value decomposition of $\mathbf{P}_{X|Y}$, we have
$\mathbf{P}_{X|Y}=\mathbf{U}\mathbf{\Sigma}\mathbf{V}^T$,
where the matrix of right eigenvectors is $\mathbf{V}=\begin{bmatrix}
\mathbf{v}_1&\mathbf{v}_2&\dots&\mathbf{v}_{|\mathcal{Y}|}
\end{bmatrix}.$
By assuming (without loss of generality) that the singular values are arranged in a descending order, only the first $\textnormal{rank}(\mathbf{P}_{X|Y})$ singular values are non-zero. Therefore, the null space of $\mathbf{P}_{X|Y}$ can be written as $\textnormal{Null}(\mathbf{P}_{X|Y})=\textnormal{Span}\{\mathbf{v}_{\textnormal{rank}(\mathbf{P}_{X|Y})+1},\mathbf{v}_{\textnormal{rank}(\mathbf{P}_{X|Y})+2},\ldots,\mathbf{v}_{|\mathcal{Y}|}\}.$

In the Markov chain $X-Y-U$, the random variables $X$ and $U$ are independent if and only if $\mathbf{P}_{X|Y}(\mathbf{p}_Y-\mathbf{p}_{Y|u})=\mathbf{0},\ \forall u\in\mathcal{U}$, which is equivalent to $ (\mathbf{p}_Y-\mathbf{p}_{Y|u})\in\textnormal{Null}(\mathbf{P}_{X|Y}),\ \forall u\in\mathcal{U}.$
Let $\mathbf{A}$ be defined as 
$\mathbf{A}\triangleq\begin{bmatrix}\mathbf{v}_1&\mathbf{v}_2&\dots&\mathbf{v}_{\textnormal{rank}(\mathbf{P}_{X|Y})}\end{bmatrix}^T.$
Therefore, we have $X\independent U$ in $X-Y-U$ if and only if $ \mathbf{A}(\mathbf{p}_Y-\mathbf{p}_{Y|u})=\mathbf{0},\ \forall u\in\mathcal{U}.$
Let $\mathbb{S}_{X,Y}$ be defined as
\begin{equation}\label{poly}
\mathbb{S}_{X,Y}\triangleq\bigg\{\mathbf{t}\in\mathbb{R}^{|\mathcal{Y}|}\bigg|\mathbf{A}\mathbf{t}=\mathbf{A}\mathbf{p}_Y\ ,\ \mathbf{t}\geq 0\bigg\},
\end{equation}
which is a convex polytope in $\mathcal{P}(\mathcal{Y})$, since it can be written as the intersection of a finite number of half-spaces in $\mathcal{P}(\mathcal{Y})$. With this definition, we have that having $X\independent U$ in $X-Y-U$ results in $\mathbf{p}_{Y|u}\in\mathbb{S}_{X,Y},\ \forall u\in\mathcal{U}$. On the other hand, for any pair $(Y,U)$, for which $\mathbf{p}_{Y|u}\in\mathbb{S}_{X,Y},\ \forall u\in\mathcal{U}$, we can simply have $X-Y-U$ and $X\independent U$.
Therefore, we can write
\begin{equation}\label{siz}
X-Y-U,\ X\independent U \Longleftrightarrow \mathbf{p}_{Y|u}\in\mathbb{S}_{X,Y},\ \forall u\in\mathcal{U}.
\end{equation}


\begin{theorem}\label{THLP}
The supremum in (\ref{def}) is attained, and hence, it is a maximum. Furthermore, in the evaluation of $g_0(X,Y)$, the optimal privacy-preserving mapping is the solution to a standard linear program (LP), and it is sufficient to have $|\mathcal{U}|\leq \textnormal{nul}(\mathbf{P}_{X|Y})+1$. Finally, if $p^*_{Y,U}$ corresponds to a maximizer $p^*_{U|Y}$, for any $u\in\mathcal{U}$, we have
\begin{equation}
    |\{y\in\mathcal{Y}|p^*(y|u)>0\}|\leq\textnormal{rank}(\mathbf{P}_{X|Y}).
\end{equation}
\end{theorem}
\begin{proof}
The proof of the attainability of the supremum, and the upper bound $|\mathcal{U}|\leq \textnormal{nul}(\mathbf{P}_{X|Y})+1$\footnote{{\color{black}The proof of this upper bound follows the application of cardinality bounding technique and taking into account the convex polytope $\mathbb{S}_{X,Y}$ in (\ref{poly}). Although we are considering perfect privacy here, i.e, $g_0(X,Y)$, in the evaluation of $g_\epsilon(X,Y), \forall\epsilon>0$, it is sufficient to have $|\mathcal{U}|\leq|\mathcal{Y}|+1$ as in \cite{info7010015}.}} are provided in Appendix \ref{app1}. We have
\begin{align}
g_0(X,Y)
&= H(Y)-\!\!\!\!\!\!\!\!\min_{\substack{p_U(\cdot),\ \mathbf{p}_{Y|u}\in\mathbb{S}_{X,Y},\ \forall u\in\mathcal{U}:\\ \sum_{u} p_U(u)\mathbf{p}_{Y|u}=\mathbf{p}_Y}}\!\!\!\!\!\!\!\!H(Y|U)\label{min},
\end{align}
where in (\ref{min}), since the minimization is over $\mathbf{p}_{Y|u}$ rather than $p_{U|Y}$, a constraint was added to preserve the marginal distribution $\mathbf{p}_Y$. The minimization of the concave functional in (\ref{min}) simplifies to an LP as stated in \cite{Witsen1975}.

In order to prove the final claim in the statement of this Theorem, we need to address the solution to this LP, whose linear algebraic analysis (i.e., characterizations of the null space, extreme points, etc.) is the basis for some of the main results obtained in this paper. We address this solution as follows.
\begin{lemma}\label{Lemex}
In minimizing $H(Y|U)$ over $\{\mathbf{p}_{Y|u}\in\mathbb{S}_{X,Y}|\sum_{u} p_U(u)\mathbf{p}_{Y|u}=\mathbf{p}_Y\}$, it is sufficient to consider only $\textnormal{nul}(\mathbf{P}_{X|Y})+1$ extreme points of $\mathbb{S}_{X,Y}$.
\end{lemma}
\begin{proof}
The proof is provided in Appendix \ref{app3}.
\end{proof}
From lemma \ref{Lemex}, the solution to the minimization in (\ref{min}) can be obtained in two phases: in phase one, the extreme points of set $\mathbb{S}_{X,Y}$ are identified, while in phase two, proper weights over these extreme points are obtained to minimize the objective function.
 
For the first phase, we proceed as follows. The extreme points of $\mathbb{S}_{X,Y}$ are the basic feasible solutions (see \cite{LP1}, \cite{LP2}) of 
$\{\mathbf{x}\in\mathbb{R}^{|\mathcal{Y}|}|\mathbf{A}\mathbf{x}=\mathbf{b}\ ,\ \mathbf{x}\geq 0\},$
where $\mathbf{b}=\mathbf{A}\mathbf{p}_Y$. 
The procedure of finding the extreme points of $\mathbb{S}_{X,Y}$ is as follows. Pick a set $\mathcal{B}\subset[|\mathcal{Y}|]$ of indices that correspond to $\textnormal{rank}(\mathbf{P}_{X|Y})$ linearly independent columns of matrix $\mathbf{A}$ defined prior to (\ref{poly}). 
Let $\mathbf{A}_{\mathcal{B}}$ be a $\textnormal{rank}(\mathbf{P}_{X|Y})\times\textnormal{rank}(\mathbf{P}_{X|Y})$ matrix whose columns are the columns of $\mathbf{A}$ indexed by the indices in $\mathcal{B}$. Also, for any $\mathbf x\in\mathbb{S}_{X,Y}$, define a corresponding vector $\tilde{\mathbf x}\triangleq\begin{bmatrix}\mathbf{x}_{\mathcal{B}}^T&\mathbf{x}_{\mathcal{N}}^T\end{bmatrix}^T$, where $\mathbf{x}_{\mathcal{B}}$ and $\mathbf{x}_{\mathcal{N}}$ are $\textnormal{rank}(\mathbf{P}_{X|Y})$-dimensional and $\textnormal{nul}(\mathbf{P}_{X|Y})$-dimensional vectors whose elements are the elements of $\mathbf{x}$ indexed by the indices in $\mathcal{B}$ and $[|\mathcal{Y}|]\backslash\mathcal{B}$, respectively.


For any basic feasible solution $\mathbf{x}^*$, there exists a set $\mathcal{B}\subset[|\mathcal{Y}|]$ of indices that correspond to a set of $\textnormal{rank}(\mathbf{P}_{X|Y})$ linearly independent columns of $\mathbf{A}$, such that the corresponding vector of $\mathbf{x}^*$, i.e. $\tilde{\mathbf{x}}^*=\begin{bmatrix}{\mathbf{x}^*_\mathcal{B}}^T&{\mathbf{x}^*_\mathcal{N}}^T\end{bmatrix}^T$, satisfies the following
\begin{equation}
\mathbf{x}_\mathcal{N}^*=\mathbf{0},\ \ \ \mathbf{x}_\mathcal{B}^*=\mathbf{A}_\mathcal{B}^{-1}\mathbf{b},\ \ \ \mathbf{x}_\mathcal{B}^*\geq 0,
\end{equation}
where the inequality is element-wise. On the other hand, for any set $\mathcal{B}\subset[|\mathcal{Y}|]$ of indices that correspond to a set of $\textnormal{rank}(\mathbf{P}_{X|Y})$ linearly independent columns of $\mathbf{A}$, if $\mathbf{A}_\mathcal{B}^{-1}\mathbf{b}\geq 0$, then $\begin{bmatrix}
\mathbf{b}^T\mathbf{A}_\mathcal{B}^{-T}&\mathbf{0}^T
\end{bmatrix}^T$ is the corresponding vector of a basic feasible solution.
Hence, the extreme points of $\mathbb{S}_{X,Y}$ are obtained as mentioned above, and {\color{black}their number is at most ${|\mathcal{Y}|\choose \textnormal{rank}(\mathbf{P}_{X|Y})}$, which is justified as follows. Since an extreme point is identified if and only if A) the $\textnormal{rank}(\mathbf{P}_{X|Y})$ selected columns are linearly independent, B) the corresponding $\mathbf{x}_\mathcal{B}$ has all non-negative elements, it is concluded that the total number of extreme points is upper bounded by the total number of ways to choose $\textnormal{rank}(\mathbf{P}_{X|Y})$ linearly independent columns out of $|\mathcal{Y}|$ columns. The latter is also upper bounded by the total number of ways to choose $\textnormal{rank}(\mathbf{P}_{X|Y})$ columns out of $|\mathcal{Y}|$ columns, which is ${|\mathcal{Y}|\choose \textnormal{rank}(\mathbf{P}_{X|Y})}$.} Furthermore, each extreme point has at most $\textnormal{rank}(\mathbf{P}_{X|Y})$ non-zero elements corresponding to $\mathbf{x}_{\mathcal{B}}$, which is equivalent to $|\{y\in\mathcal{Y}|p^*(y|u)>0\}|\leq\textnormal{rank}(\mathbf{P}_{X|Y})$ for any $u\in\mathcal{U}$.

%
For the second phase, we proceed as follows. {\color{black}Assume that $\mathbb{S}_{X,Y}$ has $K$ (a positive integer) extreme points, denoted by $\mathbf{p}_1,\mathbf{p}_2,\ldots,\mathbf{p}_K$, which were identified in the first phase.} Then, (\ref{min}) is equivalent to
\begin{align}
g_0(X,Y)=H(Y)-&\min_{\mathbf{w}\geq 0}\ \begin{bmatrix}
H(\mathbf{p}_1)&H(\mathbf{p}_2)&\dots&H(\mathbf{p}_K)
\end{bmatrix}\cdot\mathbf{w}\nonumber\\
&\ \ \ \textnormal{s.t. }\begin{bmatrix}
\mathbf{p}_1&\mathbf{p}_2&\dots&\mathbf{p}_K
\end{bmatrix}\mathbf{w}=\mathbf{p}_Y,\label{LP}
\end{align}
where $\mathbf{w}$ is a $K$-dimensional weight vector, and it can be verified that the constraint $\sum_{i=1}^Kw_i=1$ is satisfied if the constraint in (\ref{LP}) is met. The problem in (\ref{LP}) is a standard LP.
\end{proof}
\begin{corollary}\label{cor1}
In the evaluation of $g_0(X,Y)$, it is necessary to have $|\mathcal{U}|\geq \left \lceil \frac{|{\mathcal{Y}}|}{\textnormal{rank}(\mathbf{P}_{X|Y})}\right\rceil$.
\end{corollary}
\begin{proof}
In the proof of Theorem \ref{THLP}, in order to write the $|{\mathcal{Y}}|$-dimensional probability vector $\mathbf{p}_{Y}$ as a convex combination of the extreme points of $\mathbb{S}_{X,Y}$, that have at most $\textnormal{rank}(\mathbf{P}_{X|Y})$ non-zero elements, at least $\left\lceil \frac{|{\mathcal{Y}}|}{\mbox{rank}(\mathbf{P}_{X|Y})}\right\rceil$ points are needed, which results in $|\mathcal{U}|\geq\left\lceil \frac{|{\mathcal{Y}}|}{\mbox{rank}(\mathbf{P}_{X|Y})}\right\rceil$.
\end{proof}
{\color{black}
\begin{corollary}\label{cor22}
We have the following bounds on $g_0(X,Y)$.
\begin{equation}\label{lovb}
    \left(H(Y)-\log \textnormal{rank}(\mathbf{P}_{X|Y})\right)^+\leq g_0(X,Y)\leq \min\{\log \left(\textnormal{nul}(\mathbf{P}_{X|Y})+1\right),H(Y|X)\}.
\end{equation}
\end{corollary}
\begin{proof}
The first term in the upper bound is immediate from $I(Y;U)\leq H(U)\leq\log|\mathcal{U}|\leq \log \left(\textnormal{nul}(\mathbf{P}_{X|Y})+1\right)$, where the last inequality follows from Theorem \ref{THLP}. The second term in the upper bound follows from \cite{Calmon2}. The lower bound is proved as follows. As mentioned in the proof of Theorem \ref{THLP}, each extreme point of $\mathbb{S}_{X,Y}$ has at most $\textnormal{rank}(\mathbf{P}_{X|Y})$ non-zero elements, which means that the entropy of each extreme point is upper bounded by $\log(\textnormal{rank}(\mathbf{P}_{X|Y}))$. Hence,
\begin{align*}
    \min_{\substack{p_U(\cdot),\ \mathbf{p}_{Y|u}\in\mathbb{S}_{X,Y},\ \forall u\in\mathcal{U}:\\ \sum_{u} p_U(u)\mathbf{p}_{Y|u}=\mathbf{p}_Y}}\!\!\!\!\!\!\!\!H(Y|U)&\leq\log(\textnormal{rank}(\mathbf{P}_{X|Y})),
\end{align*}
which results in the lower bound in (\ref{lovb}).
\end{proof}
}
\begin{remark} 
It can be verified that in the degenerate case of $X\independent Y$, we have $\textnormal{Null}(\mathbf{P}_{X|Y})=\textnormal{Span}\{\mathbf v_1,\mathbf v_2,\ldots,\mathbf v_{|\mathcal Y|}\}$, or equivalently, $\mathbb{S}_{X,Y}=\mathcal P(\mathcal Y)$. {\color{black}In this case, the extreme points of $\mathbb{S}_{X,Y}$, which are the standard unit vectors of the $|\mathcal{Y}|$-dimensional Cartesian coordinate system denoted by $\{\mathbf{e}_i\}_{i=1}^{|\mathcal{Y}|}$, have zero entropy.} Therefore, the minimum value of $H(Y|U)$ is zero, and $\mathcal U=\{u_1,u_2,\ldots,u_{|\mathcal Y|}\}$ with $p_U(u_i)=p_Y(y_i),\forall i\in[|\mathcal Y|]$ and $\mathbf p_{Y|{u_i}}=\mathbf e_i$. As a result, $g_0(X,Y)=H(Y)$, which is also consistent with the fact that $U\triangleq Y$ is independent of $X$ and maximizes $I(Y;U)$.
\end{remark}
\begin{remark}
It can be verified that for the general scenario of $(X,Y)-W-U$, where the privacy-preserving mapping is denoted by $p_{U|W}$, with mutual information, or MMSE (i.e., $\mathds{E}[(Y-U)^2]$), or the probability of error (i.e., $\textnormal{Pr}\{Y\neq U\}$) as the utility measure, the optimization in obtaining the perfect privacy-preserving mapping also simplifies to an LP.
\end{remark}
The following example clarifies the LP solution in Theorem 1.
\begin{example}
Consider the pair $(X,Y)\in[2]\times[4]$, and the joint distribution 
\begin{equation*}
\mathbf{P}_{X,Y}=\begin{bmatrix}
0.15 & 0.2 & 0.0625 & 0.05 \\ 0.35 & 0.05 & 0.0625 & 0.075
\end{bmatrix},
\end{equation*}
which results in 
\begin{equation*}
\mathbf{p}_Y=\begin{bmatrix}
\frac{1}{2}&\frac{1}{4}&\frac{1}{8}&\frac{1}{8}\end{bmatrix}^T,\ \ \ \mathbf{P}_{X|Y}=\begin{bmatrix}
0.3 & 0.8 & 0.5 & 0.4 \\ 0.7 & 0.2 & 0.5 & 0.6
\end{bmatrix}.
\end{equation*}
Since $|\mathcal{Y}|>|\mathcal{X}|$, we have $\textnormal{nul}(\mathbf{P}_{X|Y})\neq0$; and therefore, $g_0(X,Y)>0$. The singular value decomposition of $\mathbf{P}_{X|Y}$ is
\begin{equation*}
\mathbf{P}_{X|Y}=\begin{bmatrix}
-0.7071 & -0.7071 \\ -0.7071 & 0.7071
\end{bmatrix}\begin{bmatrix}
1.4142 & 0 & 0 & 0 \\ 0 & 0.5292 & 0 & 0
\end{bmatrix}\begin{bmatrix}
-0.5 & 0.5345 & -0.4163 & -0.5394 \\ 
-0.5 & -0.8018 & -0.3154 & -0.0876\\
-0.5 & 0      &0.8452   & -0.1889 \\
-0.5 & 0.2673 &-0.1135  &  0.8159
\end{bmatrix}^T\!\!\!\!,
\end{equation*}
where it is obvious that columns 3 and 4 of the matrix of the right eigenvectors span the null space of $\mathbf{P}_{X|Y}$. Hence, matrix $\mathbf{A}$, defined before (\ref{poly}), is
\begin{equation*}
\mathbf{A}=\begin{bmatrix}
-0.5 & -0.5 & -0.5 & -0.5 \\ 0.5345 & -0.8018 & 0 & 0.2673
\end{bmatrix}.
\end{equation*}
For the first phase, i.e., finding the extreme points of $\mathbb{S}_{X,Y}$, the index set $\mathcal{B}$ (as in the proof of Theorem \ref{THLP}) can be $\{1,2\},\{1,3\},\{1,4\},\{2,3\},\{2,4\}$ or $\{3,4\}$. From $\mathbf{x}_\mathcal{B}=\mathbf{A}_\mathcal{B}^{-1}\mathbf{b}$, we get
\begin{align*}
&\mathbf{x}_{\{1,2\}}=\begin{bmatrix}
0.675&0.325
\end{bmatrix}^T,\mathbf{x}_{\{1,3\}}=\begin{bmatrix}
0.1875&0.8125
\end{bmatrix}^T,{\color{red}\mathbf{x}_{\{1,4\}}=\begin{bmatrix}
-0.625&1.625
\end{bmatrix}^T}\\
&{\color{red}\mathbf{x}_{\{2,3\}}=\begin{bmatrix}
-0.125&1.125
\end{bmatrix}^T},\mathbf{x}_{\{2,4\}}=\begin{bmatrix}
0.1563&0.8437
\end{bmatrix}^T,\mathbf{x}_{\{3,4\}}=\begin{bmatrix}
0.625&0.375
\end{bmatrix}^T.
\end{align*}
It is obvious that $\mathbf{x}_{\{1,4\}}$ and $\mathbf{x}_{\{2,3\}}$ are not feasible, since they do not satisfy $\mathbf{x}_\mathcal{B}\geq 0$. Therefore, the extreme points of $\mathbb{S}_{X,Y}$ are obtained as
\begin{equation*}
\mathbf{p}_{1}=\begin{bmatrix}
0.675\\0.325\\0\\0
\end{bmatrix},\mathbf{p}_{2}=\begin{bmatrix}
0.1875\\0\\0.8125\\0
\end{bmatrix},\mathbf{p}_{3}=\begin{bmatrix}
0\\0.1563\\0\\0.8437
\end{bmatrix},\mathbf{p}_{4}=\begin{bmatrix}
0\\0\\0.625\\0.375
\end{bmatrix},
\end{equation*}
each having at most $\textnormal{rank}(\mathbf{P}_{X|Y})(=2)$ non-zero elements. For the second phase, the standard LP in (\ref{LP}) reduces to
\begin{align*}
&\min_{\mathbf{w}\geq 0}\ \ \ \begin{bmatrix}
0.9097&0.6962&0.6254&0.9544
\end{bmatrix}.\mathbf{w}\nonumber\\
&\ \ \ \textnormal{S.t. }\begin{bmatrix}
0.675&0.1875&0&0\\
0.325&0&0.1563&0\\
0&0.8125&0&0.625\\
0&0&0.8437&0.375
\end{bmatrix}\mathbf{w}=\begin{bmatrix}
\frac{1}{2}\\\frac{1}{4}\\\frac{1}{8}\\\frac{1}{8}
\end{bmatrix},
\end{align*}
where the minimum value of the objective function is $0.8437$ bits, which is achieved by
\begin{equation*}
\mathbf{w}^*=\begin{bmatrix}
0.698&0.1538&0.1481&0
\end{bmatrix}^T.
\end{equation*}
Therefore, $g_0(X,Y)=0.9063$ bits, $\mathcal{U}=\{u_1,u_2,u_3\}$ (which is consistent with $2\leq|\mathcal{U}|\leq 3$), $\mathbf{p}_U=\begin{bmatrix}
0.698 & 0.1538 & 0.1481
\end{bmatrix}^T$ and $\mathbf{p}_{Y|u_i}=\mathbf{p}_i,\ \forall i\in[3]$. Furthermore, $p^*_{U|Y}$ corresponds to the matrix $\mathbf{P}^*_{U|Y}$ given as
\begin{equation*}
\mathbf{P}^*_{U|Y}=\begin{bmatrix}
0.9423&0.9074&0&0\\
0.0577&0&1&0\\
0&0.0926&0&1
\end{bmatrix}.
\end{equation*}
The bounds in (\ref{lovb}) are
\begin{equation*}
    0.75\leq g_0(X,Y)\leq\min\{1.585,1.6216\}.
\end{equation*}
{\color{black}Finally, as shown in Theorem \ref{thlower9}, we have
\begin{equation*}
    \lim_{\epsilon\to 0}\frac{g_\epsilon(X,Y)-g_0(X,Y)}{\epsilon}=\infty.
\end{equation*}}
\end{example}
Thus far, we have investigated perfect privacy when $|\mathcal X|,|\mathcal Y|<\infty$. 
In what follows, i.e., Theorem \ref{Guassiannotfeasible}, it is shown that perfect privacy is not feasible for the (correlated) jointly Gaussian pair. {\color{black}Part of the proof of Theorem \ref{Guassiannotfeasible} relies on using a privacy-preserving mapping $p_{U|Y}$ that quantizes the useful data $Y$ with infinitely small quantization intervals, which in turn, is based on the following lemma.}
\begin{lemma}\label{Quantize}
Let $Z$ be an r.v. distributed over an interval $[a,b]$, in which $a,b\in\mathbb{R}$ ($a<b$), with a bounded smooth pdf denoted by $f_Z(\cdot)$\footnote{{\color{black}Note that since $Z$ admits a density, its support being a segment, as $(a,b)$, or an interval, as $[a,b]$, or a mixture does not change the result in this lemma. Hence, with a slight abuse of notation, segment and interval are used interchangeably.}}. For positive integers $M,n$, define a partition $a=a_0<a_1<a_2<\ldots<a_{Mn -1}<a_{Mn}=b.$ Let $\mathcal{I}_i\triangleq[a_{i-1},a_i),\ \forall i\in[Mn-1],$ and $\mathcal{I}_{Mn}\triangleq[a_{Mn-1},b]$. Let $U$ be a function of $Z$ defined as
\begin{equation}\label{UZ}
    u(z) \triangleq (i-1)\ \textnormal{mod } M,\ \textnormal{if } z\in \mathcal{I}_i,\ \textnormal{for some }i\in[Mn].
\end{equation}
If for all $i\in[Mn]$, we have $(a_i-a_{i-1})\to 0$ with $n\to\infty$, then
\begin{equation}\label{entu}
    \lim_{n\to\infty}H(U)=\log  M.
\end{equation}
\end{lemma}
\begin{proof}
Let $p_U$ denote the pmf of $U$, whose realizations are given in (\ref{UZ}). Also, let $\mathcal{J}_i\triangleq\cup_{k=(i-1)M+1}^{iM}\mathcal{I}_k,\ \forall i\in[n].$
Let $\hat{Z}$ be an r.v. whose pdf is a {\color{black} piecewise} uniform approximation of $f_Z$ over the intervals $\mathcal{J}_i, i\in[n]$. Hence, we have
\begin{equation}
    f_{\hat{Z}}(\hat{z}) = \frac{1}{l(\mathcal{J}_i)}\int_{\mathcal{J}_i}f_Z(z)dz,\ \forall \hat{z}\in\mathcal{J}_i,\ \forall i\in[n],
\end{equation}
 where $l(\mathcal{J}_i)=a_{iM}-a_{(i-1)M}$ denotes the length of the segment $\mathcal{J}_i,\ i\in[n]$. 
 
Let $\hat{U}$ be a function of $\hat{Z}$ in exactly the same way that $U$ is defined as a function of $Z$, i.e., as in (\ref{UZ}). Since $f_{\hat{Z}}$ is flat over $\mathcal{J}_i,\ i\in[n]$, $\hat{U}$ is uniform over $[0:M-1]$. Since $(a_i-a_{i-1})\to 0$ as $n\to\infty,\ \forall i\in[Mn]$, we conclude that $l(\mathcal{J}_i)\to 0$ as $n\to\infty,\ \forall i\in[n]$. As a result, $f_{\hat{Z}}(\cdot)$ converges pointwise to $f_Z(\cdot)$, due to the smoothness of the latter.
Therefore, we have $d_{\textnormal{TV}}(f_{\hat{Z}},f_Z)=\int|f_{\hat{Z}}-f_Z|dz\to 0$ as $n\to \infty$, which is a direct consequence of Lebesgue's Dominated Convergence Theorem. Hence, by viewing $Z$ and $\hat{Z}$ as the inputs to a (deterministic) channel in (\ref{UZ}), with the corresponding outputs $U$ and $\hat{U}$, respectively, we get $\lim_{n\to\infty}d_{\textnormal{TV}}(p_U,p_{\hat{U}})=0$, which follows from the data processing inequality of f-divergences, i.e., $d_{\textnormal{TV}}(p_U,p_{\hat{U}})\leq d_{\textnormal{TV}}(f_{\hat{Z}},f_Z).$ Finally, from the fact that $H(\hat{U})=\log  M$, and by the continuity of entropy, (\ref{entu}) is proved.
\end{proof}
\begin{theorem}\label{Guassiannotfeasible}
Let $(X,Y)\sim\mathcal{N}(\mathbf{\mu},\mathbf{\Sigma})$ be a pair of jointly Gaussian random variables, where
\begin{equation}\label{Gauss}
\mathbf{\mu}=\begin{bmatrix}\mu_X\\\mu_Y\end{bmatrix}, \mathbf{\Sigma}=\begin{bmatrix}
\sigma_X^2 & \rho\sigma_X\sigma_Y\\ \rho\sigma_X\sigma_Y & \sigma_Y^2 \end{bmatrix},
\end{equation}
in which $\rho\neq 0$, since otherwise $X\independent Y$. 
We have 
\begin{equation}\label{ginfinity}
g_\epsilon(X,Y)=\left\{\begin{array}{ccc}
0&\epsilon =0\\\infty &\textnormal{o.w.}
\end{array}\right.
\end{equation}
\end{theorem}
\begin{proof}
{\color{black}First, it is shown that $g_0(X,Y)=0$.} If there exists a random variable $U$ such that $X-Y-U$  form a Markov chain and $X\independent U$, we must have $F_X(\cdot)=F_{X|U}(\cdot|u),\ \forall u\in\mathcal{U}$; and hence, $f_X(\cdot)=f_{X|U}(\cdot|u),\ \forall u\in\mathcal{U}$, since $X$ has a density. Equivalently, we must have
\begin{equation}\label{G11}
f_X(\cdot)=\int f_{X|Y}(\cdot|y)dF_{Y|U}(y|u),\ \forall u\in\mathcal{U}.
\end{equation}
Also, to have $g_0(X,Y)>0$, there must exist $u_1,u_2\in\mathcal{U}$, such that
\begin{equation}\label{G22}
F_{Y|U}(\cdot|u_1)\neq F_{Y|U}(\cdot|u_2).
\end{equation}
In what follows we show that if (\ref{G11}) holds, (\ref{G22}) cannot be satisfied; and therefore, perfect privacy is not feasible for a jointly Gaussian $(X,Y)$ pair.
 
It is known that $X$ conditioned on $\{Y=y\}$ is also Gaussian, given by
\begin{equation}\label{che}
X|\{Y=y\}\sim\mathcal{N}\bigg(\underbrace{\frac{\rho\sigma_X}{\sigma_Y}(y-\mu_Y)+\mu_X}_{\alpha y+\beta},\underbrace{(1-\rho^2)\sigma_X^2}_{\sigma^2}\bigg).
\end{equation}
From (\ref{G11}), (\ref{che}), and for $u_1,u_2\in\mathcal{U}$, we have
\begin{align*}
f_X(x)&=\int \frac{e^{-\frac{(x-\alpha y-\beta)^2}{2\sigma^2}}}{\sqrt{2\pi\sigma^2}}dF_{Y|U}(y|u_1)\\
&=\int \frac{e^{-\frac{(x-\alpha y-\beta)^2}{2\sigma^2}}}{\sqrt{2\pi\sigma^2}}dF_{Y|U}(y|u_2),\ \forall x\in\mathbb R,
\end{align*}
or, equivalently,
\begin{equation}\label{tk}
\int \frac{e^{-\frac{(x-\alpha y-\beta)^2}{2\sigma^2}}}{\sqrt{2\pi\sigma^2}}d\bigg(F_{Y|U}(y|u_1)-F_{Y|U}(y|u_2)\bigg)=0,\ \forall x\in\mathbb R.
\end{equation}
Multiplying both sides of (\ref{tk}) by $e^{j\omega x}$, and taking the integral with respect to $x$, we obtain
\begin{equation*}
\int e^{j\omega x}\bigg[\int \frac{e^{-\frac{(x-\alpha y-\beta)^2}{2\sigma^2}}}{\sqrt{2\pi\sigma^2}}d\bigg(F_{Y|U}(y|u_1)-F_{Y|U}(y|u_2)\bigg)\bigg]dx=0.
\end{equation*}
By Fubini's theorem\footnote{Note that $\int |f_{X|U}(x|u_1)-f_{X|U}(x|u_2)|dx\leq\int [|f_{X|U}(x|u_1)|+|f_{X|U}(x|u_2)|]dx=2<+\infty$.}, we can write
\begin{equation*}
\int \bigg[\int e^{j\omega x} \frac{e^{-\frac{(x-\alpha y-\beta)^2}{2\sigma^2}}}{\sqrt{2\pi\sigma^2}}dx\bigg]d\bigg(F_{Y|U}(y|u_1)-F_{Y|U}(y|u_2)\bigg)=0,
\end{equation*}
and after some manipulations, we get
\begin{equation}\label{mani}
\int e^{j\omega \alpha y}d\bigg(F_{Y|U}(y|u_1)-F_{Y|U}(y|u_2)\bigg)=0.
\end{equation}
Since $\rho\neq 0$, from (\ref{che}), we have $\alpha\neq 0$. Hence, the LHS of (\ref{mani}) is a Fourier transform. Due to the invertiblity of the Fourier transform, i.e. $\int e^{j\omega t}dg(t)=0\Longleftrightarrow dg(t)=0$, we must have $F_{Y|U}(\cdot|u_1)=F_{Y|U}(\cdot|u_2)$. Therefore, (\ref{G22}) does not hold and perfect privacy is not feasible for the (correlated) jointly Gaussian pair $(X,Y)$.

{\color{black} In order to show $g_\epsilon(X,Y)=\infty,\ \forall\epsilon>0$, two proofs/methods are provided. Both of them aim to construct a privacy-preserving mapping $p_{U|Y}$ as an $M$-level quantizer (for an arbitrary integer $M>0$), which satisfies the privacy constraint, and results in a utility that grows with $M$. Hence, the proof is completed by letting $M\to\infty$. In the first method, this is done by quantizing the support of $Y$ into equiprobable intervals, while in the second method, a uniform quantizer is employed, which partitions the support of $Y$ into intervals of the same length (denoted by $\Delta$). The advantages/disadvantages of these two methods are elaborated further in the remarks that follow the Theorem.} 

In what follows, without loss of optimality, we consider that both $X$ and $Y$ have the standard Normal distribution\footnote{This is due to the fact that for the tuple $(X,Y,U)$, $I(aX+b;U)=I(X;U)$, and $I(cY+d;U)=I(Y;U)$ for constants $a,b,c$ and $d$. Furthermore, the set of mappings from $Y$ has a one-to-one correspondence with the set of mappings from $cY+d$.}.
{\color{black}\subsection{First method for showing $g_\epsilon(X,Y)=\infty,\ \forall\epsilon>0$: Equiprobable quantizer}}
Fix $\epsilon>0$, and a positive integer $M$. For each integer $n>1$, define
\begin{align}
    \mathcal{B}_n&\triangleq \left[\Phi^{-1}(\frac{1}{n}),\Phi^{-1}(1-\frac{1}{n})\right]^2,\\
    p_n &\triangleq \textnormal{Pr}\bigg\{(X,Y)\not\in\mathcal{B}_n\bigg\},
\end{align}
{\color{black}where $\Phi^{-1}(\cdot)$ is the inverse function of standard Normal CDF $\Phi(x)\triangleq\frac{1}{\sqrt{2\pi}}\int_{-\infty}^xe^{-\frac{t^2}{2}}dt$.}

As $n\to\infty$, we have $p_n\to 0$, and hence, $\left(p_n\log M+H_b(p_n)\right)\to 0$. Therefore, there exists a positive integer $N_0$ such that $p_n\log M+H_b(p_n)\leq\frac{\epsilon}{2}$, for all $n\geq N_0$. Let $N_1(\epsilon)$ denote the minimum $N_0$ for which the previous statement holds.

Let $E\triangleq\mathds{1}_{(X,Y)\in\mathcal{B}_{N_1(\epsilon)}}$ be a binary indicator which is $1$ when $(X,Y)\in\mathcal{B}_{N_1(\epsilon)}$, and $0$ otherwise. For a positive integer $n$, let $\{\Phi^{-1}(\frac{i}{MN_1(\epsilon)n})\}_{i=1}^{MN_1(\epsilon)n-1}$ be a set of points that divide the support of a standard Normal into $MN_1(\epsilon)n$ {\color{black} equiprobable} intervals, which are denoted by $\mathcal{I}_1=\left(-\infty,\Phi^{-1}(\frac{1}{MN_1(\epsilon)n})\right)$, and $\mathcal{I}_i=\left[\Phi^{-1}(\frac{i-1}{MN_1(\epsilon)n}),\Phi^{-1}(\frac{i}{MN_1(\epsilon)n})\right)$ for $i\in[2:MN_1(\epsilon)n)$, with the convention $\Phi^{-1}(1)=\infty$. 

Define $U$ as a function of $Y$ according to
\begin{equation}\label{UZ1}
    u(y) \triangleq (i-1)\ \textnormal{mod } M,\ \textnormal{if } y\in \mathcal{I}_i,\ \textnormal{for some }i\in[MN_1(\epsilon)n].
\end{equation}
From the construction in (\ref{UZ1}), we have that $U$ is uniform over $[0:M-1]$, and $I(Y;U)=\log  M$ for any positive integer $n$. In the sequel, we show that there exists a positive integer $N$, such that $I(X;U)\leq\epsilon,\ \forall n\geq N$, which results in $g_\epsilon(X,Y)\geq\log  M$. Since $M$ is arbitrary, we get $g_\epsilon(X,Y) = \infty,$ which concludes the proof.

Conditioned on the event $\{E=1\}$, we have $H(U|E=1)=\log  M$, since $Y|\{E=1\}$ with pdf $f_{Y|E}(\cdot|1)$, which is a scaled version of $f_Y(\cdot)$, is distributed over $[\Phi^{-1}(\frac{1}{N_1(\epsilon)}),\Phi^{-1}(1-\frac{1}{N_1(\epsilon)})]$, which has been divided into $Mn(N_1(\epsilon)-2)$ {\color{black} equiprobable} intervals, i.e., $\{\mathcal{I}_i\}_{i=Mn+1}^{Mn(N_1(\epsilon)-1)}$, and from (\ref{UZ1}), $U|\{E=1\}$ becomes uniform over $[0:M-1]$.

For each realization $x\in [\Phi^{-1}(\frac{1}{N_1(\epsilon)}),\Phi^{-1}(1-\frac{1}{N_1(\epsilon)})]$ of $X$, the conditional pdf $f_{Y|X,E}(\cdot|x,1)$ is a bounded smooth density. Hence, from lemma \ref{Quantize}, there exists a positive integer $N(x,\epsilon)$, such that
\begin{equation}\label{givenX}
    H(U|X=x,E=1)\geq\log  M-\frac{\epsilon}{2},\ \forall n\geq N(x,\epsilon).
\end{equation}
Furthermore, since $[\Phi^{-1}(\frac{1}{N_1(\epsilon)}),\Phi^{-1}(1-\frac{1}{N_1(\epsilon)})]$ is a compact subset of the real line, we can define
\begin{equation*}
    N_2(\epsilon)\triangleq \max_{x\in[\Phi^{-1}(\frac{1}{N_1(\epsilon)}),\Phi^{-1}(1-\frac{1}{N_1(\epsilon)})]}N(x,\epsilon).
\end{equation*}
Therefore, for all $n\geq N_2(\epsilon)$, we have
\begin{align}
    I(X;U|E=1)&=H(U|E=1)-H(U|X,E=1)\nonumber\\
    & = \log  M -H(U|X,E=1)\label{akh1}\\
    &\leq \frac{\epsilon}{2},\label{akh2}
\end{align}
where (\ref{akh1}) and (\ref{akh2}) follow, respectively, from $U|\{E=1\}$ being uniform, and (\ref{givenX}). 

Finally, we have that for $n\geq \max\{N_1(\epsilon),N_2(\epsilon)\}$,
\begin{align}
    I(X;U)&\leq I(X;U,E)\nonumber\\
    &\leq H(E)+I(X;U|E)\label{xc1}\\
    &=H(\textnormal{Pr}\{E=0\})+\textnormal{Pr}\{E=0\}I(X;U|E=0) +\textnormal{Pr}\{E=1\}I(X;U|E=1)\nonumber\\
    &< H(\textnormal{Pr}\{E=0\})+\textnormal{Pr}\{E=0\}\log M +I(X;U|E=1)\label{kh1}\\
    &\leq \frac{\epsilon}{2}+\frac{\epsilon}{2}\label{kh2}\\
    &=\epsilon,\nonumber
\end{align}
where (\ref{kh1}) follows from the trivial upper bound $I(X;U|E=0)\leq\log M$; (\ref{kh2}) results from (\ref{akh2}) and the fact that $p_n\log M+H_b(p_n)\leq\frac{\epsilon}{2}$ for $n\geq N_1(\epsilon)$, in which $p_n=\textnormal{Pr}\{E=0\}$. Therefore, for any $\epsilon>0$, we have constructed a privacy-preserving mapping $p_{U|Y}$ for which $I(Y;U)=\log  M$, and $I(X;U)\leq\epsilon$. Finally, letting $M\to\infty$ completes the first proof.
{\color{black}\subsection{Second method for showing $g_\epsilon(X,Y)=\infty,\ \forall\epsilon>0$: Uniform quantizer}
Fix $\epsilon>0$. Fix a positive integer $M$, and set $U_M^\Delta\triangleq \left\lfloor \frac{MY}{\Delta}\right\rfloor\ \textnormal{mod } M, \ \forall \Delta>0$. 
From lemma \ref{Quantize}, we have
 \begin{equation}\label{s30}
     \lim_{\Delta\to 0} I(Y;U_M^\Delta)=\lim_{\Delta\to 0} H(U_M^\Delta)=\log  M,
\end{equation}
which follows from $H(U_M^\Delta|Y)=0$. Therefore, we have
\begin{equation}
    \textnormal{For any } \delta>0: \exists \Delta_0>0 \Longrightarrow I(Y;U_M^\Delta)\geq\log M-\delta,\ \forall \Delta\leq\Delta_0.
\end{equation}
Since the conditional distribution of $Y$ given $\{X=x\}$ still satisfies the conditions of lemma \ref{Quantize} (i.e., replace $f_Z$ with $f_{Y|X}$), we obtain
\begin{equation}\label{kf22}
    \textnormal{For any } x\in\mathbb{R}: \exists \Delta_x>0 \Longrightarrow H(U_M^\Delta|X=x)\geq\log M-\frac{\epsilon}{2},\ \forall \Delta\leq\Delta_x.
\end{equation}
Let $\mathcal{I}_0\triangleq[-\Phi^{-1}(1-\frac{\epsilon}{4\log M}),\Phi^{-1}(1-\frac{\epsilon}{4\log M})]$, and define $E_0\triangleq\mathds{1}_{\{X\in\mathcal{I}_0\}}$ as an r.v. indicating if $X$ belongs to $\mathcal{I}_0$. Finally, set $\Delta_1\triangleq\min_{x\in\mathcal{I}_0}\Delta_x$, where $\Delta_x$ is given in (\ref{kf22}). Note that since $\Delta_x>0$, and the minimization is over a compact set, i.e., $\mathcal{I}_0$, we have $\Delta_1>0$. For any $\Delta\leq\Delta_1$, we can write
\begin{align}
    I(X;U_M^\Delta)&=I(X,E_0;U_M^\Delta)\label{kf2}\\
    &=H(U_M^\Delta)-\textnormal{Pr}\{E_0=1\}H(U_M^\Delta|X,E_0=1)-\textnormal{Pr}\{E_0=0\}H(U_M^\Delta|X,E_0=0)\nonumber\\
    &\leq\log M - (1-\frac{\epsilon}{2\log M})(\log M-\frac{\epsilon}{2})\label{kf3}\\
    &=\frac{\epsilon}{2}(2-\frac{\epsilon}{2\log M})\nonumber\\
    &\leq\epsilon\nonumber,
\end{align}
where (\ref{kf2}) follows from having $E_0$ as a deterministic function of $X$; (\ref{kf3}) results from $H(U_M^\Delta)\leq\log M$, $\textnormal{Pr}\{E_0=1\}=1-\frac{\epsilon}{2\log M}$, (\ref{kf22}), and the non-negativity of entropy. Hence, it is shown that
for any $\epsilon,\delta>0$ and integer $M>0$, there exists $\Delta_2\triangleq\min\{\Delta_0,\Delta_1\}$, such that
 \begin{align*}
     I\left(X;U_M^{\Delta}\right)&\leq \epsilon,\\
     I\left(Y;U_M^{\Delta}\right)&\geq\log M - \delta,\  \forall\Delta\in(0,\Delta_2).
 \end{align*}
 Finally, by letting $M\to\infty$, the proof is completed. \footnote{{\color{black}As it can be observed, in the first method of showing $g_\epsilon(X,Y)=\infty,\ \forall\epsilon>0$ (i.e., equiprobable quantization), which is more complicated than the second one, we have $I(Y;U)=\log  M$, while in the second method (i.e., uniform quantization), $I(Y;U)$ approaches $\log  M$, since $I\left(Y;\lfloor\frac{MY}{\Delta}\rfloor\textnormal{ mod }M\right)<\log  M,\ \forall \Delta>0$ in general. This advantage of the first method is used in the proof of Remark \ref{remach}, i.e., replacing the supremum with maximum when $\mathcal{U}$ is a finite set. However, the second method has this advantage that it is simpler and has a fixed quantization interval.}}
 }
\end{proof}
{\color{black}\begin{remark}(\textbf{Generalization of $g_0(X,Y)=0.$})
It is important to note that in the first claim of (\ref{ginfinity}), i.e., $g_0(X,Y)=0$, Gaussianity of $Y$ is not used in particular, and the proof solely relies on the characteristics of the conditional pdf $f_{X|Y}(\cdot|\cdot)$. Therefore, for an arbitrary pair of random variables $(X,Y)$, in which $X = Y+N$, with $N$ being Gaussian and independent of $Y$, we have $g_0(X,Y)=0.$ This means that attaining perfect privacy of an additive Gaussian noisy version of any random variable comes at the cost of zero utility.
\end{remark}}
\begin{remark}\label{Rem3}
It is important to note that in the second claim of (\ref{ginfinity}), i.e., $g_\epsilon(X,Y)=\infty,\ \forall \epsilon>0$, Gaussianity of the pair $(X,Y)$ is not necessary. Hence, $g_\epsilon(X,Y)=\infty,\ \forall\epsilon>0$ for an arbitrary pair $(X,Y)$, where $Y$ conditioned on $\{X=x\}$, i.e., $Y|\{X=x\}$, admits a bounded smooth pdf for any $x\in\mathcal{X}$. 
\end{remark}
{\color{black}\begin{definition}
For a positive integer $M$, the utility-privacy trade-off with an $M$-ary release alphabet is defined as 
\begin{equation}\label{gme}
    g_\epsilon^M(X,Y)\triangleq \sup_{\substack{p_{U|Y}:|\mathcal{U}|\leq M\\I(X;U)\leq\epsilon\\X-Y-U}}I(Y;U),
\end{equation}
which is similar to (\ref{def}) with the addition of $|\mathcal{U}|\leq M$.
\end{definition}}
\begin{remark}
For any pair $(X,Y)$ that satisfies the condition in Remark \ref{Rem3}, from the proof of Theorem \ref{Guassiannotfeasible}, we get
\begin{equation}\label{nokn}
    g_\epsilon^M(X,Y)=\log  M,\ \forall\epsilon>0.
\end{equation}
Therefore, the fact that mutual information may not be a suitable measure of utility (or privacy\footnote{{\color{black}For example, if mutual information is only used as the privacy measure, and the total variation distance is used as the utility measure, i.e., $T(Y;U)\triangleq d_{\textnormal{TV}}(F_{Y,U},F_Y\cdot F_U)$, still maximum utility (which is 2 for TV distance) is achieved for a Gaussian pair $(X,Y)$. Hence, this phenomenon is not restricted to only the cases in which utility is measured by mutual information. This, however, is not the case when MMSE or the probability of error are used as the utility measure.}}) for the continuous alphabets scenarios is not just because the utility can be unbounded. Even constrained on a finite alphabet $\mathcal{U}$, it can reach its supremum of $\log |\mathcal{U}|$ for arbitrarily small leakage, rendering the term "trade-off" pointless. \footnote{In general, $g^M_\epsilon(X,Y)$ is not a continuous function of $\epsilon$, which results from two facts i) relative entropy is lower semi-continuous, and so is mutual information, ii) supremum of continuous functions is itself lower semi-continuous. Hence, having $g_0^M(X,Y)=0$, while $g_\epsilon^M(X,Y)=\log  M$ for $\epsilon>0$ is permissible. However, for the finite alphabet scenarios, all the probability simplices are compact, and it can be shown that $g_\epsilon$ is continuous.} {\color{black}Therefore, in order to fully capture the utility-privacy trade-off for continuous alphabets, either mutual information can be used in conjunction with certain imposed constraints (such as the restriction on the set of permissible $p_{U|Y}$ in (\ref{def}) as in \cite{Gfilter}), or a different measure needs to be adopted.}
\end{remark}
\begin{remark}\label{remach}
For an arbitrary pair $(X,Y)$, where $Y|\{X=x\}$ admits a bounded, and positive smooth pdf for any $x\in\mathcal{X}$, the supremum in (\ref{gme}) can be replaced by maximum\footnote{This holds in spite of the non-compactness of the search space.} for $\epsilon>0$ as
\begin{equation*}
    g_\epsilon^M(X,Y)= \max_{\substack{p_{U|Y}:|\mathcal{U}|\leq M\\I(X;U)\leq\epsilon\\X-Y-U}}I(Y;U)=\log  M, \ \forall \epsilon>0.
\end{equation*}
This follows similar steps as in the first method in the prrof of Theorem \ref{Guassiannotfeasible} with the modification of replacing $\Phi^{-1}(\cdot)$ with the inverse function of $F_Y(\cdot)$. 
\end{remark}
\begin{remark}
For an arbitrary pair $(X,Y)$, when $|\mathcal X|<\infty$ and $|\mathcal Y|=\infty$, we have $g_0(X,Y)=\infty$. This can be proved as follows. Let $\hat{Y}$ be an $M-$uniform (not necessarily deterministic) quantized version of $Y$ . If we take the privacy-preserving mapping from $\hat{Y}$ rather than $Y$, we have the Markov chain $X-Y-\hat{Y}-U$. Hence, $g_0(X,Y)\geq g_0(X,\hat{Y})$. From Corollary \ref{cor22}, we get the lower bound of $(\log M-\log \min\{M,|\mathcal{X}|\})^+$, which tends to infinity with $M$. Hence, $g_0(X,Y)=\infty$.
\end{remark}
\section{Non-private information vs. $g_0(X,Y)$}\label{S-Non-private}
For a pair of random variables $(X,Y)\in \mathcal X\times\mathcal Y$, the \textit{private information about $X$ carried by $Y$} is defined in \cite{Shahab1} as
\begin{equation}\label{pr}
C_X(Y)\triangleq \min_{\substack{W:X-W-Y,\\H(W|Y)=0}}H(W).
\end{equation}
Since $H(W|Y)=0$ implies that $W$ is a deterministic function of $Y$, (\ref{pr}) means that among all the functions of $Y$ that make $X$ and $Y$ conditionally independent, we want to find the one with the lowest entropy. It can be verified that $I(X;Y)\leq C_X(Y)\leq H(Y)$, 
where the first inequality is due to the data processing inequality applied on the Markov chain $X-W-Y$, i.e., $I(W;Y)\geq I(X;Y)$, and the second inequality is a direct result of the fact that $W=Y$ satisfies the constraints in (\ref{pr}).

The \textit{non-private information about $X$ carried by $Y$} is defined in \cite{Shahab1} as
\begin{equation}
D_X(Y)\triangleq H(Y)-C_X(Y).
\end{equation}
Let $T^{\mathcal{X}}:\mathcal{Y}\to\mathcal{P}(\mathcal{X})$ be a mapping from $\mathcal{Y}$ to the probability simplex on $\mathcal{X}$ defined by $y\to p_{X|Y}(\cdot|y)$. It was shown in \cite[Theorem 3]{Shahab1} that the minimizer in (\ref{pr}) is $W^*=T^{\mathcal{X}}(Y)$; and hence,
\begin{equation}\label{sinoh}
D_X(Y)=H(Y)-H(T^{\mathcal{X}}(Y)).
\end{equation}
Furthermore, it was proved in \cite[lemma 5]{Shahab1} that $C_X(Y)=H(Y)$, i.e., $D_X(Y)=0$, if and only if there do not exist $y_1,y_2\in\mathcal{Y}$ such that $p_{X|Y}(\cdot|y_1)=p_{X|Y}(\cdot|y_2)$. 

In \cite{Shahab1}, three examples were provided, where in two of them $g_0(X,Y)=D_X(Y)$, while in the last one $g_0(X,Y)>D_X(Y)$. Finally, a question was raised regarding the condition on the joint distribution $p_{X,Y}$ under which $g_0(X,Y)=D_X(Y)$ holds. In Theorem \ref{Shahabconnection}, we characterize the relation between $D_X(Y)$ and $g_0(X,Y)$. To this end, some preliminaries and two lemmas are needed, as explained in the sequel.

If $\mathbf{P}_{X|Y}$ has at least two identical columns, we define $\hat{\mathbf{P}}_{X|Y}$ as follows\footnote{If this is not the case, let $\hat{\mathbf{P}}_{X|Y}\triangleq\mathbf{P}_{X|Y}$.}. Let $\mathcal{E}_m\subset [|\mathcal{Y}|],\forall m\in[B]$, for some integer $B\geq 1$, be a set of indices corresponding to the columns in $\mathbf{P}_{X|Y}$ that are equal, i.e.,
$\mathbf{p}_{X|y_i}=\mathbf{p}_{X|y_j},\ \forall i,j\in \mathcal{E}_m, \forall m\in[B]$, and $
\mathbf{p}_{X|y_i}\neq \mathbf{p}_{X|y_k},\ \forall i\in \mathcal{E}_m,\forall k\in [|\mathcal{Y}|]\backslash \mathcal{E}_m,\ \forall m\in[B].$
Let $G\triangleq \sum_{i=1}^B|\mathcal{E}_i|$. We construct a corresponding $|\mathcal{X}|\times (|\mathcal{Y}|-G+B)$-dimensional matrix $\hat{\mathbf{P}}_{X|Y}$ from $\mathbf{P}_{X|Y}$ by eliminating all the columns in each $\mathcal{E}_{m}$, except one. For example, we have the following pair
\begin{equation}\label{mesal}
\mathbf{P}_{X|Y}=\begin{bmatrix}
0.3&0.3&0.4&0.5&0.4\\
0.2&0.2&0.5&0.5&0.5\\
0.5&0.5&0.1&0&0.1
\end{bmatrix}\ \ \textnormal{and}\ \ \hat{\mathbf{P}}_{X|Y}=\begin{bmatrix}
0.3&0.4&0.5\\
0.2&0.5&0.5\\
0.5&0.1&0
\end{bmatrix},
\end{equation}
where $B=2$, $G=4$, $\mathcal{E}_1=\{1,2\}$, and $\mathcal{E}_2=\{3,5\}$.

Since $\mathbf{p}_{X|y_i}=\mathbf{p}_{X|y_j},\forall i,j\in \mathcal{E}_m,\forall m\in[B]$, we have $T^{\mathcal{X}}(y_i)=T^{\mathcal{X}}(y_j),\forall i,j\in \mathcal{E}_m,\forall m\in[B]$.
Hence, $T^{\mathcal{X}}(Y)$ is a random variable whose support has the cardinality $|\mathcal{Y}|-G+B$ and whose mass probabilities are the elements of the following set
\begin{equation}\label{mass}
\bigg\{\sum_{i\in \mathcal{E}_1}p_Y(y_i),\sum_{i\in \mathcal{E}_2}p_Y(y_i),\ldots,\sum_{i\in \mathcal{E}_B}p_Y(y_i)\bigg\}\cup\bigg\{p_Y(y_i)\bigg|i\not\in\cup_{m=1}^B\mathcal{E}_m\bigg\}.
\end{equation}
Let $\mathbb{S}_{X,Y}'$ be a set of $\prod_{i=1}^B|\mathcal{E}_i|$ probability vectors in the simplex $\mathcal{P}(\mathcal{Y})$ given by
\begin{equation}\label{preem}
\mathbb{S}_{X,Y}'=\bigg\{\mathbf{s}_{m_{[B]}}\bigg|\forall m_{[B]}\in\prod_{i=1}^B\mathcal{E}_i\bigg\},
\end{equation}
where the tuple $(m_1,m_2,\ldots,m_B)$ is written in short as $m_{[B]}$ and the probability vectors $\mathbf{s}_{m_{[B]}}$ are defined element-wise as
\begin{equation}\label{kor}
s_{m_{[B]}}(k)=\left\{\begin{array}{ccc}
\sum_{t\in \mathcal{E}_{i}} p_Y(y_t)&k=m_i, i\in[B]\\p_Y(y_k)&k\not\in\cup_{i=1}^B\mathcal{E}_i\\0&\textnormal{otherwise}
\end{array}\right.,\forall k\in[|\mathcal{Y}|],\ \forall m_{[B]}\in\prod_{i=1}^B\mathcal{E}_i.
\end{equation}

\begin{lemma}\label{Prop6}
For the set $\mathbb{S}_{X,Y}'$ in (\ref{preem}) and the set $\mathbb{S}_{X,Y}$ in (\ref{poly}), we have
$\mathbb{S}_{X,Y}'\subseteq\mathbb{S}_{X,Y}$  and $H(\mathbf{s})=H(T^{\mathcal{X}}
(Y)),\ \forall \mathbf{s}\in\mathbb{S}_{X,Y}'.
$
Furthermore, the probability vector $\mathbf{p}_Y$ can be written as a convex combination of the points in $\mathbb{S}_{X,Y}'$, i.e.
\begin{equation}\label{lin0}
\mathbf{p}_Y=\sum_{m_{[B]}\in\prod_{i=1}^B \mathcal{E}_i}\alpha_{m_{[B]}}\mathbf{s}_{m_{[B]}},
\end{equation}
where $\alpha_{m_{[B]}}\geq 0,\forall m_{[B]}\in\prod_{i=1}^B\mathcal{E}_i$ and $\sum_{m_{[B]}\in\prod_{i=1}^B \mathcal{E}_i}\alpha_{m_{[B]}}=1.$
\end{lemma}
\begin{proof}
The proof is provided in Appendix \ref{ppo}.
\end{proof}
For example, assume that in the example in (\ref{mesal}), we have $\mathbf{p}_Y=\begin{bmatrix}
0.1&0.2&0.15&0.25&0.3
\end{bmatrix}^T$. We can write $\mathbf{p}_Y=\frac{1}{9}\mathbf{s}_{1,3}+\frac{2}{9}\mathbf{s}_{1,5}+\frac{2}{9}\mathbf{s}_{2,3}+\frac{4}{9}\mathbf{s}_{2,5}$, where $\mathbf{s}_{1,3}=\begin{bmatrix}
0.3&0&0.45&0.25&0
\end{bmatrix}^T$, $\mathbf{s}_{1,5}=\begin{bmatrix}
0.3&0&0&0.25&0.45
\end{bmatrix}^T$, $\mathbf{s}_{2,3}=\begin{bmatrix}
0&0.3&0.45&0.25&0
\end{bmatrix}^T$, and $\mathbf{s}_{2,5}=\begin{bmatrix}
0&0.3&0&0.25&0.45
\end{bmatrix}^T$. 
\begin{lemma}\label{Prop7}
{\color{black}If $\textnormal{nul}(\hat{\mathbf{P}}_{X|Y})=0$, we have $\textnormal{ext}(\mathbb{S}_{X,Y})=\mathbb{S}_{X,Y}'$.} Otherwise, none of the elements in $\mathbb{S}_{X,Y}'$ belongs to $\textnormal{ext}(\mathbb{S}_{X,Y})$, where $\textnormal{ext}(\mathbb{S}_{X,Y})$ denotes the set of extreme points of $\mathbb{S}_{X,Y}$.
\end{lemma}
\begin{proof}
The proof is provided in Appendix \ref{appE}.
\end{proof}
\begin{theorem}\label{Shahabconnection}
For a pair of random variables $(X,Y)\in\mathcal{X}\times\mathcal{Y}$, we have
\begin{equation}\label{th}
g_0(X,Y)\geq D_X(Y),
\end{equation} 
where the equality holds if and only if either of the following holds:
\begin{enumerate}
\item Perfect privacy is not feasible, i.e., $\textnormal{nul}({\mathbf{P}}_{X|Y})=0$,
\item Perfect privacy is feasible, and $\textnormal{nul}(\hat{\mathbf{P}}_{X|Y})=0$. 
\end{enumerate}
\end{theorem}
\begin{proof}
{\color{black}The proof of the inequality in (\ref{th}) is as follows.}
It is obvious that when there exist no $y_1,y_2\in\mathcal{Y}$ such that $\mathbf{p}_{X|Y}(\cdot|y_1)=\mathbf{p}_{X|Y}(\cdot|y_2)$, we have $D_X(Y)=0$, and (\ref{th}) holds from the non-negativity of $g_0(X,Y)$. Assume that there exist index sets $\mathcal{E}_m,\forall m\in[B]$, corresponding to equal columns of $\mathbf{P}_{X|Y}$, as defined before. We can write
\begin{align}
g_0(X,Y)
&=H(Y)-\min_{\substack{F_U(\cdot),\ \mathbf{p}_{Y|u}\in\mathbb{S}_{X,Y},\ \forall u\in\mathcal{U}:\\ \int_{\mathcal{U}} \mathbf{p}_{Y|u}dF(u)=\mathbf{p}_Y}}H(Y|U)\label{pr1}\\
&\geq H(Y) -\!\!\!\!\!\!\!\!\! \sum_{m_{[B]}\in\prod_{i=1}^B \mathcal{E}_i}\alpha_{m_{[B]}}H(\mathbf{s}_{m_{[B]}})\label{pr2}\\
&=H(Y) -\!\!\!\!\!\!\!\!\! \sum_{m_{[B]}\in\prod_{i=1}^B \mathcal{E}_i}\alpha_{m_{[B]}}H(T^{\mathcal{X}}(Y))\label{pr3}\\
&=H(Y)-H(T^{\mathcal{X}}(Y))\nonumber\\
&=D_X(Y),\label{pr5}
\end{align}
where (\ref{pr1}) is from (\ref{min}); (\ref{pr2}) is justified as follows.
According to lemma \ref{Prop6}, $\mathbb{S}_{X,Y}'\subseteq\mathbb{S}_{X,Y}$, and $\mathbf{p}_Y$ is preserved from (\ref{lin0}). Hence, the vectors in $\mathbb{S}_{X,Y}'$ belong to the constraint of the minimization in (\ref{pr1}), and the inequality follows. (\ref{pr3}) is from lemma \ref{Prop6}, and (\ref{pr5}) is due to (\ref{sinoh}). This proves the inequality (\ref{th}).

{\color{black}The proof of the sufficient conditions for the equality in (\ref{th}) is as follows.}
If $\textnormal{nul}(\hat{\mathbf{P}}_{X|Y})=0$, from lemma \ref{Prop7}, we can say that for any vector $\mathbf{s}$ that is an extreme point of $\mathbb{S}_{X,Y}$, we have $H(\mathbf{s})=H(T^{\mathcal{X}}(Y))$, which means that
\begin{equation*}
\min_{\substack{F_U(\cdot),\ \mathbf{p}_{Y|u}\in\mathbb{S}_{X,Y},\ \forall u\in\mathcal{U}:\\ \int_{\mathcal{U}} \mathbf{p}_{Y|u}dF(u)=\mathbf{p}_Y}}H(Y|U)=H(T^{\mathcal{X}}(Y)).
\end{equation*} 
This is equivalent to $g_0(X,Y)=D_X(Y)$, from (\ref{min}) and (\ref{sinoh}).

{\color{black}The proof of the necessary conditions for the equality in (\ref{th}) is as follows.}
Assume that $g_0(X,Y)=D_X(Y).$ If $g_0(X,Y)=0$, we have that perfect privacy is not feasible and the proof is complete. However, if $g_0(X,Y)>0$, we must have $D_X(Y)>0$, according to our assumption of $g_0(X,Y)=D_X(Y).$ In this case, as in \cite{Shahab1}, there must exist index sets $\mathcal{E}_m,\forall m\in[B]$, corresponding to equal columns of $\mathbf{P}_{X|Y}$. We prove that $\textnormal{nul}(\hat{\mathbf{P}}_{X|Y})=0$ by contradiction. Assume that $\textnormal{nul}(\hat{\mathbf{P}}_{X|Y})\neq 0$. From Proposition \ref{Prop7}, we conclude that none of the elements in $\mathbb{S}_{X,Y}'$ is an extreme point of $\mathbb{S}_{X,Y}$. In other words, for any $\mathbf{s}$ in $\mathbb{S}_{X,Y}'$, {\color{black}which is also a member of $\mathbb{S}_{X,Y}$ according to lemma \ref{Prop6}}, we can find the triplet $(\mathbf{s}',\mathbf{s}'',\beta)$, such that $\mathbf{s}=\beta\mathbf{s}'+(1-\beta)\mathbf{s}''$, 
where $\mathbf{s}',\mathbf{s}''\in\mathbb{S}_{X,Y}$ ($\mathbf{s}'\neq\mathbf{s}''$) and $\beta\in(0,1)$. Therefore,
\begin{align}
H(T^{\mathcal{X}}(Y))&=\sum_{m_{[B]}\in\prod_{i=1}^B \mathcal{E}_i}\alpha_{m_{[B]}}H(\mathbf{s}_{m_{[B]}})\nonumber\\
&=\sum_{m_{[B]}\in\prod_{i=1}^B \mathcal{E}_i}\alpha_{m_{[B]}}H\bigg(\beta_{m_{[B]}}\mathbf{s}'_{m_{[B]}}+(1-\beta_{m_{[B]}})\mathbf{s}''_{m_{[B]}}\bigg)\nonumber\\
&>\sum_{m_{[B]}\in\prod_{i=1}^B \mathcal{E}_i}\beta_{m_{[B]}}\alpha_{m_{[B]}}H(\mathbf{s}'_{m_{[B]}})+\sum_{m_{[B]}\in\prod_{i=1}^B \mathcal{E}_i}(1-\beta_{m_{[B]}})\alpha_{m_{[B]}}H(\mathbf{s}''_{m_{[B]}})\label{konca}\\
&\geq\min_{\substack{F_U(\cdot),\ \mathbf{p}_{Y|u}\in\mathbb{S}_{X,Y},\ \forall u\in\mathcal{U}:\\ \int_{\mathcal{U}} \mathbf{p}_{Y|u}dF(u)=\mathbf{p}_Y}}H(Y|U)\label{ol},
\end{align}  
where (\ref{konca}) is due to the strict concavity of the entropy; (\ref{ol}) comes from the fact that $\mathbf{s}'_{m_{[B]}}$ and $\mathbf{s}''_{m_{[B]}}$ with corresponding mass probabilities $\beta_{m_{[B]}}\alpha_{m_{[B]}}$ and $(1-\beta_{m_{[B]}})\alpha_{m_{[B]}},\ \forall m_{[B]}\in\prod_{i=1}^B\mathcal{E}_i$, belong to the constraints of minimization in (\ref{ol}). This results in $g_0(X,Y)>D_X(Y)$, which is a contradiction. Hence, we must have $\textnormal{nul}(\hat{\mathbf{P}}_{X|Y})=0$. 

\end{proof}

\section{Full data observation model}\label{S-Full data}
In this section, we assume that the curator has access to both $X$ and $Y$, and investigate $G_\epsilon(X,Y)$, as defined in (\ref{Geps}), at $\epsilon = 0.$ We start with a lemma, which is used in the sequel.

{\color{black}
\begin{lemma}\label{lemcardi}
In the evaluation of $G_0(X,Y)$, for any $u\in\mathcal{U}$, we have
\begin{equation}
    |\{y\in\mathcal{Y}|p(x,y|u)>0\}|=1,\ \forall x \in\mathcal{X}.
\end{equation}
\end{lemma}
\begin{proof}
The proof is provided in Appendix \ref{AppendixCC}.
\end{proof}
Define the support of a given pair $(X,Y)$ as 
\begin{equation*}
    \textnormal{supp}(X,Y)\triangleq\bigg\{(x,y)\in\mathcal{X}\times\mathcal{Y}\bigg|p_{X,Y}(x,y)>0\bigg\}.
\end{equation*}
\begin{proposition}\label{G0bound}
In the evaluation of $G_0(X,Y)$, we must have
\begin{equation}\label{upperU}
  \max_x|\{y\in\mathcal{Y}|p(y|x)>0\}|\leq|\mathcal{U}|\leq|\textnormal{supp}(X,Y)|-|\mathcal{X}|+1,   
\end{equation}
where the first and second inequalities are necessary and sufficient conditions, respectively.
\end{proposition}
\begin{proof}
The proof is provided in Appendix \ref{AppendixDD}.
\end{proof}
}
\begin{theorem}\label{G0feas}
Perfect privacy is feasible in the full data observation model, i.e., $G_0(X,Y)>0$, if and only if $Y$ is not a deterministic function of $X$. 
\end{theorem} 
\begin{proof}
If $Y$ is a deterministic function of $X$, we have $Y-X-U$ form a Markov chain. From data processing inequality, $I(X;U)=0$ results in $I(Y;U)=0$. This proves the first direction of the theorem.

For the second direction, we proceed as follows. If $Y$ is not a deterministic function of $X$, there must exist $x_1\in\mathcal{X}$ and $y_1,y_2\in\mathcal{Y}$ ($y_1\neq y_2$) such that $p_{X,Y}(x_1,y_1)>0$ and $p_{X,Y}(x_1,y_2)>0$. Let $\mathcal{U}=\{u_1,u_2\}$ and $p_U(u_1)=\frac{1}{2}$. Choose a sufficiently small $\epsilon>0$ and let
\begin{align}
p_{X,Y|U}(x,y|u_1)&=\left\{\begin{array}{ccc}
p_{X,Y}(x_1,y_1)+\epsilon& (x,y)=(x_1,y_1)\\p_{X,Y}(x_1,y_2)-\epsilon& (x,y)=(x_1,y_2)\\p_{X,Y}(x,y)&\textnormal{otherwise}
\end{array}\right.,\nonumber\\
p_{X,Y|U}(x,y|u_2)&=2p_{X,Y}(x,y)-p_{X,Y|U}(x,y|u_1),\ \forall (x,y)\in\mathcal{X}\times\mathcal{Y}.\nonumber
\end{align}
It can be verified that $p_{X,Y}$ is preserved in $p_{X,Y,U}$. Also, $p_{X|U}(\cdot|u)=p_X(\cdot),\ \forall u\in\mathcal{U}$, and $p_{Y|U}(y_1|u_1)\neq p_Y(y_1)$, where the former indicates that $X\independent U$, and the latter shows that $Y\not\independent U$.

{\color{black} In the light of Proposition \ref{G0bound}, an alternative proof for this Theorem is provided as follows. If $Y$ is a deterministic function of $X$, we have $|\textnormal{supp}(X,Y)|=|\mathcal{X}|$, which results in $|\mathcal{U}|\leq 1$ according to Proposition \ref{G0bound}, which in turn results in $G_0(X,Y)=0$. If $Y$ is not a deterministic function of $X$, we have $|\{y\in\mathcal{Y}|p(y|x)>0\}|\geq 2$ for some $x\in\mathcal{X}$, which results in the necessity of having $|\mathcal{U}|\geq 2$ according to Proposition \ref{G0bound}, which in turn results in $G_0(X,Y)>0$ (since otherwise, $|\mathcal{U}|=1$ is sufficient, and $|\mathcal{U}|\geq 2$ is not necessary, which is a contradiction).}
\end{proof}

%
In what follows, a lower bound is provided for the utility of the full data observation model. Prior to that, a quantity, which is used in the sequel, needs to be defined and investigated.
\begin{definition}
For a given pair $(X,Y)$, define the mapping $J:\mathcal{X}\times\textnormal{int}(\mathcal{P}(\mathcal{Y}))\longrightarrow(0,1]$ as
\begin{equation}\label{Jdef}
    J(x,q_Y)\triangleq \frac{1}{\max_{y\in\mathcal{Y}}\frac{p_{Y|X}(y|x)}{q_Y(y)}}.
\end{equation}
\end{definition}
Therefore, we have that $J(x,q_Y)=q_Y(x)$, if $X=Y$, and $J(x,q_Y)=\min_y \frac{q_Y(y)}{p_Y(y)}$, if $X\independent Y$. {\color{black}The following Proposition relates the above quantity to the \textit{maximal leakage} from $Y$ to $X$ defined as (\cite{Issa})
\begin{equation}\label{maxl}
\mathcal{L}(Y\to X)\triangleq\log\sum_{x\in\mathcal{X}}\max_{\substack{y\in\mathcal{Y}\\p_Y(y)>0}}p_{X|Y}(x|y).
\end{equation}
\begin{proposition}
For a given pair $(X,Y)$, we have
\begin{equation}\label{Jlowerb}
    \mathds{E}_X[J(X,p_Y)]\geq 2^{-\mathcal{L}(Y\to X)},
\end{equation}
where equality holds if and only if $\max_{y\in\mathcal{Y}}\frac{p_{Y|X}(y|x)}{p_Y(y)}$ does not vary with $x\in\mathcal{X}$, which includes the special cases of i) $X\independent Y$, and ii) $X=Y$ and uniformly distributed.
\end{proposition}
\begin{proof}
The proof is provided in Appendix \ref{AppendixEE}.
\end{proof}}
\begin{theorem}\label{thbound5}
For a given pair $(X,Y)\in\mathcal{X}\times\mathcal{Y}$, we have
\begin{equation}\label{G0lowerbound}
   G_0(X,Y)\geq \left(\mathds{E}_X[J(X,q^*)]\log |\mathcal{Y}|-1\right)^+,
\end{equation}
where $q^*$ denotes the uniform pmf over $\mathcal{Y}$.
\end{theorem}
\begin{proof}
{\color{black}Fix an arbitrary pmf in the interior of $\mathcal{P}(\mathcal{Y})$ and denote it by $q^*$. A privacy-preserving mapping $p_{U|X,Y}$ is designed such that the conditional pmf of $U$ conditioned on $\{X=x\}$ is the same as $q^*$ for any $x\in\mathcal{X}$. Therefore, for this privacy-preserving mapping, we have that $X\independent U$, and the resulting $I(Y;U)$ serves as a lower bound on $G_0(X,Y)$. The only reason for selecting $q^*$ as the uniform pmf over $\mathcal{Y}$ is that its corresponding $I(Y;U)$ can be further lower bounded in a closed form way.} 

Let $\Theta_X\in\{0,1\}$ be a Bernoulli r.v., parametrized by $X$, with $\textnormal{Pr}\{\Theta_X=1|X=x\}=J(x,q^*),\ \forall x\in\mathcal{X}.$ The privacy-preserving mapping is designed as $U\triangleq \Theta_X Y+(1-\Theta_X)\tilde{Y}_X$, where for each $x\in\mathcal{X}$, {\color{black} $\tilde{Y}_x$ is an r.v. over $\mathcal{Y}$, which is distributed according to $\frac{q^*(\cdot)-J(x,q^*)p_{Y|X}(\cdot|x)}{1-J(x,q^*)}$, when $J(x,q^*)<1$, and arbitrarily distributed when $J(x,q^*)=1$.}\footnote{{\color{black}Note that when $J(x,q^*)=1$, we have $\Theta_x=1$, and therefore, the coefficient of $\tilde{Y}_x$ in $U$ becomes zero.}} {\color{black}Conditioned on each realization $x\in\mathcal{X}$, we have that $p_{U|X}(\cdot|x)$ is a convex combination of $p_{Y|X}(\cdot|x)$ and $\frac{q^*(\cdot)-J(x,q^*)p_{Y|X}(\cdot|x)}{1-J(x,q^*)}$ with weights $J(x,q^*)$, and $1-J(x,q^*)$, respectively.} Hence, $U$ conditioned on $\{X=x\}$, i.e., $U|\{X=x\}$, is distributed according to $q^*$ for each $x\in\mathcal{X}$, and therefore, $X\independent U$. Since $q^*$ is an arbitrary point in $\textnormal{int}(\mathcal{P}(\mathcal{Y}))$, we have
\begin{align}
    G_0(X,Y)&\geq \max_{q^*}I(Y;U)\nonumber\\
    &=\max_{q^*} \{H(U)-H(U|Y)\}\nonumber\\
    &\geq \max_{q^*}  \{H(q^*)-H(U|Y,\Theta_X)-H(\Theta_X|Y)\}\nonumber\\
    &\geq \max_{q^*} \{H(q^*)-H(U|Y,\Theta_X)\}-1\label{tbi}\\
    &=\max_{q^*} \{H(q^*)-H(\Theta_X Y+(1-\Theta_X)\tilde{Y}_X|Y,\Theta_X)\}-1\nonumber\\
    &=\max_{q^*} \{H(q^*)-\textnormal{Pr}\{\Theta_X=1\}H(\tilde{Y}_X|Y,\Theta_X=0)\}-1\label{tbtb}\\
    &\geq \textnormal{Pr}\{\Theta_X=1\}\log |\mathcal{Y}|-1\label{tb2}\\
    &=\mathds{E}_X[J(X,q^*)]\log |\mathcal{Y}|-1,\nonumber
\end{align}
where (\ref{tbi}) follows from $\Theta_X$ being binary, and (\ref{tbtb}) results from $H(U|Y,\Theta_X=1)=0$. In (\ref{tb2}), we pick the uniform $q^*$, and use the fact that $H(\tilde{Y}_X|Y,\Theta_X=0)\leq\log |\mathcal{Y}|.$ \footnote{Note that if $\tilde{Y}_x$ and $Y$ are not independent, the bound $H(\tilde{Y}_X|Y,\Theta_X=0)\leq\log |\mathcal{Y}|$ can be further tightened. This, in turn, calls for an algorithmic approach to this problem that aims to maximize $I(Y;\tilde{Y}_x)$ over the joint distribution for fixed marginals.}

\end{proof}

Considering the output perturbation model, Theorem \ref{Guassiannotfeasible} proved that perfect privacy is not feasible for the (correlated) jointly Gaussian pair. This is not the case in the full data observation model.

Consider a jointly Gaussian pair $(X,Y)$ with correlation coefficient $\rho\in(0,1)$. 
Denoting the variances of $X$ and $Y$ by $\sigma_X^2$ and $\sigma_Y^2$, respectively, it is already known that we can write
\begin{equation}\label{usss}
Y=\frac{\rho\sigma_Y}{\sigma_X}X+\sigma_Y\sqrt{1-\rho^2}N,
\end{equation}
where $N\sim\mathcal{N}(0,1)$ is independent of $X$. At a first glance, by letting $U\triangleq\frac{1}{\sigma_Y\sqrt{1-\rho^2}}(Y-\frac{\rho\sigma_Y}{\sigma_X}X)$, i.e. $U\triangleq N$, we have $X\independent U$, and 
\begin{align}
I(Y;U)&=h(Y)-h(Y|U)\nonumber\\
&=\frac{1}{2}\log 2\pi e\sigma_Y^2-h\left(\frac{\rho\sigma_Y}{\sigma_X}X+\sigma_Y\sqrt{1-\rho^2}N\bigg|N\right)\label{suuu}
\end{align}
\begin{align}
&= \frac{1}{2}\log 2\pi e\sigma_Y^2 - h(\frac{\rho\sigma_Y}{\sigma_X}X)\label{su2}\\
&=\frac{1}{2}\log 2\pi e\sigma_Y^2 -\frac{1}{2}\log 2\pi e\rho^2\sigma_Y^2\nonumber \\
&=-\log \rho,\label{rho}
\end{align}
which provides a (non-zero) lower bound on $G_0(X,Y)$ meaning that perfect privacy is feasible in this case. A natural question arises whether this bound is tight. By using the following Theorem, which includes a broad range of joint distributions on $\mathcal{X}\times\mathcal{Y}$, we show that $G_0(X,Y)$ is actually unbounded, which is stated in corollary \ref{co5}.

\begin{theorem}\label{densityuniform}
For the class of additive noise, i.e., when $Y=X+N$ (where $N$ is not necessarily independent of $X$), if there exists $\alpha,\beta\in\mathds{R}$, and $\Delta>0$ such that $\mathcal{I}\triangleq(\mathcal{X}\cap[\alpha,\alpha+\Delta])\times[\beta,\beta+\Delta]\subset\mathcal{X}\times\mathcal{N}$, and $N|\{X=x\}$ admits a bounded smooth density over $[\beta,\beta+\Delta]$ for each $x\in\mathcal{X}\cap[\alpha,\alpha+\Delta]$, we have
\begin{equation}\label{Ginf}
    G_0(X,Y) = \infty.
\end{equation}
\end{theorem}
\begin{proof}
{\color{black} The sketch of the proof is as follows. A privacy-preserving mapping $p_{U|X,Y}$ is designed whose output, i.e., $U$ is independent of $X$. Therefore, we conclude that $G_0(X,Y)$ is lower bounded by the utility, i.e., $I(Y;U)$, of this privacy-preserving mapping. By showing that the latter can grow unboundedly, the proof of (\ref{Ginf}) is complete.}

Define the Bernoulli r.v. $E$, which is $1$ when $(X,N)\in\mathcal{I}$, and $0$ elsewhere. Define
\begin{equation}\label{fstar}
f^*_x\triangleq \max_{n\in[\beta,\beta+\Delta]}f_{N|X,E}(n|x,1),\ \forall x\in\mathcal{X}\cap[\alpha,\alpha+\Delta].
\end{equation}
which is defined by the assumption of having a bounded $f_{N|X,E}(\cdot|x,1)$ over $[\beta,\beta+\Delta]$ for each $x\in\mathcal{X}\cap[\alpha,\alpha+\Delta]$. Moreover, we have $f^*_x\geq \frac{1}{\Delta}$, since otherwise $f_{N|X,E}(\cdot|x,1)$ will not integrate to $1$ over its support, i.e., $[\beta,\beta+\Delta].$ {\color{black}Also, we have $f^*_x= \frac{1}{\Delta}$ if and only if $f_{N|X,E}(\cdot|x,1)$ is uniform.

Let $\tilde{N}_x$ be a continuous r.v., independent of $(X,N)$, with the following density
\begin{equation*}
    f_{\tilde{N}_x}(t)=\frac{f^*_x-f_{N|X,E}(t|x,1)}{\Delta f^*_x -1},\ \forall (x,t)\in(\mathcal{X}\cap[\alpha,\alpha+\Delta])\times[\beta,\beta+\Delta],
\end{equation*}
when  $f^*_x> \frac{1}{\Delta}$, and arbitrarily distributed when $f^*_x= \frac{1}{\Delta}$.}

Define the Bernoulli r.v. $\Theta_X\in\{0,1\}$, with $\textnormal{Pr}\{\Theta_X=1|X=x\}=\frac{1}{\Delta f^*_x}$, and set $R\triangleq\Theta_X N+(1-\Theta_X)\tilde{N}_X$. We have $R\sim\textnormal{Uniform}[\beta,\beta+\Delta]$, since for each $x\in\mathcal{X}\cap[\alpha,\alpha+\Delta]$, the pdf of $R$, which is a convex combination of $f_{N|X,E}(\cdot|x,1)$ and $f_{\tilde{N}_x}(\cdot)$, with the corresponding weights of $\frac{1}{\Delta f^*_x}$ and $1-\frac{1}{\Delta f^*_x}$, is equal to $\frac{1}{\Delta}$ over $[\beta,\beta+\Delta]$.

Let $M$ be an arbitrary positive integer and set
\begin{equation}
   U\triangleq E\left(  \bigg\lfloor\frac{M(X+R)}{\Delta}\bigg\rfloor \textnormal{ mod } M\right) + (1-E)\tilde{U},
\end{equation}
where $\tilde{U}$ is a uniform pmf over $[0:M-1]$.

With some simple calculations, it can be verified that the conditional distribution of $U$ conditioned on $\{X=x\}$ remains uniform over $[0:M-1]$ for any realization $x\in\mathcal{X}$.\footnote{The fact that $U|\{X=x,E=1\}$ is uniform over $[0:M-1]$ is immediate from noting that $\left(\lfloor \frac{MA}{\Delta}\rfloor\textnormal{ mod } M\right)$ is uniform if $A$ is uniform. The uniform distribution of $U|\{X=x,E=0\}$ is immediate from construction.} Hence, $U\independent X$.

We can write
\begin{align}
    I(X+N;U)&=I(X+N,\Theta_X;U|E)+I(E;U)-I(E;U|X+N,\Theta_X)-I(\Theta_X;U|X+N)\nonumber\\
    &\geq I(X+N,\Theta_X;U|E)-2\label{EBernouli}\\
    &\geq p_E(1)I(X+N,\Theta_X;U|E=1)-2\label{MP}\\
    &=p_E(1)\bigg(H(U|E=1)-H(U|X+N,\Theta_X,E=1)\bigg)-2\nonumber\\
    &\geq p_E(1)\textnormal{Pr}\{\Theta_X=1\}\bigg(\log M-H(U|X+N,\Theta_X=1,E=1)\bigg)-2\label{finf0}\\
    &\geq p_E(1)\textnormal{Pr}\{\Theta_X=1\}\log  M -2\label{finf},
\end{align}
where (\ref{EBernouli}), and (\ref{MP}) follow, respectively, from the facts that $E,\Theta_X$ are binary (having a maximum entropy of $1$), and mutual information is non-negative; (\ref{finf0}) and (\ref{finf}) result, respectively, from having $H(U|X+N,\Theta_X=0,E=1)\leq\log M$, and $H(U|X+N,\Theta_X=1,E=1) =0$.

Finally, as $\textnormal{Pr}\{\Theta_X=1\}=\int_{x\in\mathcal{X}\cap[\alpha,\alpha_\Delta]}\frac{1}{\Delta f^*_x}dF_X(x)>0$, by letting $M\to\infty$ in (\ref{finf}), (\ref{Ginf}) is proved \footnote{Note that the set of all mappings $p_{U|\{X,N\}}$ can be put into a one-to-one correspondence with the set of all mappings $q_{U|\{X,X+N\}}$.}.

\end{proof}
\begin{corollary}\label{co5}
For the jointly Gaussian pair $(X,Y)$, $G_0(X,Y)$ is unbounded. 
\end{corollary}

Since $G_\epsilon(X,Y), \forall\epsilon\geq 0$ is unbounded for a jointly Gaussain pair, a natural question arises to investigate the optimal trade-off when the released data has a finite alphabet.
Let $G_\epsilon^M(X,Y)$ be as in (\ref{Geps}) with an added constraint of $|\mathcal{U}|\leq M$. From definition, we have $G_\epsilon^M(X,Y)\geq g_\epsilon^M(X,Y)$. Hence, from Remark \ref{Rem3}, we get $G_\epsilon^M(X,Y)=\log  M,\ \forall\epsilon>0$ for a jointly Gaussian $(X,Y)$; however, the exact value of $G_0^M(X,Y)$ is not known.
\begin{conjecture} 
For a jointly Gaussian pair $(X,Y)$, we have $G_0^M(X,Y)<\log  M$.\footnote{This, however, is not true in general. Consider a pair $(X,Y)$, for which we have a collection of disjoint subsets of $\mathcal{Y}$, denoted as $\mathcal{Y}_1,\mathcal{Y}_2,\ldots,\mathcal{Y}_M$, that satisfy $\textnormal{Pr}\{Y\in\mathcal{Y}_i|X=x\}=\frac{1}{M},\ \forall (x,i)\in\mathcal{X}\times[M]$. We have $G_0^M(X,Y)=\log  M$, and $U\triangleq \sum_{i=1}^Mi\cdot\mathds{1}_{y\in\mathcal{Y}_i}$ achieves it.}.
\end{conjecture}

\section{Asymptotic analysis}\label{asymp}
In the previous sections, we have mainly focused on one extreme point of the utility-privacy trade-off curve, corresponding to perfect privacy either in the output perturbation or full data observation models. In general, characterizing the whole of this trade-off curve is analytically challenging. Therefore, to better understand the fundamental trade-off between utility and privacy, we will next consider the output perturbation model, and study the slope of $g_\epsilon(X,Y)$ as $\epsilon\to 0$. This will reveal us how much utility we can gain at the expense of a small amount of privacy leakage. The analysis depends on whether perfect privacy is feasible or not.

Consider a pair of random variables $(X,Y)\in\mathcal{X}\times\mathcal{Y}$ distributed according to $\mathbf{P}_{X,Y}$, with the marginals $\mathbf{p}_X$ and $\mathbf{p}_Y$. The matrix $\mathbf{P}_{X|Y}$ can be viewed as a channel with input $Y$ and output $X$. When the input of this channel is distributed according to $\mathbf{q}_Y$, the output is distributed according to $\mathbf{q}_X=\mathbf{P}_{X|Y}\mathbf{q}_Y$. 

Define $r:\mathcal{P}(\mathcal{Y})\backslash\{\mathbf{p}_Y\}\to[0,1]$ as
\begin{equation}\label{D}
r(\mathbf{q}_Y)\triangleq\frac{D(\mathbf{q}_X||\mathbf{p}_X)}{D(\mathbf{q}_Y||\mathbf{p}_Y)}.
\end{equation}

Let $V^*\in[1,+\infty]$ be defined as
\begin{equation}\label{vdef}
V^*\triangleq\sup_{\substack{\mathbf{q}_Y:\\\mathbf{q}_Y\neq\mathbf{p}_Y}}\frac{1}{r(\mathbf{q}_Y)}=\sup_{\substack{\mathbf{q}_Y:\\\mathbf{q}_Y\neq\mathbf{p}_Y}}\frac{D(\mathbf{q}_Y||\mathbf{p}_Y)}{D(\mathbf{q}_X||\mathbf{p}_X)},
\end{equation}
with the convention that if for some $\mathbf{q}_Y(\neq\mathbf{p}_Y)$, we have $\mathbf{q}_X=\mathbf{p}_X$, then $V^*=+\infty$.
\begin{proposition}
We have $g_0(X,Y)=0$ if and only if $V^*<+\infty$.\footnote{A claim, similar to this proposition, is provided in \cite{Calmon2}; however, the proof is incomplete as follows. In the proof of \cite[Theorem 2]{Calmon2}, on page 1799, it is not clear how (13) is obtained immediately after (12) (where these numbers refer to \cite{Calmon2}). In other words, in order to show that having $v^*(p_{S,X})=0$ results in $\delta(p_{S,X})=0$, the authors proceed as follows. Since in \cite[Lemma 4]{Calmon2}, $v^*(p_{S,X})$ is shown to be equal to $\inf_{q_X:q_X\neq p_X}\frac{D(q_S||p_S)}{D(q_X||p_X)}$, they get that if $v^*(p_{S,X})=0$, then for any $\epsilon>0$, there exists $\delta>0$, such that $D(q_X||p_X)\geq\delta>0$,
and $D(q_S||p_S)<\epsilon$.
Then, a sequence is constructed as $q_X^1,q_X^2,q_X^3,\ldots$ such that $q_X^i\neq p_X$, $D(q_S^k||p_S)\leq\epsilon_k$, and $\lim_{k\to\infty}\epsilon_k=0.$ For the probability vector $\mathbf{q}_S^k$, Pinsker's inequality is applied to obtain $\epsilon_k\geq\frac{1}{2}||\mathbf{q}_S^k-\mathbf{p}_S||_1^2\geq \frac{1}{2}||\mathbf{q}_S^k-\mathbf{p}_S||_2^2.
$
Defining $\mathbf{x}^k\triangleq\mathbf{q}_X^k-\mathbf{p}_X$, it is observed that $0<||\mathbf{x}^k||_2^2\leq2$, and $||\mathbf{P}_{S|X}\mathbf{x}^k||_2\leq\sqrt{2\epsilon_k}$. Just after this, it is written: \textit{"Hence,} $\lim_{k\to\infty}\frac{||\mathbf{P}_{S|X}\mathbf{x}^k||_2^2}{||\mathbf{x}^k||_2^2}=0." 
$ The issue appears in obtaining the latter just based on the arguments prior to it. The reasoning is provided in a way to show that the numerator becomes vanishingly small, while the denominator is bounded away from zero. However, this has not been discussed. More specifically, the fact that for any $\epsilon>0$, there exists $\delta>0$, such that $D(q_X||p_X)\geq\delta>0$ maps to: For any $\epsilon_k$, there exists $\delta_k>0$, such that $D(q_X^k||p_X)\geq\delta_k>0$, and it is not shown that $\delta_k$ is bounded away from zero for all $k$. 
} 
\end{proposition} 
\begin{proof}
The proof is provided in Appendix \ref{app5}.
\end{proof}

\subsection{Perfect privacy is not feasible.}
If perfect privacy is not feasible, i.e., $g_0(X,Y)=0$, then the slope of $g_\epsilon(X,Y)$ at $\epsilon=0$ is equal to $V^*$ as shown in \cite{Calmon2}. However, $V^*$ itself is written as a supremization, and hence, practical approximations of the this slope based on the properties of the joint distribution $p_{X,Y}$ is of interest. The following Theorem provides a lower bound on this slope. 

{\color{black}Let $\hat{\mathcal{Y}}$ denote the set of all the subsets of $\mathcal{Y}$ excluding the empty set and $\mathcal{Y}$, i.e., $\hat{\mathcal{Y}}\triangleq\{\mathcal{V}|\mathcal{V}\subset\mathcal{Y}\}-\{\mathcal{Y},\emptyset\}$.
}
{\color{black}
\begin{theorem}\label{thg0} 
We have
\begin{align}
\lim_{\epsilon\to 0}\frac{g_\epsilon(X,Y)}{\epsilon}&\geq\max\{A(X,Y),B_0(X,Y)\}\label{sheeb}\\&\geq\frac{H(Y)}{I(X;Y)}\label{shib2},
\end{align}
where
\begin{align}
    A(X,Y)&\triangleq \max_{\mathcal{B}:\mathcal{B}\in\hat{\mathcal{Y}}}\frac{-\log \left(\sum_{y\in\mathcal{B}}p_Y(y)\right)}{D\bigg(\frac{\sum_{y\in\mathcal{B}}p_Y(y)p_{X|Y}(\cdot|y)}{\sum_{y\in\mathcal{B}}p_Y(y)}\bigg|\bigg|p_X(\cdot)\bigg)},\label{defineA}\\
    B_{\alpha}(X,Y)&\triangleq  \frac{(H(Y)-1)^+-\alpha}{I(X;Y)-\max_{x\in\mathcal{X}}p(x)D(p_{Y|X}(\cdot|x)||p_Y(\cdot))},\ \alpha\geq 0.\label{defineB}
\end{align}
\begin{proof}
The equality in (\ref{sheeb}) is proved in \cite{Calmon2}, the proof of the lower bound is as follows. Fix an arbitrary $\mathcal{B}\in\hat{\mathcal{Y}}$, and define the pmf $q'_Y$ as
\begin{equation}
    q'_Y(y)\triangleq\frac{p_Y(y)}{\sum_{t\in\mathcal{B}}p_Y(t)}\cdot\mathds{1}_{\{y\in\mathcal{B}\}}.
\end{equation}
From (\ref{vdef}), we have
\begin{align}
   V^*&\geq \frac{D(\mathbf{q}'_Y||\mathbf{p}_Y)}{D(\mathbf{P}_{X|Y}\mathbf{q}'_Y||\mathbf{p}_X)}\label{inc}\\
   &=\frac{-\log \left(\sum_{y\in\mathcal{B}}p_Y(y)\right)}{D\bigg(\frac{\sum_{y\in\mathcal{B}}p_Y(y)p_{X|Y}(\cdot|y)}{\sum_{y\in\mathcal{B}}p_Y(y)}\bigg|\bigg|p_X(\cdot)\bigg)},\label{incakh}
\end{align}
where (\ref{inc}) follows from the definition in (\ref{vdef}) and the fact that $q'_Y(\cdot)\neq p_Y(\cdot)$, since $\mathcal{Y}\not\in\hat{\mathcal{Y}}$. Since (\ref{incakh}) is valid for any $\mathcal{B}\in\hat{\mathcal{Y}}$, by taking its maximum over $\mathcal{B}\in\hat{\mathcal{Y}}$, we have
\begin{equation}\label{zss1}
    \lim_{\epsilon\to 0}\frac{g_\epsilon(X,Y)}{\epsilon}\geq A(X,Y).
\end{equation}
Let $x^*\triangleq \argmax_{x\in\mathcal{X}}p(x)D(p_{Y|X}(\cdot|x)||p_Y(\cdot))$. Define the binary r.v. $\hat{X}$ as a deterministic function of $X$ given by $\hat{x}(x)\triangleq\mathds{1}_{\{x=x^*\}}$. Hence, we have $\hat{X}-X-Y$ form a Markov chain. From Corollary \ref{cor22}, we have $g_0(\hat{X},Y)\geq (H(Y)-\log\textnormal{rank}(\mathbf{P}_{\hat{X}|Y}))^+=(H(Y)-1)^+$. Therefore, there exists a privacy-preserving mapping $p_{U|Y}$, such that $\hat{X}-X-Y-U$ form a Markov chain, $I(Y;U)\geq (H(Y)-1)^+$, and $I(\hat{X};U)=0$. We have
\begin{align}
    I(X;U)&=\sum_{x\in\mathcal{X}}p(x)D\left(p_{U|X}(\cdot|x)||p_U(\cdot)\right)\nonumber
\end{align}
\begin{align}
    &=p(x^*)D\left(p_{U|X}(\cdot|x^*)||p_U(\cdot)\right)+\sum_{x\in\mathcal{X}\backslash\{x^*\}}p(x)D\left(p_{U|X}(\cdot|x)||p_U(\cdot)\right)\nonumber\\
    &=p_{\hat{X}}(1)D\left(p_{U|\hat{X}}(\cdot|1)||p_U(\cdot)\right)+\sum_{x\in\mathcal{X}\backslash\{x^*\}}p(x)D\left(p_{U|X}(\cdot|x)||p_U(\cdot)\right)\label{bv1}\\
    &=\sum_{x\in\mathcal{X}\backslash\{x^*\}}p(x)D\left(p_{U|X}(\cdot|x)||p_U(\cdot)\right)\label{bv11}\\
    &\leq \sum_{x\in\mathcal{X}\backslash\{x^*\}}p(x)D\left(p_{Y|X}(\cdot|x)||p_Y(\cdot)\right)\label{bv2}\\
    &=I(X;Y)-p(x^*)D\left(p_{Y|X}(\cdot|x^*)||p_Y(\cdot)\right)\label{bv3},
\end{align}
where (\ref{bv1}) follows from $\hat{x}(x)\triangleq\mathds{1}_{x=x^*}$, which results in $p_{\hat{X}}(1)=p_X(x^*)$, and $p_{U|\hat{X}}(\cdot|1)=p_{U|X}(\cdot|x^*)$; (\ref{bv11}) follows from having $U\independent \hat{X}$, and hence, $p_{U|\hat{X}}(\cdot|1)=p_U(\cdot)$, and (\ref{bv2}) results from the data processing inequality by viewing two pmfs $p_{Y|X}(\cdot|x)$ and $p_Y(\cdot)$ entering the channel $p_{U|Y}$. 

In this construction, we have $X-Y-U$ form a Markov chain and a point with utility of at least $(H(Y)-1)^+$, and privacy leakage of at most $I(X;Y)-p(x^*)D\left(p_{Y|X}(\cdot|x^*)||p_Y(\cdot)\right)$ is achievable in the utility-privacy trade-off curve. By noting the concavity of $g_\epsilon(X,Y)$ in $\epsilon$ (see \cite[lemma 2]{info7010015}), the slope at $(0,0)$ is lower bounded by the slope of the straight line connecting this point to the origin. Hence,
\begin{equation}\label{zss2}
    \lim_{\epsilon\to 0}\frac{g_\epsilon(X,Y)}{\epsilon}\geq B_0(X,Y).
\end{equation}
From (\ref{zss1}), (\ref{zss2}), the lower bound in (\ref{sheeb}) is proved. As a special case, if in (\ref{defineA}), $\mathcal{B}$ is restricted to the space of singletons, i.e., subsets of $\mathcal{Y}$ with only one element, we get a lower bound on the slope at origin as
\begin{equation*}
    \max_{y\in\mathcal{Y}}\frac{-\log \left(p_Y(y)\right)}{D\left(p_{X|Y}(\cdot|y)||p_X(\cdot)\right)},
\end{equation*}
which is proved differently in \cite[lemma 19]{info7010015}, and shown to satisfy the inequality in (\ref{shib2}).
\end{proof}
\end{theorem}
}


Thus far, we have observed that when perfect privacy is not feasible, the slope of the trade-off curve is finite. In other words, for a vanishingly small privacy leakage, only a linearly proportional vanishingly small utility can be attained. This is not necessarily the case when perfect privacy is feasible, which is discussed next.
\subsection{Perfect privacy is feasible.}
For a given pair $(X,Y)$, assume that $g_0(X,Y)$, obtained through the LP formulation in Theorem 1, is achieved by
\begin{equation}\label{maxdef}
U^*\in\mathcal{U}^*=\{u^*_1,u^*_2,\ldots,u^*_{|\mathcal{U}^*|}\},\ \mathbf{p}_{Y|u^*},\forall u^*\in\mathcal{U}^*,
\end{equation}
where the vectors $\mathbf{p}_{Y|u^*},\ \forall u^*\in\mathcal{U}^*$ belong to the extreme points of the set $\mathbb{S}_{X,Y}$, as in (\ref{poly}).
\begin{definition}
Define
\begin{equation}\label{nav}
\psi(u^*)\triangleq\sup_{\substack{\mathbf{q}_Y:\\0<D(\mathbf{q}_Y||\mathbf{p}_{Y|u^*})<+\infty}}\frac{D(\mathbf{q}_Y||\mathbf{p}_{Y|u^*})}{D(\mathbf{q}_X||\mathbf{p}_X)},\ \forall u^*\in\mathcal{U}^*,
\end{equation}
and if for some $u^*$, there is no $\mathbf{q}_Y$ for which $0<D(\mathbf{q}_Y||\mathbf{p}_{Y|u^*})<+\infty$ (which happens exactly when $\mathbf{p}_{Y|u^*}$ is a corner point of the probability simplex), then let $\psi(u^*)\triangleq0$. {\color{black}Therefore, in order to evaluate $\psi(u^*)$ for some $u^*\in\mathcal{U}^*$, the search space in (\ref{nav}), includes the set of all probability vectors in $\mathcal{P}(\mathcal{Y})$ such that i) they are not equal to the extreme point $\mathbf{p}_{Y|u^*}$ (equivalent to $0<D(\mathbf{q}_Y||\mathbf{p}_{Y|u^*})$), ii) if $\mathbf{p}_{Y|u^*}$ has a zero entry, they will also have a zero in the same entry (equivalent to $D(\mathbf{q}_Y||\mathbf{p}_{Y|u^*})<+\infty$).}
\end{definition}
The following lemma is needed in the sequel.
\begin{lemma}\label{psiinf}
We have $\psi(u^*)<+\infty,\ \forall u^*\in\mathcal{U}^*$.
\end{lemma}
\begin{proof}
The proof is provided in Appendix \ref{app6}.
\end{proof}

\begin{theorem}
For a given pair $(X,Y)$, if perfect privacy is feasible, we have
\begin{equation}\label{Lboun}
\lim_{\epsilon\to 0}\frac{g_\epsilon(X,Y)-g_0(X,Y)}{\epsilon}\geq \max\bigg\{L(X,Y),B_{g_0(X,Y)}(X,Y),\frac{H(Y)-g_0(X,Y)}{I(X;Y)}\bigg\},
\end{equation}
where
\begin{equation}\label{ma}
L(X,Y)\triangleq\max_{u^*\in\mathcal{U}^*}\psi(u^*),
\end{equation}
and $B_\alpha(X,Y)$ is defined in (\ref{defineB}).
\end{theorem}
\begin{proof}
First, we note that from lemma \ref{psiinf}, $L(X,Y)$ is well defined. Denote a/the maximizer of (\ref{ma}) by $u^*_j$ for some $j\in[|\mathcal{U}^*|]$. 
From (\ref{nav}) and (\ref{ma}), for an arbitrary fixed $\delta>0$, we have
\begin{equation}\label{LOW}
\exists\ \mathbf{q}_Y\neq\mathbf{p}_{Y|u^*_j},\ D(\mathbf{q}_Y||\mathbf{p}_{Y|u^*_j})<+\infty:\ L(X,Y)-\delta<\frac{D(\mathbf{q}_Y||\mathbf{p}_{Y|u^*_j})}{D(\mathbf{q}_X||\mathbf{p}_X)}\leq L(X,Y).
\end{equation}
Construct the pair $(Y,U)$ as follows. Let $\mathcal{U}=\{u_1,u_2,\ldots,u_{|\mathcal{U}^*|},\hat{u}_j\}$, and for sufficiently small $\gamma>0$, let
\begin{align}
p_{U}(u_i)=p_{U^*}(u^*_i),&\ \mathbf{p}_{Y|u_i}=\mathbf{p}_{Y|u^*_i},\ \forall i\in[|\mathcal{U}^*|], i\neq j\label{yek1},\\
\ p_{U}(u_j)=\gamma p_{U^*}(u^*_j),\ p_{U}(\hat{u}_j)=(1-\gamma) p_{U^*}(u^*_j),&\ \mathbf{p}_{Y|u_j}=\mathbf{q}_{Y},\ \mathbf{p}_{Y|\hat{u}_j}=\frac{1}{1-\gamma}(\mathbf{p}_{Y|u^*_j}-\gamma\mathbf{q}_{Y}).\label{do2}
\end{align}
Note that for sufficiently small $\gamma>0$, $\mathbf{p}_{Y|\hat{u}_j}$ in (\ref{do2}) is a probability vector, as we have $D(\mathbf{q}_Y||\mathbf{p}_{Y|u^*_j})<+\infty$. In other words, for any entry of the vector $\mathbf{p}_{Y|u^*_j}$ that is zero (note that it is an extreme point of $\mathbb{S}_{X,Y}$), the corresponding entry in $\mathbf{q}_Y$ is also zero. Finally, it can be verified from (\ref{yek1}) and (\ref{do2}) that the marginal probability vector $\mathbf{p}_Y$ is also preserved.

{\color{black} With $I_\gamma(Y;U)$, and $I_\gamma(X;U)$ denoting the corresponding mutual information terms in this construction, and from the concavity of $g_\epsilon(X,Y)$ in $\epsilon$ (see \cite{info7010015}), the LHS of (\ref{Lboun}) is lower bounded by
\begin{align}
\frac{I_\gamma(Y;U)-g_0(X,Y)}{I_\gamma(X;U)}&=\frac{\sum_{u\in\mathcal{U}}p_U(u)D(\mathbf{p}_{Y|u}||\mathbf{p}_Y)-\sum_{u^*\in\mathcal{U}^*}p_{U^*}(u^*)D(\mathbf{p}_{Y|u^*}||\mathbf{p}_Y)}{\sum_{u\in\mathcal{U}}p_U(u)D(\mathbf{p}_{X|u}||\mathbf{p}_X)}\nonumber\\
&=\frac{\sum_{u\in\{u_j,\hat{u}_j\}}p_U(u)D(\mathbf{p}_{Y|u}||\mathbf{p}_Y)-p_{U^*}(u^*_j)D(\mathbf{p}_{Y|u^*_j}||\mathbf{p}_Y)}{\sum_{u\in\{u_j,\hat{u}_j\}}p_U(u)D(\mathbf{p}_{X|u}||\mathbf{p}_X)}\label{pp1}\\
&=\frac{\gamma D(\mathbf{q}_{Y}||\mathbf{p}_Y)+(1-\gamma)D\bigg(\frac{1}{1-\gamma}(\mathbf{p}_{Y|u^*_j}-\gamma\mathbf{q}_{Y})\bigg|\bigg|\mathbf{p}_Y\bigg)-D(\mathbf{p}_{Y|u^*_j}||\mathbf{p}_Y)}{\gamma D(\mathbf{q}_{X}||\mathbf{p}_X)+(1-\gamma)D\bigg(\frac{1}{1-\gamma}(\mathbf{p}_{X}-\gamma\mathbf{q}_{X})\bigg|\bigg|\mathbf{p}_X\bigg)},\label{pp2}
\end{align}
where the numerator in (\ref{pp1}) follows from (\ref{yek1}); the denominator in (\ref{pp1}) is from the fact that $\mathbf{p}_{X|u_i}=\mathbf{p}_{X|u^*_i}, \forall i\in[|\mathcal{U}^*|], i\neq j$ and $\mathbf{p}_{X|u^*}=\mathbf{P}_{X|Y}\mathbf{p}_{Y|u^*}=\mathbf{p}_X,\forall u^*\in\mathcal{U}^*$; (\ref{pp2}) follows from (\ref{do2}). 

For three generic pmfs $p$ on $\mathcal{Y}$, and $q,r$ on $\tilde{Y}\subset\mathcal{Y}$, when $\gamma\to 0$, we can write  
\begin{align}
D\bigg(\frac{1}{1-\gamma}(\mathbf{r}-\gamma\mathbf{q})\bigg|\bigg|\mathbf{p}\bigg)&=\sum_{y\in\tilde{\mathcal{Y}}} \frac{r(y)-\gamma q(y)}{1-\gamma}\bigg[\log\frac{1}{1-\gamma}+\log\frac{r(y)-\gamma q(y)}{p(y)}\bigg]\nonumber\\
&=-\log(1-\gamma)+\sum_{y\in\tilde{\mathcal{Y}}} \frac{r(y)-\gamma q(y)}{1-\gamma}\log\frac{r(y)}{p(y)}\left(1-\gamma\frac{ q(y)}{r(y)}\right)\nonumber
\end{align}
\begin{align}
&\approx-\log(1-\gamma)+\sum_{y\in\tilde{\mathcal{Y}}} \frac{r(y)-\gamma q(y)}{1-\gamma}\bigg[\log\frac{r(y)}{p(y)}-\gamma\frac{ q(y)}{r(y)}\bigg]\label{finap}\\
&=\frac{1}{1-\gamma}\left(D(r||p)-\gamma\sum_{y\in\tilde{\mathcal{Y}}}q(y)\log\frac{r(y)}{p(y)}+O(\gamma^2)\right),\label{finfin}
\end{align}
where in (\ref{finap}), the first order approximation, i.e., $\log(1+x)\approx x$ for $x\to 0$, is used\footnote{{\color{black}Note that this is true only if the logarithm is natural, i.e., to the base of the mathematical constant $e$; however, since in this section, we are dealing with the ratios of mutual information terms, and hence the ratios of logarithms, this has no effect on the results, as we can multiply both the numerator and denominator by $\log_e2$.}}.
Using the approximation in (\ref{finfin}) for both of the second terms in the numerator and denominator of (\ref{pp2}), after some manipulation, we get
\begin{align}
\lim_{\epsilon\to 0}\frac{g_\epsilon(X,Y)-g_0(X,Y)}{\epsilon}&\geq\lim_{\gamma\to 0}\frac{I_{\gamma}(Y;U)-g_0(X,Y)}{I_{\gamma}(X;U)}\nonumber,\\
&=\lim_{\gamma\to 0}\frac{\gamma D(\mathbf{q}_Y||\mathbf{p}_{Y|u^*_j})+O(\gamma^2)}{\gamma D(\mathbf{q}_X||\mathbf{p}_{X})+O(\gamma^2)}\nonumber,\\
&=\frac{D(\mathbf{q}_Y||\mathbf{p}_{Y|u^*_j})}{D(\mathbf{q}_X||\mathbf{p}_{X})}\nonumber,\\
&>L(X,Y)-\delta,\label{lwo}
\end{align}
where (\ref{lwo}) follows from (\ref{LOW}). Since $\delta>0$ was chosen arbitrarily, we have
\begin{equation*}
    \lim_{\epsilon\to 0}\frac{g_\epsilon(X,Y)-g_0(X,Y)}{\epsilon}\geq L(X,Y).
\end{equation*}
The second term in the RHS of (\ref{Lboun}) follows similarly to the discussion in the proof of Theorem \ref{thg0}. In other words, it is a lower bound for the slope of a straight line that connect $(0,g_0(X,Y))$ to a point with utility of at least $(H(Y)-1)^+$, and privacy leakage of at most $I(X;Y)-p(x^*)D\left(p_{Y|X}(\cdot|x^*)||p_Y(\cdot)\right)$. Finally, the second term in the RHS of (\ref{Lboun}) is the slope of the straight line connecting the end points of the curve of $g_\epsilon(X,Y)$ vs. $\epsilon$, i.e., $(0,g_0(X,Y))$ and $(I(X;Y),H(Y))$. The fact that these are lower bounds on the slope at origin follow from the concavity of $g_\epsilon(X,Y)$ in $\epsilon$.}
\end{proof}
{\color{black} Assume that the maximizer in $g_0(X,Y)$, i.e., (\ref{maxdef}), induces the joint distribution $p_{Y,U}^*(\cdot,\cdot)$. In what follows, it is shown that under certain conditions, when perfect privacy is feasible, the slope at origin is infinite. The following lemmas are needed in the sequel.
\begin{lemma}\label{ren}
Let $p,q$ denote two pmfs on $\mathcal{Y}$, and assume that $p(y)>0,\ \forall y\in\mathcal{Y}$. We have
\begin{equation}\label{ppo12}
    \sum_y\frac{q^2(y)}{p(y)}\geq 1,
\end{equation}
with equality if and only if $q=p$.
\end{lemma}
\begin{proof}
We have $\sum_y\frac{q^2(y)}{p(y)} =1+\sum_y\frac{(q(y)-p(y))^2}{p(y)}\geq 1$, with equality if and only if $p=q$.\footnote{{\color{black}Alternatively, the LHS of (\ref{ppo12}) is $2^{D_2(q||p)}$, where $D_\alpha(\cdot||\cdot)$ is the R\'{e}nyi divergence of order $\alpha$. By noting that $D_\alpha(p||q)\geq 0$, with equality if and only if $p=q$, the proof is complete.}}
\end{proof}}
{\color{black}
\begin{lemma}\label{ins}
For a given pair $(X,Y)$, if there exists $y_0\in\mathcal{Y}$, for which $p_{X|Y}(\cdot|y_0)=p_X(\cdot)$, there must exist $u^*_0\in\mathcal{U}^*$, defined in (\ref{maxdef}), such that
\begin{align*}
    p_{Y|U^*}(y_0|u^*_0) = 1,\ p_{Y|U^*}(y_0|u^*) = 0,\ \forall u^*\in\mathcal{U}^*\backslash\{u^*_0\}.
\end{align*}
\end{lemma}
\begin{proof}
The proof is provided in Appendix \ref{AppendixFD}
\end{proof}
}
\begin{theorem}\label{thlower9}
If there exist $y_0\in\mathcal{Y}$, and $u_0,u_1\in\mathcal{U}^*$, such that $p^*(y_0,u_0),p^*(y_0,u_1)>0$, and $p^*(y_0|u_0)\neq p^*(y_0|u_1)$, then
\begin{equation}\label{slopeinf}
    \lim_{\epsilon\to 0}\frac{g_\epsilon(X,Y)-g_0(X,Y)}{\epsilon}=\infty.
\end{equation}
\end{theorem}
\begin{proof}
Without loss of generality, assume that $p^*(y_0|u_0)> p^*(y_0|u_1)$. Consider the tuple $(X,Y,U)$ distributed according to $p_{X|Y}\cdot p'_{Y,U}$, where $p'(y,u)=p^*(y,u)+\eta\cdot i(y,u)$, in which $i(y,u)$ is non-zero only for two cases: $i(y_0,u_0)=-i(y_0,u_1)=1$. The value of $\eta>0$ is chosen arbitrarily small such that $p'(y,u)$ is a pmf\footnote{To this end, a necessary condition is to have $\eta\leq\min\{p^*(y_0,u_1),1-p^*(y_0,u_0)\}$}. Therefore, we have the marginal pmf $p'(u)=p^*(u)+\eta\cdot i(y_0,u),\ \forall u\in\mathcal{U}$. It can be verified that with this construction, the marginal pmf $p_{X,Y}$ is preserved in the tuple $(X,Y,U)$. 

{\color{black}Let $I_\eta(Y;U)$ and $I_\eta(X;U)$ denote the mutual information terms induced by $p_{X|Y}\cdot p'_{Y,U}$. When $\eta\to 0$, we have
\begin{align}
    I_\eta(Y;U)&=D(p'_{Y,U}||p_Y\cdot p'_U)=D\bigg(p^*(y,u)+\eta \cdot i(y,u)\bigg|\bigg|p(y)\left(p^*(u)+\eta\cdot i(y_0,u)\right)\bigg),\nonumber\\
    &=\sum_{y,u}\left(p^*(y,u)+\eta \cdot i(y,u)\right)\log \frac{p^*(y,u)+\eta \cdot i(y,u)}{p(y)\left(p^*(u)+\eta\cdot i(y_0,u)\right)},\nonumber\\
    &=\sum_{y,u}\left(p^*(y,u)+\eta \cdot i(y,u)\right)\log \frac{p^*(y,u)\left(1+\eta \frac{i(y,u)}{p^*(y,u)}\right)}{p(y)p^*(u)\left(1+\eta\frac{i(y_0,u)}{p^*(u)}\right)}\nonumber,\\
    &\approx \sum_{y,u}\left(p^*(y,u)+\eta \cdot i(y,u)\right)\bigg[\log \frac{p^*(y,u)}{p(y)p^*(u)}+\eta \frac{i(y,u)}{p^*(y,u)}-\eta\frac{i(y_0,u)}{p^*(u)}\bigg],\label{logapprox}\\
    &=D(p^*(y,u)||p(y)p^*(u))+\eta\bigg(\sum_{y,u}i(y,u)-i(y_0,u)p^*(y|u)\bigg)\nonumber\\
    &\ \ \ +\eta\sum_{y,u}i(y,u)\log \frac{p^*(y,u)}{p(y)p^*(u)}+O(\eta^2),\nonumber\\
    &=g_0(X,Y)+0+\eta(\log p^*(y_0|u_0)-\log p^*(y_0|u_1))+O(\eta^2),\label{fin}
\end{align}
where all the summations above are over the support of $(Y,U)$, i.e., $\textnormal{supp}(Y,U)$; in (\ref{logapprox}), the first order approximation, i.e., $\log (1+x)\approx x$ for $x\to 0$, has been used. Also, (\ref{fin}) follows from having $g_0(X,Y) = D(p^*(y,u)||p(y)p^*(u))$, and the properties of $i(\cdot,\cdot)$.

Similarly, when $\eta\to 0$, we can write
\begin{align}
    I_\eta(X;U)&=D(p'_{X,U}||p_X\cdot p'_U)=D\bigg(p^*(x,u)+\eta \cdot p(x|y_0)i(y_0,u)\bigg|\bigg|p(x)(p^*(u)+\eta \cdot i(y_0,u))\bigg),\nonumber\\
    &=\sum_{x,u}\left(p^*(x,u)+\eta \cdot p(x|y_0)i(y_0,u)\right)\log \frac{p^*(x,u)+\eta \cdot p(x|y_0)i(y_0,u)}{p(x)(p^*(u)+\eta \cdot i(y_0,u))},\nonumber\\
    &=\sum_{x,u}\left(p^*(x,u)+\eta \cdot p(x|y_0)i(y_0,u)\right)\log \frac{p^*(x,u)\left(1+\eta \frac{p(x|y_0)i(y_0,u)}{p^*(x,u)}\right)}{p(x)p^*(u)\left(1+\eta\frac{i(y_0,u)}{p^*(u)}\right)},\nonumber\\
    &\approx\sum_{x,u}\left(p^*(x,u)+\eta \cdot p(x|y_0)i(y_0,u)\right)\bigg[\log\frac{p^*(x,u)}{p(x)p^*(u)}+\eta \frac{p(x|y_0)i(y_0,u)}{p^*(x,u)}\nonumber\\&\ \ \ -\eta^2\frac{p^2(x|y_0)i^2(y_0,u)}{2{p^*}^2(x,u)} -\eta\frac{i(y_0,u)}{p^*(u)}+\eta^2\frac{i^2(y_0,u)}{2{p^*}^2(u)}\bigg],\label{logapp}\\
    &=\eta^2\sum_{x,u}\left(\frac{p^2(x|y_0)i^2(y_0,u)}{2p^*(x,u)}-\frac{p(x|y_0)i^2(y_0,u)}{p^*(u)}+\frac{p(x)i^2(y_0,u)}{2{p^*}(u)}\right) + O(\eta^3),\label{simp1}\\
    &=\eta^2\sum_{u}\frac{i^2(y_0,u)}{2p^*(u)}\sum_{x}\left(\frac{p^2(x|y_0)}{p(x)}-2p(x|y_0)+p(x)\right) + O(\eta^3),\nonumber\\
    &=\eta^2\cdot\underbrace{\sum_{u\in\{u_0,u_1\}}\frac{1}{2p^*(u)}\left(\sum_{x}\frac{p^2(x|y_0)}{p(x)}-1\right)}_{\triangleq A} + O(\eta^3),\label{simp2}
\end{align}
where (\ref{logapp}) follows from the second order approximation $\log(1+x)\approx x-\frac{x^2}{2}$ for $x\to 0$; (\ref{simp1}) follows from having $p^*(x,u)=p(x)p^*(u)$, due to the condition of perfect privacy, and the equality $\sum_{x,u}i(y_0,u)\left(p(x|y_0)-p(x)\right)=0$. In (\ref{simp2}), we make use of the fact that $i^2(y_0,u)=1$ for $u=u_0,u_1$, and zero otherwise. Hence, from the construction of the tuple $(X,Y,U)$, we obtain a lower bound for the LHS of (\ref{slopeinf}) as
\begin{align}
     \lim_{\epsilon\to 0}\frac{g_\epsilon(X,Y)-g_0(X,Y)}{\epsilon}&\geq\lim_{\eta\to 0}\frac{I_\eta(Y;U)-g_0(X,Y)}{I_\eta(X;U)}\nonumber\\
     &= \lim_{\eta\to 0}\frac{\eta(\log (p^*(y_0|u_0))-\log (p^*(y_0|u_1)))+O(\eta^2)}{A\eta^2+O(\eta^3)}\nonumber\\
     &=+\infty,\label{lb}
\end{align}
which is valid if it can be shown that $A$ is a positive real number. From lemma \ref{ren}, we have that $A\geq0$. However, since $p_{X|Y}(\cdot|y_0)\neq p_X(\cdot)$\footnote{since otherwise, from lemma \ref{ins}, this will contradict the assumption in the statement of Theorem \ref{thlower9}. In other words, if $p_{X|Y}(\cdot|y_0)= p_X(\cdot)$, no $u_0,u_1\in\mathcal{U}$ exist such that $p^*(y_0,u_0),p^*(y_0,u_1)>0$.}, the inequality is strict, and we have $A>0$. This proves (\ref{slopeinf}).


}
\end{proof}

As a byproduct of the asymptotic analysis in section \ref{asymp}, we obtained an alternative proof for the characterization of maximal correlation previously given in \cite{Makur}. This is provided in the next (and also last) section of this paper.
\section{Maximal correlation}

Consider a pair of random variables $(X,Y)\in\mathcal{X}\times\mathcal{Y}$ distributed according to $p_{X,Y}$, with $|\mathcal{X}|,|\mathcal{Y}|<\infty$. Let $\tilde{\mathcal{F}}$ denote the set of all real-valued functions of $X$, and define
\begin{equation*}
\mathcal{F}\triangleq\bigg\{f(\cdot)\in\tilde{\mathcal{F}}\bigg|\mathds{E}[f(X)]=0,\ \mathds{E}[f^2(X)]=1\bigg\}.
\end{equation*} 
Let $\tilde{\mathcal{G}}$ and $\mathcal{G}$ be defined similarly for the random variable $Y$. The maximal correlation of $(X,Y)$ is defined as (\!\!\cite{Hirschfeld, Gebelein,Reny}):
\begin{equation*}
\rho_m(X;Y)\triangleq\max_{f\in\mathcal{F},g\in\mathcal{G}}\mathds{E}[f(X)g(Y)].
\end{equation*}
If $\mathcal{F}$ (and/or $\mathcal{G}$) is empty\footnote{When $X$ (and/or $Y$) is constant almost surely.}, then $\rho_m$ is defined to be zero.

An alternative characterization of the maximal correlation is given by Witsenhausen in \cite{Witsenhausen} as follows\footnote{For other characterizations, see \cite{Anantharam}.}. Let the matrix $\mathbf{Q}$ be defined as 
\begin{equation}\label{Q}
\mathbf{Q}\triangleq\mathbf{P}_X^{-\frac{1}{2}}\mathbf{P}_{X,Y}\mathbf{P}_Y^{-\frac{1}{2}},
\end{equation}
with singular values $\sigma_1\geq\sigma_2\geq\cdots$. It is shown in \cite{Witsenhausen} that $\sigma_1=1$, and the maximal correlation of $(X,Y)$, i.e., $\rho_m(X;Y)$, is equal to the second largest singular value of matrix $\mathbf{Q}$, i.e., $\sigma_2$. 

We consider the singular values of matrix $\mathbf{Q}$. It is shown in \cite[Theorem 3]{Makur} that the maximal correlation can be written as the following limit
\begin{equation*}
\rho_m^2(X;Y)=\lim_{\eta\to 0}\sup_{\substack{\mathbf{q}_Y:\mathbf{q}_Y\neq\mathbf{p}_Y\\D(\mathbf{q}_Y||\mathbf{p}_Y)=\eta}}r(\mathbf{q}_Y),
\end{equation*}
where $r(\mathbf{q}_Y)$ is defined in (\ref{D}). The proof is based on Courant-Fischer min-max principle. 

In what follows, we provide an alternative proof for the above equation, which is based on \cite{Golub}.
The following preliminaries from \cite{Golub} are needed in the sequel.
\subsection{Preliminaries}
Assume that $\mathbf{R}$ is an $n$-by-$n$ real symmetric matrix, and $\mathbf{c}$ is an $n$-dimensional vector satisfying $\|\mathbf{c}\|_2=1$. Assume that we are interested in finding the stationary values of 
\begin{align}\label{prob}
\mathbf{x}^T\mathbf{R}\mathbf{x},
\end{align}
subject to the constraints
\begin{align}
\mathbf{c}^T\mathbf{x}&=0,\nonumber\\
\|\mathbf{x}\|_2&=1.
\end{align}
Letting $\lambda$ and $\mu$ be the Lagrange multipliers, we have
\begin{equation*}
L(\mathbf{x},\lambda,\mu)=\mathbf{x}^T\mathbf{R}\mathbf{x}-\lambda(\mathbf{x}^T\mathbf{x}-1)+2\mu\mathbf{x}^T\mathbf{c}.
\end{equation*}
Differentiating the Lagrangian with respect to $\mathbf{x}$, we obtain
\begin{equation}\label{con}
\mathbf{R}\mathbf{x}-\lambda\mathbf{x}+\mu\mathbf{c}=\mathbf{0},
\end{equation}
which results in $\mu = -\mathbf{c}^T\mathbf{R}\mathbf{x}$, after multiplying both sides by $\mathbf{c}^T$ and noting that $\|\mathbf{c}\|_2=1$. By substituting this value of $\mu$ in (\ref{con}), we get
\begin{equation*}
\mathbf{P}\mathbf{R}\mathbf{x}=\lambda\mathbf{x},
\end{equation*}
where $\mathbf{P}=\mathbf{I}-\mathbf{c}\mathbf{c}^T$. Since $\mathbf{P}$ is a projection matrix, i.e. $\mathbf{P}^2=\mathbf{P}$, the stationary values of $\mathbf{x}^T\mathbf{R}\mathbf{x}$ are the singular values of the matrix $\mathbf{PR}$ that occur at the corresponding eigenvectors.

Finally, assume that the vector $\mathbf{c}$ in the constraints is replaced with an $n\times r$ matrix $\mathbf{C}$ with $r\leq n$. Also, assume that the columns of matrix $\mathbf{C}$ are orthonormal. It can be verified that the results remain the same after having $\mathbf{P}$ modified as $\mathbf{P}=\mathbf{I}-\mathbf{CC}^T$.
\subsection{Alternative characterization of $\rho_m(X;Y)$}

We write $\mathbf{q}_Y\to\mathbf{p}_Y$ when $\|\mathbf{q}_Y-\mathbf{p}_Y\|_2\to 0$ and $\mathbf{q}_Y\neq\mathbf{p}_Y$. We are interested in finding the stationary values of $r(\mathbf{q}_Y)$ when $\mathbf{q}_Y\to\mathbf{p}_Y$.

\begin{theorem}\label{th6}
The stationary values of (\ref{D}), when $\mathbf{q}_Y\to\mathbf{p}_Y$, are the squared singular values of matrix $\mathbf{Q}\triangleq\mathbf{P}_X^{-\frac{1}{2}}\mathbf{P}_{X,Y}\mathbf{P}_Y^{-\frac{1}{2}}$, and in particular,
\begin{equation*}
\rho_m^2(X;Y)=\lim_{\eta\to 0}\sup_{\substack{\mathbf{q}_Y:\\0<\|\mathbf{q}_Y-\mathbf{p}_Y\|_2\leq\eta}}r(\mathbf{q}_Y).
\end{equation*}
\end{theorem}
\begin{proof}
Having $\mathbf{q}_Y\to\mathbf{p}_Y$, we can write
\begin{equation*}
\mathbf{q}_Y=\mathbf{p}_Y+\boldsymbol{\epsilon},\ 
\mathbf{1}_{|\mathcal{Y}|}^T\cdot\boldsymbol{\epsilon}=0,\  \|\boldsymbol{\epsilon}\|_2\to 0, \boldsymbol{\epsilon}\neq\mathbf{0},
\end{equation*}
where $\boldsymbol{\epsilon}$ is an auxiliary vector. From the relationship $\mathbf{q}_X=\mathbf{P}_{X|Y}\mathbf{q}_Y$, we have
\begin{equation}\label{tasav}
r(\mathbf{q}_Y)=\frac{D(\mathbf{p}_X+\mathbf{P}_{X|Y}\boldsymbol{\epsilon}||\mathbf{p}_X)}{D(\mathbf{p}_Y+\boldsymbol{\epsilon}||\mathbf{p}_Y)}.
\end{equation}
Assume that $\mathbf{p}_0$ and $\mathbf{p}$ are two probability vectors in the interior of $\mathcal P(\mathcal Y)$. Let $p_0(\cdot)$ and $p(\cdot)$ denote their corresponding probability mass functions. We can write the Taylor series expansion of the relative entropy as
\begin{equation}\label{tod}
D(\mathbf{p}_0+\boldsymbol{\epsilon}||\mathbf{p})=D(\mathbf{p}_0||\mathbf{p})+\boldsymbol{\epsilon}^T\cdot\mathbf{\nabla}D|_{\mathbf{p}_0}+\frac{1}{2}\boldsymbol{\epsilon}^T\mathbf{\nabla}^2D|_{\mathbf{p}_0}\boldsymbol{\epsilon}+\cdots,
\end{equation}
where
\begin{align*}
\mathbf{\nabla}D|_{\mathbf{p}_0}&=\begin{bmatrix}\log \frac{p_0(y_1)}{p(y_1)}+1&\log \frac{p_0(y_2)}{p(y_2)}+1&\dots&\log \frac{p_0(y_{|\mathcal{Y}|})}{p(y_{|\mathcal{Y}|})}+1\end{bmatrix}^T,\\
\mathbf{\nabla}^2D|_{\mathbf{p}_0}&=\textnormal{diag}\bigg(\begin{bmatrix}\frac{1}{p_0(y_1)}&\frac{1}{p_0(y_2)}&\dots&\frac{1}{p_0(y_{|\mathcal{Y}|})}\end{bmatrix}\bigg),
\end{align*}
are the gradient and the Hessian of $D(\cdot||\mathbf{p})$ at $\mathbf{p}_0$, respectively, and the higher order terms of $\mathbf{\epsilon}$ are denoted by dots in (\ref{tod}). Therefore, (\ref{tasav}) reduces to
\begin{align}\label{nog}
r(\mathbf{q}_Y)&=\frac{D(\mathbf{p}_X||\mathbf{p}_X)+\boldsymbol{\epsilon}^T\mathbf{P}^T_{X|Y}\mathbf{1}_{|\mathcal{X}|}+\boldsymbol{\epsilon}^T\mathbf{P}^T_{X|Y}\mathbf{P}^{-1}_X\mathbf{P}_{X|Y}\boldsymbol{\epsilon}+\ldots}{D(\mathbf{p}_Y||\mathbf{p}_Y)+\boldsymbol{\epsilon}^T\cdot\mathbf{1}_{|\mathcal{Y}|}+\boldsymbol{\epsilon}^T\mathbf{P}^{-1}_Y\boldsymbol{\epsilon}+\ldots}\nonumber\\
&=\frac{\boldsymbol{\epsilon}^T\mathbf{P}^T_{X|Y}\mathbf{P}^{-1}_X\mathbf{P}_{X|Y}\boldsymbol{\epsilon}+\ldots}{\boldsymbol{\epsilon}^T\mathbf{P}^{-1}_Y\boldsymbol{\epsilon}+\ldots},
\end{align}
where we have used the facts that $D(\mathbf{p}||\mathbf{p})=0$, $\mathbf{P}^T_{X|Y}\mathbf{1}_{|\mathcal{X}|}=\mathbf{1}_{|\mathcal{Y}|}$ and $\boldsymbol{\epsilon}^T\cdot\mathbf{1}_{|\mathcal{Y}|}=0$. When $\|\boldsymbol{\epsilon}\|_2\to 0$, the higher order terms of $\boldsymbol{\epsilon}$ in (\ref{nog}), shown with dots, can be ignored. Hence, we are interested in finding the stationary values of
\begin{equation}\label{kasr}
\frac{\boldsymbol{\epsilon}^T\mathbf{P}^T_{X|Y}\mathbf{P}^{-1}_X\mathbf{P}_{X|Y}\boldsymbol{\epsilon}}{\boldsymbol{\epsilon}^T\mathbf{P}^{-1}_Y\boldsymbol{\epsilon}},
\end{equation}
when $\mathbf{1}_{|\mathcal{Y}|}^T\cdot\boldsymbol{\epsilon}=0,\ \boldsymbol{\epsilon}\neq\mathbf{0}$. Note that the condition $\|\boldsymbol{\epsilon}\|_2\to 0$ is redundant as the norm $\|\boldsymbol{\epsilon}\|_2$ cancels out from both the numerator and the denominator of (\ref{kasr}).
We can equivalently write (\ref{kasr}) as
\begin{equation*}
\frac{\mathbf{v}^T\mathbf{P}^{\frac{1}{2}}_Y\mathbf{P}^T_{X|Y}\mathbf{P}^{-1}_X\mathbf{P}_{X|Y}\mathbf{P}^{\frac{1}{2}}_Y\mathbf{v}}{\mathbf{v}^T\cdot\mathbf{v}},
\end{equation*}
where $\mathbf{v}\triangleq\mathbf{P}^{-\frac{1}{2}}_Y\boldsymbol{\epsilon}$, $\mathbf{v}\neq\mathbf{0}$, $\mathbf{c}^T.\mathbf{v}=0$ with $\mathbf{c}=\mathbf{P}^{\frac{1}{2}}_Y\mathbf{1}_{|\mathcal{Y}|}$, and it is obvious that $\|\mathbf{c}\|_2=1$. Without loss of generality, we assume that $\|\mathbf{v}\|_2=1.$ Therefore, we are led to finding the stationary values of
\begin{equation}\label{B}
\mathbf{v}^T\mathbf{R}\mathbf{v},
\end{equation}
where $\mathbf{R}=\mathbf{P}^{\frac{1}{2}}_Y\mathbf{P}^T_{X|Y}\mathbf{P}^{-1}_X\mathbf{P}_{X|Y}\mathbf{P}^{\frac{1}{2}}_Y$, subject to the constraints
\begin{align}\label{B2}
\mathbf{c}^T\cdot\mathbf{v}&=0,\nonumber\\
\|\mathbf{v}\|_2&=1.
\end{align}
Note that $\mathbf{R}$ is a $|\mathcal{Y}|$-by-$|\mathcal{Y}|$ real symmetric matrix, and $\mathbf{c}$ is a $|\mathcal{Y}|$-dimensional vector satisfying $\|\mathbf{c}\|_2=1$. Therefore, (\ref{B}) is the same problem as in (\ref{prob}) whose stationary values are the eigenvalues of the matrix $(\mathbf{I}-\mathbf{c}\mathbf{c}^T)\mathbf{R}$, which occur at their corresponding eigenvectors.

We have
\begin{equation*}
\mathbf{R}=\mathbf{P}^{\frac{1}{2}}_Y\mathbf{P}^T_{X|Y}\mathbf{P}^{-1}_X\mathbf{P}_{X|Y}\mathbf{P}^{\frac{1}{2}}_Y=\bigg(\mathbf{P}^{-\frac{1}{2}}_X\mathbf{P}_{X,Y}\mathbf{P}^{-\frac{1}{2}}_Y\bigg)^T\bigg(\mathbf{P}^{-\frac{1}{2}}_X\mathbf{P}_{X,Y}\mathbf{P}^{-\frac{1}{2}}_Y\bigg)=\mathbf{Q}^T\mathbf{Q},
\end{equation*}
where $\mathbf{Q}$ is defined in (\ref{Q}). Also, $\mathbf{c}$ is the eigenvector of $\mathbf{R}$ corresponding to the eigenvalue of 1, which follows from:
\begin{align*}
\mathbf{R}\mathbf{c}&=\mathbf{P}^{\frac{1}{2}}_Y\mathbf{P}^T_{X|Y}\mathbf{P}^{-1}_X\mathbf{P}_{X|Y}\mathbf{P}^{\frac{1}{2}}_Y\mathbf{c}\\
&=\mathbf{P}^{\frac{1}{2}}_Y\mathbf{P}^T_{X|Y}\mathbf{P}^{-1}_X\mathbf{P}_{X|Y}\mathbf{P}_Y\mathbf{1}_{|\mathcal{Y}|}\\
&=\mathbf{P}^{\frac{1}{2}}_Y\mathbf{P}^T_{X|Y}\mathbf{P}^{-1}_X\mathbf{P}_{X|Y}\mathbf{p}_Y\\
&=\mathbf{P}^{\frac{1}{2}}_Y\mathbf{P}^T_{X|Y}\mathbf{P}^{-1}_X\mathbf{p}_X\\
&=\mathbf{P}^{\frac{1}{2}}_Y\mathbf{P}^T_{X|Y}\mathbf{1}_{|\mathcal{X}|}\\
&=\mathbf{P}^{\frac{1}{2}}_Y\mathbf{1}_{|\mathcal{Y}|}\\
&=\mathbf{c}.
\end{align*}
Therefore, the eigenvalues of the matrix $(\mathbf{I}-\mathbf{c}\mathbf{c}^T)\mathbf{R}$ are $\lambda_1=\sigma_2^2\geq\lambda_2=\sigma_3^2\geq\cdots$ and 0, where $\sigma_i$'s are the singular values of matrix $\mathbf{Q}$, and hence, $\lambda_1=\rho_m^2$. This leads us to the following equality for the maximal correlation
\begin{equation*}
\rho_m^2(X,Y)=\limsup_{\substack{\mathbf{q}_Y\to\mathbf{p}_Y\\\mathbf{q}_Y\neq\mathbf{p}_Y}}r(\mathbf{q}_Y).
\end{equation*}
The other eigenvalues of $(\mathbf{I}-\mathbf{cc}^T)\mathbf{R}$ (or equivalently, the other singular values of matrix $\mathbf{Q}$, except the largest one) can be interpreted  in a similar way. Assume that $\mathbf{v}_1$ is the maximizer of (\ref{B}), i.e., $\mathbf{v}_1$ is the eigenvector of $(\mathbf{I}-\mathbf{cc}^T)\mathbf{R}$ that corresponds to the eigenvalue $\lambda_1=\rho_m^2$. Equivalently, when $\mathbf{q}_Y\to\mathbf{p}_Y\ (\mathbf{q}_Y\neq\mathbf{p}_Y)$, the ratio in (\ref{D}) is maximized if $\mathbf{q}_Y$ converges to $\mathbf{p}_Y$ in the direction of $\boldsymbol{\epsilon}_1=\mathbf{P}_Y^{\frac{1}{2}}\mathbf{v}_1$. If besides the constraints in (\ref{B2}), we also impose the constraint that $\mathbf{v}$ should be orthogonal to $\mathbf{v}_1$, i.e., replacing $\mathbf{c}$ by matrix $\mathbf{C}$ whose first and second columns are, respectively, $\mathbf{c}$ and $\mathbf{v}_1$, the maximum of (\ref{B}) would be $\lambda_2=\sigma_3^2$, achieved by its corresponding eigenvector $\mathbf{v}_2$. 
Equivalently, when $\mathbf{q}_Y\to\mathbf{p}_Y\ (\mathbf{q}_Y\neq\mathbf{p}_Y)$ and $(\mathbf{q}_Y-\mathbf{p}_Y)\perp \mathbf{P}_Y^{\frac{1}{2}}\mathbf{v}_1$, the ratio in (\ref{D}) is maximized if $\mathbf{q}_Y$ converges to $\mathbf{p}_Y$ in the direction of $\boldsymbol{\epsilon}_2=\mathbf{P}_Y^{\frac{1}{2}}\mathbf{v}_2$. This procedure can be continued to cover all the singular values of matrix $\mathbf{Q}$, from the second largest to the smallest. 
\end{proof} 
\begin{remark}
A natural question arises whether the largest singular value, which is one, has a similar interpretation. If in (\ref{B2}), the constraint $\mathbf{c}^T\cdot\mathbf{v}=0$ is omitted, the maximum of (\ref{B}) would be 1, which occurs at $\mathbf{v}=\mathbf{c}$. The constraint $\mathbf{c}^T\cdot\mathbf{v}=0$ is due to $\mathbf{1}^T_{|\mathcal{Y}|}\cdot\boldsymbol{\epsilon}=0$, which in turn results from the fact that $\mathbf{q}_Y$ is a probability vector. 
Therefore, if the definition of relative entropy is extended to the vectors with positive elements, when $\mathbf{q}_Y\to\mathbf{p}_Y\ (\mathbf{q}_Y\neq\mathbf{p}_Y)$ and $\mathbf{q}_Y$ can be any vector with positive elements, the ratio in (\ref{D}) is maximized if $\mathbf{q}_Y$ converges to $\mathbf{p}_Y$ in the direction of $\boldsymbol{\epsilon}_0=\mathbf{P}_Y^{\frac{1}{2}}\mathbf{c}=\mathbf{p}_Y$.
\end{remark}  
\begin{remark}
It can be readily verified that
\begin{align}
\frac{1}{V^*}&=\lim_{\delta\to 0}\sup_{\substack{U:X-Y-U\\ \mathds{E}_U\left[ D\left(p_{X|U}(\cdot|U)||p_X(\cdot)\right)\right]\leq\delta}}\frac{I(X;U)}{I(Y;U)},\label{fed1}\\
\rho_m^2(X;Y)&=\lim_{\delta\to 0}\sup_{\substack{U:X-Y-U\\ \max_u D\left(p_{X|U}(\cdot|u)||p_X(\cdot)\right)\leq \delta}}\frac{I(X;U)}{I(Y;U)},\label{fed2}
\end{align}
where $V^*$ is given in (\ref{vdef}). In (\ref{fed1}), we use the convention $\frac{1}{\infty}=0$; in (\ref{fed1}), the expectation term $\mathds{E}_U[D\left(p_{X|U}(\cdot|U)||p_X(\cdot)\right)]$ is equal to $I(X;U)$, and it is written in this form to emphasize its relation to maximal correlation: while we impose an average constraint in (\ref{fed1}), a per-realization constraint is imposed in (\ref{fed2}). 
\end{remark} 
\section{Conclusions}
This paper addresses the problem of perfect privacy, where the goal is to find the maximum $I(Y;U)$, while guaranteeing $I(X;U)=0$. This problem boils down to a standard linear program when the utility is measured by the mutual information between $Y$ and $U$, as well as other utility measures such as mean-square error and probability of error. By solving this LP, upper and lower bounds for the cardinality of the disclosed data and the maximum utility are obtained. It is shown that when the private variable and the useful data form a jointly Gaussian pair, utility can be obtained only at the expense of privacy; that is, perfect privacy is not feasible. On the other hand, when the privacy-preserving mapping has direct access to both the useful data $Y$ and the latent variable $X$, perfect privacy is feasible and the utility is actually unbounded. Finally, we have investigated the slope of the optimal utility-privacy trade-off curve as we approach to the perfect privacy point, i.e., $I(X;U)=0$. We observe that if perfect privacy is not feasible, this slope is finite, and provide two lower bounds on it. However, when perfect privacy is feasible, under mild conditions, this slope is infinite, i.e., the rate of disclosing information is infinite for a vanishingly small privacy leakage. 
\appendices
 \section{}\label{app1}
Let $\mathcal{U}$ be an arbitrary set. 
Let $\mathbb{S}_{X,Y}$ be the set of probability vectors defined in (\ref{poly}). Let $\mathcal{Q}$ denote an index set of $\mbox{rank}(\mathbf{P}_{X|Y})$ linearly independent columns of $\mathbf{P}_{X|Y}$. Hence, the columns corresponding to the index set $\mathcal{Q}^c=[|{\mathcal{Y}}|]\backslash \mathcal{Q}$ can be written as a linear combination of the columns indexed by $\mathcal{Q}$. Let $\pi:[\mbox{nul}(\mathbf{P}_{X|Y})]\to\mathcal{Q}^c$ such that $\pi(i)<\pi(j)$ for $i<j,\forall i,j\in [\mbox{nul}(\mathbf{P}_{X|Y})]$.  Let $\mathbf r :\mathbb{S}_{X,Y}\to\mathbb{R}^{\mbox{nul}(\mathbf{P}_{X|Y})+1}$ be a vector-valued mapping defined element-wise as
\begin{align*}
r_{i}(\mathbf{p})&=\mathbf{p}(\pi(i)),\ \forall i\in[\mbox{nul}(\mathbf{P}_{X|Y})]\nonumber\\
r_{\mbox{nul}(\mathbf{P}_{X|Y})+1}(\mathbf{p})&=H(\mathbf{p}),
\end{align*}
where $\mathbf{p}(\pi(i))$ denotes the $\pi(i)$-th element of the probability vector $\mathbf{p}$. Since $\mathbb{S}_{X,Y}$ is a closed and bounded subset of $\mathcal{P}({\mathcal{Y}})$, it is compact. Also, $\mathbf{r}$ is a continuous mapping from $\mathbb{S}_{X,Y}$ to $\mathbb{R}^{\mbox{nul}(\mathbf{P}_{X|Y})+1}$. Therefore, from the support lemma \cite{Elgamal}, for every $U\sim F(u)$ defined on $\mathcal{U}$, there exists a random variable $U'\sim p(u')$ with $|\mathcal{U'}|\leq \mbox{nul}(\mathbf{P}_{X|Y})+1$ and a collection of conditional probability vectors $\mathbf{p}_{Y|u'}\in\mathbb{S}_{X,Y}$ indexed by $u'\in\mathcal{U}'$, such that
\begin{equation*}
\int_{\mathcal{U}}r_i(\mathbf{p}_{Y|u})dF(u)=\sum_{u'\in\mathcal{U'}}r_i(\mathbf{p}_{Y|u'})p(u'),\ i\in[\mbox{nul}(\mathbf{P}_{X|Y})+1].
\end{equation*}
It can be verified that by knowing the marginal $\mathbf{p}_{X}$, and the $\mbox{nul}(\mathbf{P}_{X|Y})$ elements of $\mathbf{p}_{Y}$ corresponding to index set $\mathcal{Q}^c$, the remaining $\mbox{rank}(\mathbf{P}_{X|Y})$ elements of $\mathbf{p}_{Y}$ can be uniquely identified by solving $\mathbf{p}_X=\mathbf{P}_{X|Y}\mathbf{p}_Y$. Therefore, for an arbitrary $U$ in $X-Y-U$, that satisfies $X\independent U$, the terms $p_Y(\cdot)$, and $I(Y;U)$ are preserved if $U$ is replaced with $U'$. So are the condition of independence $X\independent U'$ as $\mathbf{p}_{Y|u'}\in\mathbb{S}_{X,Y}, \forall u'\in\mathcal{U}^*$. Since we can simply construct the Markov chain $X-Y-U'$, there is no loss of optimality in considering $|\mathcal{U}|\leq\mbox{nul}(\mathbf{P}_{X|Y})+1$.

The attainability of the supremum follows from the continuity of $I(Y;U)$, and the compactness of $\mathbb{S}_{X,Y}$, since $\mathcal{X,Y}$ are finite.
 \section{}\label{app3}
Assume that the minimum in (\ref{min}) is achieved by $K$($\leq \textnormal{nul}(\mathbf{P}_{X|Y})+1$) points in $\mathbb{S}_{X,Y}$. We prove that all of these $K$ points must belong to the extreme points of $\mathbb{S}_{X,Y}$. Let $\mathbf{p}$ be an arbitrary point among these $K$ points. $\mathbf{p}$ can be written as\footnote{The convex polytope $\mathbb{S}_{X,Y}$ is a $(\textnormal{nul}(\mathbf{P}_{X|Y}))$-dimensional convex set. Therefore, any point in $\mathbb{S}_{X,Y}$ can be written as a convex combination of at most $\textnormal{nul}(\mathbf{P}_{X|Y})+1$ extreme points of $\mathbb{S}_{X,Y}$.}
\begin{equation}\label{coco}
\mathbf{p}=\sum_{i=1}^{\textnormal{nul}(\mathbf{P}_{X|Y})+1}\alpha_i\mathbf{p}_i,
\end{equation}
where $\alpha_i\geq 0\ (\forall i\in[\textnormal{nul}(\mathbf{P}_{X|Y})+1])$ and $\sum_{i=1}^{\textnormal{nul}(\mathbf{P}_{X|Y})+1}\alpha_i=1$; points $\mathbf{p}_i \ (\forall i\in[\textnormal{nul}(\mathbf{P}_{X|Y})+1])$ belong to the extreme points of $\mathbb{S}_{X,Y}$ and $\mathbf{p}_i\neq\mathbf{p}_j$ ($i\neq j$).
From the concavity of entropy, we have
\begin{equation}\label{strict}
H(\mathbf{p})\geq\sum_{i=1}^{\textnormal{nul}(\mathbf{P}_{X|Y})+1}\alpha_iH(\mathbf{p}_i),
\end{equation}
where the equality holds if and only if all of the $\alpha_i$s but one are zero. From the definition of an extreme point, if $\mathbf{p}$ is not an extreme point of $\mathbb{S}_{X,Y}$, it can be written as in (\ref{coco}) with at least two non-zero $\alpha_i$s, which makes the inequality in (\ref{strict}) strict. However, this violates the assumption that the $K$ points achieve the minimum. Hence, all of the $K$ points of the minimizer must belong to the set of extreme points of $\mathbb{S}_{X,Y}$.
{\color{black}
\section{}\label{AppendixCC}
Fix an arbitrary $x_0\in\mathcal{X}$. We show that in the evaluation of $G_0(X,Y)$, for any $u\in\mathcal{U}$, there exists $y'\in\mathcal{Y}$ such that $p(x_0,y'|u)>0$, and $p(x_0,y|u)=0,\ \forall y\neq y'$. 

If $p(x_0,y|u)=0,\ \forall y\in\mathcal{Y}$, we obtain $p(x_0|u)=\sum_yp(x_0,y|u)=0$, which contradicts the condition $p(x_0|u)=p(x_0)$. Hence, the first claim is proved, i.e., there exists $y'\in\mathcal{Y}$ such that $p(x_0,y'|u)>0$, for any $u\in\mathcal{U}$.

Let $\mathcal{U}\triangleq\{u_1,\ldots,u_{|\mathcal{U}|}\}$. Assume that for some $u_j,\ j\in[|\mathcal{U}|]$, we have that $p(x_0,y'|u_j)>0$, and $p(x_0,y''|u_j)>0$ with $y'\neq y''$. It is shown that this cannot be optimal by construction. Assume the random variable $\hat{U}\in(\mathcal{U}\backslash\{u_j\})\cup\{u_j',u_j''\}$ such that 
\begin{align*}
    p_{X,Y|\hat{U}}(x_0,y'|u_j'),p_{X,Y|\hat{U}}(x_0,y''|u_j'')&\triangleq p_{X,Y|U}(x_0,y'|u_j)+p_{X,Y|U}(x_0,y''|u_j)\\p_{X,Y|\hat{U}}(x_0,y''|u_j'),p_{X,Y|\hat{U}}(x_0,y'|u_j'')&\triangleq 0,\\
    p_{\hat{U}}(u_j')&\triangleq p_{U}(u_j)\frac{p_{X,Y|U}(x_0,y'|u_j)}{p_{X,Y|U}(x_0,y'|u_j)+p_{X,Y|U}(x_0,y''|u_j)}\nonumber\\
    p_{\hat{U}}(u_j'')&\triangleq p_{U}(u_j)\frac{p_{X,Y|U}(x_0,y''|u_j)}{p_{X,Y|U}(x_0,y'|u_j)+p_{X,Y|U}(x_0,y''|u_j)}\nonumber\\
    p_{X,Y,\hat{U}}(x,y,u)&\triangleq p_{X,Y,U}(x,y,u),\ \forall (x,y,u)\in\mathcal{X}\times\mathcal{Y}\times(\mathcal{U}\backslash\{u_j,u_j',u_j''\})\nonumber\\
    p_{X,Y,\hat{U}}(x,y|u)&\triangleq p_{X,Y,U}(x,y|u_j),\ \forall u\in\{u_j',u_j''\}, \forall (x,y)\neq (x_0,y'),(x_0,y'').
\end{align*}
It can be verified that with this construction, the marginal $p_{X,Y}$ is preserved in $(X,Y,\hat{U})$, $X\independent \hat{U}$, and $H(Y|\hat{U})<H(Y|U)$ due to strict concavity of entropy, which in turn, contradicts the attainability of $G_0(X,Y)$ by $p_{U|X,Y}$. Hence, we must have $p(x_0,y|u)=0,\ \forall y\neq y'$. The proof is complete by noting that $x_0$ was chosen arbitrarily.
\section{}\label{AppendixDD}
Let $W\triangleq(X,Y)$. Hence, in the full data observation model, we have that $X-W-U$ form a Markov chain. For the binary matrix $\mathbf{P}_{X|W}$ (all elements being 0 or 1), which has $|\mathcal{X}|$ rows and $|\textnormal{supp}(X,Y)|$ columns, we have $\textnormal{rank}(\mathbf{P}_{X|W})=|\mathcal{X}|$. Hence, the upper bound in (\ref{upperU}) follows similarly to the analysis in Appendix \ref{app1}. 
Fix an arbitrary $x_0\in\mathcal{X}$. We have that for each $y'\in\{y\in\mathcal{Y}|p(x_0,y)>0\}$, there must exist a corresponding $u'\in\mathcal{U}$ such that $p(x_0,y'|u')>0$, since otherwise, we get $p(x_0,y')=0$, which is a contradiction. Moreover, from lemma \ref{lemcardi}, for this $u'$, we have $p(x_0,y|u')=0,\ \forall y\neq y'$, which results in $|\mathcal{U}|\geq|\{y\in\mathcal{Y}|p(x_0,y)>0\}|$. Finally, by noting that $x_0$ is chosen arbitrarily, and $p(x_0)>0,\ \forall x_0\in\mathcal{X}$, the lower bound in (\ref{upperU}) is obtained.

\section{}\label{AppendixEE}
We have
\begin{align}
    \mathds{E}_X[J(X,p_Y)]&=\mathds{E}_X\bigg[\frac{1}{\max_{y\in\mathcal{Y}}\frac{p_{Y|X}(y|X)}{p_Y(y)}}\bigg]\nonumber\\
    &\geq \frac{1}{\mathds{E}_X\bigg[\max_{y\in\mathcal{Y}}\frac{p_{Y|X}(y|X)}{p_Y(y)}\bigg]}\label{convexity}\\
    &=\frac{1}{\mathds{E}_X\bigg[\max_{y\in\mathcal{Y}}\frac{p_{X|Y}(X|y)}{p_X(X)}\bigg]}\label{mani}\\
    &=\frac{1}{\sum_{x\in\mathcal{X}}\max_{y\in\mathcal{Y}}p_{X|Y}(x|y)}\nonumber\\
    &=2^{-\mathcal{L}(Y\to X)},\label{aakk}
\end{align}
where (\ref{convexity}) follows from Jensen's inequality and the convexity of $f(t)=\frac{1}{t}$ for $t>0$; (\ref{mani}) follows from Bayes's rule, and $p_X(x)>0,\ \forall x\in\mathcal{X}$; in (\ref{aakk}), we note that $p_Y(y)>0,\ \forall y\in\mathcal{Y}$. Since the function $f(t)=\frac{1}{t}$ for $t>0$ is strictly convex, the inequality in (\ref{convexity}) is tight if and only if the term inside the expectation does not vary with $x$. 
}
{\color{black}
\section{}\label{AppendixFD}
Without loss of generality, assume that $y_0$ is the first element of $\mathcal{Y}$. Hence, $\mathbf{e}_1\triangleq\begin{bmatrix}
1\ \mathbf{0}_{|\mathcal{Y}|-1}^T
\end{bmatrix}^T$ is an extreme point of $\mathds{S}_{X,Y}$, since we have $\mathbf{p}_X=\mathbf{P}_{X|Y}\mathbf{e}_1$, and $\mathbf{e}_1$ cannot be written as a convex combination of two distinct probability vectors in $\mathds{S}_{X,Y}$. Furthermore, the first element of all the other extreme points of $\mathds{S}_{X,Y}$ is zero as proved by contradiction in what follows. If $\mathbf{r}$ ($\neq \mathbf{e}_1$) is an extreme point of $\mathds{S}_{X,Y}$ whose first element is non-zero, we can write it as $\mathbf{r}=\begin{bmatrix}
\alpha\ \mathbf{v}^T
\end{bmatrix}^T$, in which $\alpha\neq 0$, and $\mathbf{v}$ is a vector of probability masses that sum to $1-\alpha$. Since $\mathbf{r}\in\mathds{S}_{X,Y}$, we must have $\mathbf{p}_X=\mathbf{P}_{X|Y}\mathbf{r}$, which, from $p_{X|Y}(\cdot|y_0)=p_X(\cdot)$, results in $\mathbf{p}_X=\mathbf{P}_{X|Y}\mathbf{r}_0$, where $\mathbf{r}_0=\begin{bmatrix}
0\ \frac{1}{1-\alpha}\mathbf{v}^T
\end{bmatrix}^T$ is a probability vector. Therefore, $\mathbf{r}_0\in\mathds{S}_{X,Y}$. However, since $\mathbf{r}$ can be written as a convex combination of $\mathbf{e}_1$ and $\mathbf{r}_0$, i.e., $\mathbf{r}=\alpha\mathbf{e}_1+(1-\alpha)\mathbf{r}_0$, it is concluded that $\mathbf{r}$ cannot be an extreme point of $\mathds{S}_{X,Y}$. Finally, by noting that in the evaluation of $g_0(X,Y)$, only the extreme points of $\mathds{S}_{X,Y}$ are involved, the proof is complete.
}
\section{}\label{ppo}
From the construction in (\ref{kor}), we have $\forall m_{[B]}\in\prod_{i=1}^B\mathcal{E}_i$,
\begin{align*}
\mathbf{P}_{X|Y}\mathbf{s}_{m_{[B]}}&=\sum_{k=1}^{|\mathcal{Y}|}\mathbf{p}_{X|y_k}s_{m_{[B]}}(k)\nonumber\\
&=\sum_{k\in\cup_{i=1}^B\mathcal{E}_i}\mathbf{p}_{X|y_k}s_{m_{[B]}}(k)+\sum_{k\not\in\cup_{i=1}^B\mathcal{E}_i}\mathbf{p}_{X|y_k}s_{m_{[B]}}(k)\nonumber\\
&=\sum_{k\in[B]}\mathbf{p}_{X|y_{m_k}}\left(\sum_{t\in \mathcal{E}_{k}}p_Y(y_t)\right)+\sum_{k\not\in\cup_{i=1}^B\mathcal{E}_i}\mathbf{p}_{X|y_k}p_Y(y_k)\\
&=\sum_{k\in[B]}\sum_{t\in \mathcal{E}_{k}}\mathbf{p}_{X|y_t}p_Y(y_t)+\sum_{k\not\in\cup_{i=1}^B\mathcal{E}_i}\mathbf{p}_{X|y_k}p_Y(y_k)\\
&=\sum_{k\in\cup_{i=1}^B\mathcal{E}_i}\mathbf{p}_{X|y_k}p_Y(y_k)+\sum_{k\not\in\cup_{i=1}^B\mathcal{E}_i}\mathbf{p}_{X|y_k}p_Y(y_k)\\
&=\sum_{k=1}^{|\mathcal{Y}|}\mathbf{p}_{X|y_k}p_Y(y_k)\\
&=\mathbf{p}_X.
\end{align*}
Hence, for all $\mathbf{s}\in\mathbb{S}_{X,Y}'$, we have $\mathbf{s}\in\mathbb{S}_{X,Y}$, which means that $\mathbb{S}_{X,Y}'\subseteq\mathbb{S}_{X,Y}$.

From (\ref{kor}), the non-zero entries of any probability vector $\mathbf{s}\in\mathbb{S}_{X,Y}'$ are the same as the elements of the set in (\ref{mass}), {\color{black} i.e., the mass probabilities of $T^{\mathcal{X}}(Y)$}. Therefore, $H(\mathbf{s})=H(T^{\mathcal{X}}(Y)),\ \forall\mathbf{s}\in\mathbb{S}_{X,Y}'$.

Finally, let the set of $|\mathcal{Y}|$-dimensional probability vectors $\{\mathbf{s}_{m_1}\}_{m_1\in \mathcal{E}_1}$ on $\mathcal{Y}$ be defined element-wise as
\begin{equation}\label{ind0}
s_{m_1}(y_k)=\left\{\begin{array}{ccc}
p_Y(y_k)&\forall k\not\in \mathcal{E}_1\\\sum_{j\in \mathcal{E}_1}p_Y(y_j)&k=m_1\\0&k\neq m_1,k\in \mathcal{E}_1 
\end{array}\right.,\ \ \forall m_1\in \mathcal{E}_1,\forall k\in[|\mathcal{Y}|].
\end{equation}
By induction, {\color{black}define the probability vector $\mathbf{s}_{m_{[n]}}$ on $\mathcal{Y}$ as}
\begin{equation}\label{ind}
s_{m_{[n]}}(y_k)=\left\{\begin{array}{ccc}
s_{m_{[n-1]}}(y_k)&\forall k\not\in \mathcal{E}_n\\\sum_{j\in \mathcal{E}_n}p_Y(y_j)&k=m_n\\0&k\neq m_n,k\in \mathcal{E}_n 
\end{array}\right.,\ \forall m_{[n]}\in \prod_{i=1}^n\mathcal{E}_i,\ 
\forall n\in[2:B],\forall k\in[|\mathcal{Y}|],
\end{equation}
where it can be verified that (\ref{ind}) and (\ref{kor}) are equivalent for $n=B$.
By constructions in (\ref{ind0}) and (\ref{ind}), we can, respectively, write
\begin{equation}\label{prt}
\mathbf{p}_Y=\sum_{m_1\in \mathcal{E}_1}\frac{p_Y(y_{m_1})}{\sum_{k\in \mathcal{E}_1}p_Y(y_k)}\mathbf{s}_{m_1},
\end{equation}
and
\begin{equation}\label{prt2}
\mathbf{s}_{m_{[n-1]}}=\sum_{m_n\in \mathcal{E}_n}\frac{p_Y(y_{m_n})}{\sum_{k\in \mathcal{E}_n}p_Y(y_k)}\mathbf{s}_{m_{[n]}},\ \forall m_{[n-1]}\in \prod_{i=1}^{n-1}\mathcal{E}_i,\ 
\forall n\in[2:B].
\end{equation}
Therefore, $\mathbf{p}_Y$ can be written from (\ref{prt}) and (\ref{prt2}) as
\begin{equation}\label{lin}
\mathbf{p}_Y=\sum_{m_{[B]}\in\prod_{i=1}^B \mathcal{E}_i}\frac{p_Y(m_1)p_Y(m_2)\ldots p_Y(m_B)}{\sum_{k\in \mathcal{E}_1}p_Y(y_k)\sum_{k\in \mathcal{E}_2}p_Y(y_k)\ldots\sum_{k\in \mathcal{E}_B}p_Y(y_k)}\mathbf{s}_{m_{[B]}}.
\end{equation}
By letting
\begin{equation*}
\alpha_{m_{[B]}}=\frac{p_Y(m_1)p_Y(m_2)\ldots p_Y(m_B)}{\sum_{k\in \mathcal{E}_1}p_Y(y_k)\sum_{k\in \mathcal{E}_2}p_Y(y_k)\ldots\sum_{k\in \mathcal{E}_B}p_Y(y_k)},\ \forall m_{[B]}\in \prod_{i=1}^B\mathcal{E}_i,
\end{equation*}
we obtain (\ref{lin0}).
\section{}\label{appE}
Without loss of generality, by an appropriate labelling of the elements in $\mathcal{Y}$, we can assume that $\mathcal{E}_1=\bigg[|\mathcal{E}_1|\bigg]$, $\mathcal{E}_2=\bigg[|\mathcal{E}_1|+1:|\mathcal{E}_1|+|\mathcal{E}_2|\bigg]$, and so on. Let the common column vector corresponding to $\mathcal{E}_m$ be denoted by $\mathbf{p}_m$, i.e. $\mathbf{p}_m\triangleq\mathbf{p}_{X|y_i},\forall i\in \mathcal{E}_m,\forall m\in[B]$. We can write
\begin{equation*}
\mathbf{P}_{X|Y}=\bigg[\underbrace{\mathbf{p}_1,\ldots,\mathbf{p}_1}_{|\mathcal{E}_1|\textnormal{ times}},\underbrace{\mathbf{p}_2,\ldots,\mathbf{p}_2}_{|\mathcal{E}_2|\textnormal{ times}},\ldots,\underbrace{\mathbf{p}_B,\ldots,\mathbf{p}_B}_{|\mathcal{E}_B|\textnormal{ times}},\mathbf{p}_{X|y_{G+1}},\mathbf{p}_{X|y_{G+2}},\ldots,\mathbf{p}_{X|y_{|\mathcal{Y}|}}\bigg],
\end{equation*}
and
\begin{equation*}
\hat{\mathbf{P}}_{X|Y}=\bigg[\mathbf{p}_1,\mathbf{p}_2,\ldots,\mathbf{p}_B,\mathbf{p}_{X|y_{G+1}},\mathbf{p}_{X|y_{G+2}},\ldots,\mathbf{p}_{X|y_{|\mathcal{Y}|}}\bigg].
\end{equation*}
Define the vectors $\forall m\in[B]$ as
\begin{equation}\label{ham3}
{\mathbf{e}}_m^1=\begin{bmatrix}
\mathbf{0}_{\sum_{i=1}^{m-1}|\mathcal{E}_i|}\\1\\-1\\\mathbf{0}_{(|\mathcal{Y}|-\sum_{i=1}^{m-1}|\mathcal{E}_i|-2)}
\end{bmatrix},{\mathbf{e}}_m^2=\begin{bmatrix}
\mathbf{0}_{\sum_{i=1}^{m-1}|\mathcal{E}_i|}\\\frac{1}{2}\\\frac{1}{2}\\-1\\\mathbf{0}_{(|\mathcal{Y}|-\sum_{i=1}^{m-1}|\mathcal{E}_i|-3)}
\end{bmatrix},\ldots,{\mathbf{e}}_m^{|\mathcal{E}_m|-1}=\begin{bmatrix}
\mathbf{0}_{\sum_{i=1}^{m-1}|\mathcal{E}_i|}\\\frac{1}{|\mathcal{E}_m|-1}\mathbf{1}_{|\mathcal{E}_m|-1}\\-1\\\mathbf{0}_{(|\mathcal{Y}|-\sum_{i=1}^{m}|\mathcal{E}_i|)}
\end{bmatrix},
\end{equation}
where $\sum_{i=1}^{m-1}|\mathcal{E}_i|=\emptyset$ when $m=1$.	

\begin{lemma}\label{Prop10}
Let 
\begin{equation*}
\mathbb{N}\triangleq\textnormal{Span}\bigg\{\mathbf{e}_m^{i}\bigg|\forall i\in[|\mathcal{E}_m|-1],\forall m\in[B] \bigg\}.
\end{equation*}
We have
\begin{equation}\label{sube}
\mathbb{N}\subseteq \textnormal{Null}(\mathbf{P}_{X|Y}),
\end{equation}
where (\ref{sube}) holds with equality if and only if $\textnormal{nul}(\hat{\mathbf{P}}_{X|Y})=0$.
\end{lemma}
\begin{proof}
The proof is provided in Appendix \ref{mash1}.
\end{proof}
{\color{black}\subsection{Proof of $\textnormal{nul}(\hat{\mathbf{P}}_{X|Y})=0\Longrightarrow\textnormal{ext}(\mathbb{S}_{X,Y})=\mathbb{S}_{X,Y}'$}}
If $\textnormal{nul}(\hat{\mathbf{P}}_{X|Y})=0$, any element in $\mathbb{S}_{X,Y}'$ is an extreme point of $\mathbb{S}_{X,Y}$. The reasoning is as follows. Note that $\mathbb{S}_{X,Y}'\subseteq\mathbb{S}_{X,Y}$. Hence, it remains to show that no point of $\mathbb{S}_{X,Y}'$ can be written as a convex combination of two different points of $\mathbb{S}_{X,Y}$. Pick an arbitrary point $\mathbf{s}$ in $\mathbb{S}_{X,Y}'$. It can be verified that no $\epsilon>0$ and $\mathbf{e}\in\mathbb{N}$ exist such that both $\mathbf{s}+\epsilon\mathbf{e}$ and $\mathbf{s}-\epsilon\mathbf{e}$ remain probability vectors. This is due to having a negative element in either or both of them. From lemma \ref{Prop10}, we have $\mathbb{N}=\textnormal{Null}(\mathbf{P}_{X|Y})$ in this case, which in turn means that $\mathbf{s}$ cannot be written as a convex combination of two different points of $\mathbb{S}_{X,Y}$. Therefore, the elements in $\mathbb{S}_{X,Y}'$ belong to the extreme points of $\mathbb{S}_{X,Y}$. 

On the other hand, when $\textnormal{nul}(\hat{\mathbf{P}}_{X|Y})=0$, assume that there exists $\mathbf{s}^*\not\in\mathbb{S}_{X,Y}'$ that is an extreme point of $\mathbb{S}_{X,Y}$. We show that this leads to a contradiction. Firstly, note that among the elements of $\mathbf{s}^*$ that correspond to $\mathcal{E}_m,\forall m\in[B]$, there must be at most one non-zero element. This is justified as follows. Assume that $s^*_i$ and $s^*_j$ ($i\neq j$) are two non-zero elements of $\mathbf{s}^*$, where $i,j\in \mathcal{E}_m$ for some $m\in[B]$. Construct the $|\mathcal{Y}|$-dimensional vector $\mathbf{f}$ where $f_i=-f_j=1$, and the remaining terms are zero. Obviously, $\mathbf{f}\in\textnormal{Null}(\mathbf{P}_{X|Y})$, as $\mathbf{P}_{X|Y}\mathbf{f}=\mathbf{0}.$ Let $\epsilon=\min\{s^*_i,s^*_j\}$. It is obvious that $\mathbf{s}^*$ can be written as a convex combination of the vectors $\mathbf{s}^*+\epsilon\mathbf{f}$ and $\mathbf{s}^*-\epsilon\mathbf{f}$, where both are in $\mathbb{S}_{X,Y}$. However, this contradicts the assumption of $\mathbf{s}^*$ being an extreme point of $\mathbb{S}_{X,Y}$. Hence, among the elements of $\mathbf{s}^*$ that correspond to $\mathcal{E}_m,\forall m\in[B]$, at most one element is non-zero. As a result, we can find a point $\mathbf{s}\in\mathbb{S}_{X,Y}'$ whose positions of its non-zero elements in $\cup_{i=1}^B\mathcal{E}_i$ matches those of $\mathbf{s}^*$. Since $\mathbf{s}^*\not\in\mathbb{S}_{X,Y}'$, $\mathbf{s}^*$ must differ with this $\mathbf{s}$ in at least one element. Assume that for some $m\in[B]$, there exists $j\in \mathcal{E}_m$ such that $s^*_j\neq s_j$. Then, the elements of $\Delta\mathbf{s}=\mathbf{s}^*-\mathbf{s}$ that correspond to $\mathcal{E}_m$ are all zero, except $\Delta s_j=s^*_j-s_j\neq 0$. It can then be verified that $\Delta\mathbf{s}$ cannot be written as a linear combination of the vectors in $\mathbb{N}$, as no linear combination of the vectors $\mathbf{e}_m^i,\forall i\in[|\mathcal{E}_m|-1]$ can produce a vector whose elements corresponding to $\mathcal{E}_m$ are all zero except one. Since $\textnormal{nul}(\hat{\mathbf{P}}_{X|Y})=0$, we have from lemma \ref{Prop10} that $\mathbb{N}=\textnormal{Null}(\mathbf{P}_{X|Y})$, which in turn means that $\mathbf{s}^*-\mathbf{s}\not\in\textnormal{Null}(\mathbf{P}_{X|Y})$. Therefore, {\color{black}by noting that $\mathbf{s}\in\mathbb{S}_{X,Y}'\subset \mathbb{S}_{X,Y}$}, we get $\mathbf{s}^*\not\in\mathbb{S}_{X,Y}$, which is a contradiction. If $s^*_j=s_j,\forall j\in\cup_{i=1}^B\mathcal{E}_i$, then we must have $s^*_j\neq s_j$ for some $j\in[G+1:|\mathcal{Y}|]$. Still, $\Delta\mathbf{s}$ cannot be written as a linear combination of the vectors in $\mathbb{N}$, as for any vector $\mathbf{n}\in\mathbb{N}$, we have $n_k=0,\forall k\in[G+1:|\mathcal{Y}|]$. This results in $\mathbf{s}^*\not\in\mathbb{S}_{X,Y}$, which is again a contradiction. Therefore, we conclude that when $\textnormal{nul}(\hat{\mathbf{P}}_{X|Y})=0$, the extreme points of $\mathbb{S}_{X,Y}$ are the elements of $\mathbb{S}_{X,Y}'$. 
{\color{black}\subsection{Proof of $\textnormal{nul}(\hat{\mathbf{P}}_{X|Y})\neq0\Longrightarrow\textnormal{ext}(\mathbb{S}_{X,Y})\cap\mathbb{S}_{X,Y}'=\emptyset$}}
If $\textnormal{nul}(\hat{\mathbf{P}}_{X|Y})\neq0$, from lemma \ref{Prop10}, there must exist a non-zero vector $\mathbf{v}$ such that $\mathbf{v}\in\textnormal{Null}(\mathbf{P}_{X|Y})$ and $\mathbf{v}\not\in\mathbb{N}$. Pick an arbitrary point of $\mathbb{S}_{X,Y}'$. In order to make the analysis simple, let the picked vector be $\hat{\mathbf{s}}_0$, which is 
\begin{equation*}
\hat{\mathbf{s}}_0\triangleq\mathbf{s}_{1,|\mathcal{E}_1|+1,|\mathcal{E}_1|+|\mathcal{E}_2|+1,\ldots,G-|\mathcal{E}_B|+1}=
\begin{bmatrix}
\sum_{i\in \mathcal{E}_1}p_Y(y_i)\\\mathbf{0}_{|\mathcal{E}_1|-1}\\\sum_{i\in \mathcal{E}_2}p_Y(y_i)\\\mathbf{0}_{|\mathcal{E}_2|-1}\\\vdots\\\sum_{i\in \mathcal{E}_B}p_Y(y_i)\\\mathbf{0}_{|\mathcal{E}_B|-1}\\p_Y(y_{G+1})\\\vdots\\p_Y(y_{|\mathcal{Y}|})
\end{bmatrix}
\end{equation*}
From $\mathbf{v}$, we can construct a non-zero vector $\tilde{\mathbf v}\in\textnormal{Null}(\mathbf{P}_{X|Y})$, as done in (\ref{hij}). Then, it is obvious that for sufficiently small $\epsilon>0$, $\hat{\mathbf{s}}_0$ can be written as a convex combination of $\hat{\mathbf{s}}_0+\epsilon\tilde{\mathbf v}$ and $\hat{\mathbf{s}}_0-\epsilon\tilde{\mathbf v}$, where both are in $\mathbb{S}_{X,Y}$. This shows that $\hat{\mathbf{s}}_0$ cannot be an extreme point of $\mathbb{S}_{X,Y}$. A similar approach\footnote{The only difference is in constructing a vector $\tilde{\mathbf{v}}$, such that when a point of $\mathbb{S}_{X,Y}'$ is perturbed along the direction of $\tilde{\mathbf{v}}$, it still lies in $\mathbb{S}_{X,Y}$. This can be done by noting that it is sufficient to construct a $\tilde{\mathbf{v}}$ whose positions of zero elements in $[G]$ match those of the arbitrary element from $\mathbb{S}_{X,Y}$. Similarly to how $\tilde{\mathbf v}(\in\textnormal{Null}(\mathbf{P}_{X|Y}))$ was constructed from $\mathbf{v}$ in (\ref{hij}), by using other orthogonal vectors in $\mathbb{N}$, instead of $\mathbf{e}_m^i$, a new $\tilde{\mathbf v}$ can be constructed whose positions of zero elements in $[G]$ match those of the arbitrary element from $\mathbb{S}_{X,Y}$.} can be applied to show that the other points of $\mathbb{S}_{X,Y}'$ do not belong to the set of extreme points of $\mathbb{S}_{X,Y}$. Hence, from $\textnormal{nul}(\hat{\mathbf{P}}_{X|Y})\neq0$, we conclude that none of the elements in $\mathbb{S}_{X,Y}'$ is an extreme point of $\mathbb{S}_{X,Y}$.
\section{}\label{mash1}
The fact that $\mathbb{N}\subseteq \textnormal{Null}(\mathbf{P}_{X|Y})$ can be verified by observing that $\mathbf{P}_{X|Y}\mathbf{e}_m^i=\mathbf{0},\forall i\in[|\mathcal{E}_m|-1],\forall m\in[B]$.
{\color{black}\subsection{Proof of $\textnormal{nul}(\hat{\mathbf{P}}_{X|Y})=0\Longrightarrow\mathbb{N}= \textnormal{Null}(\mathbf{P}_{X|Y})$}}
If $\textnormal{nul}(\hat{\mathbf{P}}_{X|Y})=0$, we must have $\mathbb{N}= \textnormal{Null}(\mathbf{P}_{X|Y})$. If this is not true, from (\ref{sube}), there must exist a non-zero vector $\mathbf{v}$ such that $\mathbf{v}\in\textnormal{Null}(\mathbf{P}_{X|Y})$ and $\mathbf{v}\not\in\mathbb{N}$. Let $v_i$ denote the $i^{\textnormal{th}}$ element in $\mathbf{v}$. We can write
\begin{equation}\label{hij}
\mathbf{v}+\sum_{m=1}^B\sum_{i=1}^{|\mathcal{E}_m|-1}\alpha_m^i\mathbf{e}_m^i=\begin{bmatrix}
\sum_{i\in \mathcal{E}_1}v_i\\\mathbf{0}_{|\mathcal{E}_1|-1}\\\sum_{i\in \mathcal{E}_2}v_i\\\mathbf{0}_{|\mathcal{E}_2|-1}\\\vdots\\\sum_{i\in \mathcal{E}_B}v_i\\\mathbf{0}_{|\mathcal{E}_B|-1}\\v_{G+1}\\\vdots\\v_{|\mathcal{Y}|}
\end{bmatrix}=\tilde{\mathbf{v}},
\end{equation}
where it can be verified that the coefficients $\alpha_m^i,\forall i\in[|\mathcal{E}_m|-1],\forall m\in[B]$ are obtained uniquely, as the vectors $\mathbf{e}_m^i$ are mutually orthogonal. Since $\mathbf{v},\mathbf{e}_m^i\in\textnormal{Null}(\mathbf{P}_{X|Y}),\forall i\in[|\mathcal{E}_m|-1],\forall m\in[B]$, we have $\tilde{\mathbf{v}}\in\textnormal{Null}(\mathbf{P}_{X|Y})$. Also, note that $\tilde{\mathbf{v}}$ is a non-zero vector, since otherwise (\ref{hij}) would result in $\mathbf{v}\in\mathbb{N}$. Finally, from the structure of $\tilde{\mathbf{v}}$ and $\hat{\mathbf{P}}_{X|Y}$, we observe that $\hat{\mathbf{P}}_{X|Y}\tilde{\mathbf{v}}'={\mathbf{P}}_{X|Y}\tilde{\mathbf{v}}=\mathbf{0}$, where $\tilde{\mathbf{v}}'$ is obtained from eliminating the zero vectors of $\tilde{\mathbf{v}}$, denoted by $\mathbf{0}_{|\mathcal{E}_i|-1},\forall i\in[B]$, in (\ref{hij}). Since $\tilde{\mathbf{v}}$ is a non-zero vector, so must be $\tilde{\mathbf{v}}'$. Hence, $\textnormal{nul}(\hat{\mathbf{P}}_{X|Y})\neq0$, which is a contradiction. Therefore, we must have $\mathbb{N}= \textnormal{Null}(\mathbf{P}_{X|Y})$.
{\color{black}\subsection{Proof of $\mathbb{N}= \textnormal{Null}(\mathbf{P}_{X|Y})\Longrightarrow\textnormal{nul}(\hat{\mathbf{P}}_{X|Y})=0$}}
If $\mathbb{N}= \textnormal{Null}(\mathbf{P}_{X|Y})$, we must have $\textnormal{nul}(\hat{\mathbf{P}}_{X|Y})=0$. If this is not true, there exists a non-zero vector $\tilde{\mathbf{r}}'$ such that $\hat{\mathbf{P}}_{X|Y}\tilde{\mathbf{r}}'=\mathbf{0}$, and correspondingly a non-zero vector $\tilde{\mathbf{r}}$ such that ${\mathbf{P}}_{X|Y}\tilde{\mathbf{r}}=\mathbf{0}$, where the relation between $\tilde{\mathbf{r}}'$ and $\tilde{\mathbf{r}}$ is similar to that between $\tilde{\mathbf{v}}'$ and $\tilde{\mathbf{v}}$ in the previous paragraph.
Therefore, we have $\tilde{\mathbf{r}}\in\textnormal{Null}(\mathbf{P}_{X|Y})$. However, it can be verified that due to the structure of the vectors $\mathbf{e}_m^i$, i.e. the positions of the zero and non-zero elements in $[G]$, $\tilde{\mathbf{r}}$ cannot be written as a linear combination of the vectors $\mathbf{e}_m^i$. This results in $\tilde{\mathbf{r}}\not\in\mathbb{N}$, which is a contradiction, as we assumed $\mathbb{N}= \textnormal{Null}(\mathbf{P}_{X|Y})$. This proves that $\textnormal{nul}(\hat{\mathbf{P}}_{X|Y})=0$.
\section{}\label{app5}
When $g_0(X,Y)=0$, there is no $\mathbf{q}_Y\neq\mathbf{p}_Y$, such that $\mathbf{q}_X=\mathbf{p}_X$, since otherwise we could have constructed a random variable $U\in\{u_1,u_2\}$, and a sufficiently small $\alpha>0$, such that
\begin{equation*}
p_U(u_1)=\alpha,\ \mathbf{p}_{Y|u_1}=\mathbf{q}_Y,\ \mathbf{p}_{Y|u_2}=\frac{1}{1-\alpha}(\mathbf{p}_Y-\alpha\mathbf{q}_Y),
\end{equation*} 
where the sufficiently small $\alpha$ makes $\mathbf{p}_{Y|u_2}$ still a probability vector. With this construction, it can be verified that $X-Y-U$, $X\independent U$ and $Y\not\independent U$ which contradicts $g_0(X,Y)=0$. Hence, the only way to have $V^*>M, \forall M\in\mathbb{R}$ is through the existence of a sequence of distributions, i.e. $\{\mathbf{q}^n_Y\}_n$, where $\mathbf{q}^n_Y\neq\mathbf{p}_Y,\forall n$ and $\mathbf{q}^n_X\to\mathbf{p}_X$. Since perfect privacy is not feasible, we must have $|\mathcal{Y}|\leq|\mathcal{X}|$ and $\sigma_i(\mathbf{P}_{X|Y})\neq 0,\forall i\in\bigg[\min\{|\mathcal{X}|,|\mathcal{Y}|\}\bigg]$. This means that in order to have $\mathbf{q}^n_X\to\mathbf{p}_X$, we must have $\mathbf{q}^n_Y\to\mathbf{p}_Y$. We know that when $\mathbf{q}_Y\to\mathbf{p}_Y (\mathbf{q}_Y\neq\mathbf{p}_Y)$, the ratio in (\ref{D}) is bounded below by the minimum eigenvalue of the matrix $\mathbf{Q}^T\mathbf{Q}$. If for an arbitrary non-zero vector $\mathbf{e}$, we have $\mathbf{Q}\mathbf{e}=\mathbf{0}$, then we must have 
\begin{align*}
\mathbf{Q}\mathbf{e}=\mathbf{P}_X^{-\frac{1}{2}}\mathbf{P}_{X,Y}\mathbf{P}_Y^{-\frac{1}{2}}\mathbf{e}&=\mathbf{P}_X^{-\frac{1}{2}}\mathbf{P}_{X|Y}\underbrace{\mathbf{P}_Y^{\frac{1}{2}}\mathbf{e}}_{\mathbf{e}'}=\mathbf{0},
\end{align*}
which is not possible, since $\mathbf{e}'$ is a non-zero vector, and so is $\mathbf{P}_{X|Y}\mathbf{e}'$ due to the fact that $\sigma_i(\mathbf{P}_{X|Y})\neq 0,\forall i\in\bigg[\min\{|\mathcal{X}|,|\mathcal{Y}|\}\bigg]$ and $|\mathcal{Y}|\leq|\mathcal{X}|$, i.e. the null space of $\mathbf{P}_{X|Y}$ is only the all-zero vector. Therefore, the minimum eigenvalue of the matrix $\mathbf{Q}^T\mathbf{Q}$ is bounded away from zero. Equivalently, the inverse of (\ref{D}) is bounded above by the inverse of the minimum eigenvalue of $\mathbf{Q}^T\mathbf{Q}$. Hence, $V^*<+\infty$.

The proof of the second direction is immediate, since having $g_0(X,Y)>0$ leads to the existence of $\mathbf{q}_Y\neq\mathbf{p}_Y$, such that $\mathbf{q}_X=\mathbf{p}_X$, which in turn violates $V^*<+\infty$.
\section{}\label{app6}
Firstly, note that for any point $\mathbf{q}_Y$ that satisfies $0<D(\mathbf{q}_Y||\mathbf{p}_{Y|u^*})<+\infty$, we have $\mathbf{q}_Y\not\in\mathbb{S}_{X,Y}$ (i.e., $\mathbf{q}_X\neq\mathbf{p}_X$), where $\mathbb{S}_{X,Y}$ is defined in (\ref{poly}), since otherwise from the fact that $D(\mathbf{q}_Y||\mathbf{p}_{Y|u^*})<+\infty$, for sufficiently small $\epsilon$, we can make the probability vector $\mathbf{q}_Y'=\frac{1}{1-\epsilon}(\mathbf{p}_{Y|u^*}-\epsilon\mathbf{q}_Y)$ which also belongs to $\mathbb{S}_{X,Y}$, as $\mathbf{P}_{X|Y}\mathbf{q}_Y'=\mathbf{p}_X$. However, this violates the fact that $\mathbf{p}_{Y|u^*}$ is an extreme point of $\mathbb{S}_{X,Y}$, since it can be written as a convex combination of two points of $\mathbb{S}_{X,Y}$, i.e. $\mathbf{q}_Y$ and $\mathbf{q}_Y'$ ($\mathbf{q}_Y\neq\mathbf{q}_Y'$). Alternatively, we can say that for any $\mathbf{q}_Y$ that satisfies $0<D(\mathbf{q}_Y||\mathbf{p}_{Y|u^*})<+\infty$, we have $(\mathbf{q}_Y-\mathbf{p}_{Y|u^*})\not\in\textnormal{Null}(\mathbf{P}_{X|Y})$.
Hence, the only way to have $\psi(u^*)$ possibly unbounded is through the existence of a sequence of distributions, i.e. $\{\mathbf{q}^n_Y\}_n$, where $0<D(\mathbf{q}^n_Y||\mathbf{p}_{Y|u^*})<+\infty,\forall n$ and $\mathbf{q}^n_X\to\mathbf{p}_X$, which requires $\mathbf{q}_Y^n$ converging to a point of $\mathbb{S}_{X,Y}$. Let $\mathcal{I}_{u^*}$ denote the set of indices corresponding to the zero elements of $\mathbf{p}_{Y|u^*}$. 
Let $\mathbb{T}$ denote the set of probability vectors $\mathbf{p}$, such that $D(\mathbf{p}||\mathbf{p}_{Y|u^*})<+\infty$. In other words, $\mathbb{T}$ is the set of probability vectors whose elements corresponding to the indices in $\mathcal{I}_{u^*}$ are zero. Since $\mathbb{T}$ is a closed set, we conclude that if $\mathbf{q}_Y^n$ ($\in\mathbb{T}$) converges to $\mathbf{p}_0$ (a point of $\mathbb{S}_{X,Y}$), $\mathbf{p}_0$ must also be in $\mathbb{T}$, i.e. it satisfies $D(\mathbf{p}_0||\mathbf{p}_{Y|u^*})<+\infty$. If $D(\mathbf{p}_0||\mathbf{p}_{Y|u^*})>0$, from what mentioned before, we have $(\mathbf{p}_0-\mathbf{p}_{Y|u^*})\not\in\textnormal{Null}(\mathbf{P}_{X|Y})$, which contradicts the fact that $\mathbf{p}_0\in\mathbb{S}_{X,Y}$, hence, we must have $\mathbf{p}_0=\mathbf{p}_{Y|u^*}$. Therefore, it suffices to consider the following problem
\begin{equation*}
\liminf_{\substack{\mathbf{q}_Y:\mathbf{q}_Y\to\mathbf{p}_{Y|u^*}\\0<D(\mathbf{q}_Y||\mathbf{p}_{Y|u^*})<+\infty}}\frac{D(\mathbf{q}_X||\mathbf{p}_{X})}{D(\mathbf{q}_Y||\mathbf{p}_{Y|u^*})}.
\end{equation*}
Similarly to the analysis after (\ref{D}), the above becomes equal to the minimum eigenvalue of $\tilde{\mathbf{Q}}^T\tilde{\mathbf{Q}}$, where
\begin{equation}\label{new}
\tilde{\mathbf{Q}}=\mathbf{P}_X^{-\frac{1}{2}}\tilde{\mathbf{P}}_{X|Y}\mathbf{P}^{\frac{1}{2}}_{Y|u^*},
\end{equation}
and $\tilde{\mathbf{P}}_{X|Y}$ is an $|\mathcal{X}|\times(|\mathcal{Y}|-|\mathcal{I}_{u^*}|)$-dimensional matrix obtained by eliminating the columns of $\mathbf{P}_{X|Y}$ that correspond to the indices in $\mathcal{I}_{u^*}$; $\mathbf{P}_{Y|u^*}$ is a $(|\mathcal{Y}|-|\mathcal{I}_{u^*}|)\times(|\mathcal{Y}|-|\mathcal{I}_{u^*}|)$ diagonal matrix whose diagonal elements are the corresponding non-zero elements of $\mathbf{p}_{Y|u^*}$.

In what follows, we show that the minimum eigenvalue of the matrix $\tilde{\mathbf{Q}}^T\tilde{\mathbf{Q}}$ is bounded away from zero, since otherwise there must exists a non-zero $(|\mathcal{Y}|-|\mathcal{I}_{u^*}|)$-dimensional vector $\tilde{\mathbf{e}}$, such that $\tilde{\mathbf{Q}}\tilde{\mathbf{e}}=\mathbf{0}$. Let $\tilde{\mathbf{e}}'\triangleq \mathbf{P}^{\frac{1}{2}}_{Y|u^*}\tilde{\mathbf{e}}$. From (\ref{new}), since $\mathbf{P}^{\frac{1}{2}}_{Y|u^*}$ and $\mathbf{P}_X^{-\frac{1}{2}}$ are full rank matrices, we must have $\tilde{\mathbf{e}}'\in\textnormal{Null}(\tilde{\mathbf{P}}_{X|Y})$, and therefore, $\mathbf{1}_{|\mathcal{Y}|-|\mathcal{I}_{u^*}|}^T.\tilde{\mathbf{e}}'=0$. Construct the $|\mathcal{Y}|$- dimensional vector $\mathbf{e}'$ as follows. Let its elements corresponding to the indices in $\mathcal{I}_{u^*}$ be zero, and its other terms be equal to the elements of $\tilde{\mathbf{e}}'$. It is obvious that $\mathbf{e}'\in\textnormal{Null}({\mathbf{P}}_{X|Y})$, since the elements of $\mathbf{e}'$ corresponding to the columns of $\mathbf{P}_{X|Y}$ that are not in $\tilde{\mathbf{P}}_{X|Y}$ are zero and we have $\tilde{\mathbf{e}}'\in\textnormal{Null}(\tilde{\mathbf{P}}_{X|Y})$. {\color{black} Having  $\mathbf{e}'\in\textnormal{Null}({\mathbf{P}}_{X|Y})$ results in $\mathbf{1}_{|\mathcal{Y}|}^T.\mathbf{e}'=\mathbf{1}_{|\mathcal{X}|}^T\mathbf{P}_{X|Y}\mathbf{e}'=0$.} For sufficiently small $\epsilon>0$, let $\mathbf{q}_Y=\mathbf{p}_{Y|u^*}+\epsilon\mathbf{e}'$. Since $\epsilon\neq 0$ and $\mathbf{e}'\neq\mathbf{0}$, we have $D(\mathbf{q}_Y||\mathbf{p}_{Y|u^*})>0$. Moreover, since the elements in $\mathbf{q}_Y$ corresponding to the indices in $\mathcal{I}_{u^*}$ are zero, we have $D(\mathbf{q}_Y||\mathbf{p}_{Y|u^*})<+\infty$. Therefore, from the reasoning at the beginning of this Appendix, we have $\mathbf{q}_Y\not\in\mathbb{S}_{X,Y}$. However, since ${\mathbf{e}}'\in\textnormal{Null}({\mathbf{P}}_{X|Y})$ and $\mathbf{P}_{X|Y}\mathbf{p}_{Y|u^*}=\mathbf{p}_X$, we have $\mathbf{q}_Y\in\mathbb{S}_{X,Y}$, which is a contradiction. Therefore, the minimum eigenvalue of the matrix $\tilde{\mathbf{Q}}^T\tilde{\mathbf{Q}}$ is bounded away from zero. This in turn means that the inverse of $\psi(u^*)$ is bounded away from zero, and therefore, $\psi(u^*)<+\infty, \forall u^*$. 
\bibliography{REFERENCE}

\begin{thebibliography}{10}
\providecommand{\url}[1]{#1}
\csname url@samestyle\endcsname
\providecommand{\newblock}{\relax}
\providecommand{\bibinfo}[2]{#2}
\providecommand{\BIBentrySTDinterwordspacing}{\spaceskip=0pt\relax}
\providecommand{\BIBentryALTinterwordstretchfactor}{4}
\providecommand{\BIBentryALTinterwordspacing}{\spaceskip=\fontdimen2\font plus
\BIBentryALTinterwordstretchfactor\fontdimen3\font minus
  \fontdimen4\font\relax}
\providecommand{\BIBforeignlanguage}[2]{{%
\expandafter\ifx\csname l@#1\endcsname\relax
\typeout{** WARNING: IEEEtran.bst: No hyphenation pattern has been}%
\typeout{** loaded for the language `#1'. Using the pattern for}%
\typeout{** the default language instead.}%
\else
\language=\csname l@#1\endcsname
\fi
#2}}
\providecommand{\BIBdecl}{\relax}
\BIBdecl

\bibitem{RG18}
B.~{Rassouli} and D.~{Gunduz}, ``On perfect privacy,'' in \emph{2018 IEEE
  International Symposium on Information Theory (ISIT)}, 2018, pp. 2551--2555.

\bibitem{Nar}
A.~Narayanan and V.~Shmatikov, ``Robust de-anonymization of large sparse
  datasets,'' in \emph{IEEE Symp. on Security and Privacy (SP)}, 2008, pp.
  111--125.

\bibitem{Ding}
X.~Ding, L.~Zhang, and W.~Zhiguo, ``A brief survey on de-anonymization attacks
  in online social networks,'' in \emph{International Conf. on Computational
  Aspects of Social Networks (CASoN)}, 2010, pp. 611--615.

\bibitem{mhealth}
S.~Kumar, W.~Nilsen, M.~Pavel, and M.~Srivastava, ``Mobile health:
  Revolutionizing healthcare through transdisciplinary research,''
  \emph{Computer}, vol.~46, pp. 28--35, 2013.

\bibitem{Gomez}
J.~Gomez-Vilardebo and D.~G\"{u}nd\"{u}z, ``Smart meter privacy for multiple
  users in the presence of an alternative energy source,'' \emph{IEEE Trans. on
  Information Forensics and Security}, pp. 132--141, 2015.

\bibitem{Abol}
A.~Motahari, G.~Bresler, and D.~Tse, ``Information theory of {DNA} shotgun
  sequencing,'' \emph{IEEE Trans. on Information Theory}, vol.~59, no.~10, pp.
  6273--6289, Oct. 2013.

\bibitem{Dwork}
C.~Dwork, F.~McSherry, K.~Nissim, and A.~Smith, ``Calibrating noise to
  sensitivity in private data analysis,'' \emph{Theory of Cryptography,
  Springer}, pp. 265--284, 2006.

\bibitem{Sweeney}
L.~Sweeney, ``k-anonymity: A model for protecting privacy,'' \emph{Intl.
  Journal of Uncertainty, Fuzziness and Knowledge-Based Systems}, vol.~10,
  no.~5, pp. 557--570, 2002.

\bibitem{LD}
A.~Machanavajjhala, D.~Kifer, J.~Gehrke, and M.~Venkitasubramaniam,
  ``l-diversity: Privacy beyond k-anonymity,'' \emph{ACM Trans. on Knowledge
  Discovery from Data}, vol.~1.

\bibitem{LL}
N.~Li, T.~Li, and S.~Venkatasubramanian, ``t-closeness: Privacy beyond
  k-anonymity and l-diversity,'' \emph{IEEE Intl. Conf. on Data Eng.}, 2007.

\bibitem{SG19}
S.~{Sreekumar} and D.~{Gündüz}, ``Optimal privacy-utility trade-off under a
  rate constraint,'' in \emph{2019 IEEE International Symposium on Information
  Theory (ISIT)}, 2019, pp. 2159--2163.

\bibitem{Makhdoumi}
A.~Makhdoumi, S.~Salamatian, N.~Fawaz, and M.~M\'{e}dard, ``From the
  information bottleneck to the privacy funnel,'' in \emph{IEEE Information
  Theory Workshop (ITW)}, 2014, pp. 501--505.

\bibitem{Tishby}
N.~Tishby, F.~Pereira, and W.~Bialek, ``The information bottleneck method,''
  \emph{arXiv preprint physics/0004057}, 2000.

\bibitem{Calmon2}
F.~Calmon, A.~Makhdoumi, and M.~M\'{e}dard, ``Fundamental limits of perfect
  privacy,'' in \emph{IEEE Int. Symp. Inf. Theory (ISIT)}, 2015, pp.
  1796--1800.

\bibitem{info7010015}
\BIBentryALTinterwordspacing
S.~Asoodeh, M.~Diaz, F.~Alajaji, and T.~Linder, ``Information extraction under
  privacy constraints,'' \emph{Information}, vol.~7, no.~1, 2016. [Online].
  Available: \url{https://www.mdpi.com/2078-2489/7/1/15}
\BIBentrySTDinterwordspacing

\bibitem{Shahab1}
S.~Asoodeh, F.~Alajaji, and T.~Linder, ``Notes on information-theoretic
  privacy,'' in \emph{52nd Annual Allerton Conference}, Illinois, USA, Oct.
  2014, pp. 1272--1278.

\bibitem{HuangK}
C.~{Huang}, P.~{Kairouz}, and L.~{Sankar}, ``Generative adversarial privacy: A
  data-driven approach to information-theoretic privacy,'' in \emph{2018 52nd
  Asilomar Conference on Signals, Systems, and Computers}, 2018, pp.
  2162--2166.

\bibitem{Tripathy}
A.~{Tripathy}, Y.~{Wang}, and P.~{Ishwar}, ``Privacy-preserving adversarial
  networks,'' in \emph{2019 57th Annual Allerton Conference on Communication,
  Control, and Computing (Allerton)}, 2019, pp. 495--505.

\bibitem{advers}
\BIBentryALTinterwordspacing
M.~Bertran, N.~Martinez, A.~Papadaki, Q.~Qiu, M.~Rodrigues, G.~Reeves, and
  G.~Sapiro, ``Adversarially learned representations for information
  obfuscation and inference,'' in \emph{Proceedings of the 36th International
  Conference on Machine Learning}, ser. Proceedings of Machine Learning
  Research, K.~Chaudhuri and R.~Salakhutdinov, Eds., vol.~97.\hskip 1em plus
  0.5em minus 0.4em\relax Long Beach, California, USA: PMLR, 09--15 Jun 2019,
  pp. 614--623. [Online]. Available:
  \url{http://proceedings.mlr.press/v97/bertran19a.html}
\BIBentrySTDinterwordspacing

\bibitem{Issa}
I.~{Issa}, A.~B. {Wagner}, and S.~{Kamath}, ``An operational approach to
  information leakage,'' \emph{IEEE Transactions on Information Theory},
  vol.~66, no.~3, pp. 1625--1657, 2020.

\bibitem{Berger}
T.~Berger and R.~Yeung, ``Multiterminal source encoding with encoder
  breakdown,'' \emph{IEEE Trans. Inf. Theory}, pp. 237--244, 1989.

\bibitem{Ishwar}
Y.~Wang, Y.~Basciftci, and P.~Ishwar, ``Privacy-utility tradeoffs under
  constrained data release mechanisms,''
  \emph{https://arxiv.org/pdf/1710.09295.pdf}, Oct. 2017.

\bibitem{Witsen1975}
H.~{Witsenhausen} and A.~{Wyner}, ``A conditional entropy bound for a pair of
  discrete random variables,'' \emph{IEEE Transactions on Information Theory},
  vol.~21, no.~5, pp. 493--501, 1975.

\bibitem{LP1}
D.~Bertsimas and J.~N. Tsitsiklis, \emph{Introduction to linear
  optimization}.\hskip 1em plus 0.5em minus 0.4em\relax Athena Scientic, 1997.

\bibitem{LP2}
K.~G. Murty, \emph{Linear Programming}.\hskip 1em plus 0.5em minus 0.4em\relax
  John Wiley and Sons, 1983.

\bibitem{Gfilter}
S.~Asoodeh, F.~Alajaji, and T.~Linder, ``Almost perfect privacy for additive
  gaussian privacy filters,'' in \emph{Information Theoretic Security}, A.~C.
  Nascimento and P.~Barreto, Eds.\hskip 1em plus 0.5em minus 0.4em\relax Cham:
  Springer International Publishing, 2016, pp. 259--278.

\bibitem{Makur}
A.~Makur and L.~Zheng, ``Bounds between contraction coefficients,'' in
  \emph{53nd Annual Allerton Conference}, Illinois, USA, Sep. 2015, pp.
  1422--1429.

\bibitem{Hirschfeld}
H.~Hirschfeld, ``A connection between correlation and contingency,'' in
  \emph{Proc. Cambridge Philosophical Soc. 31}, 1935, pp. 520--524.

\bibitem{Gebelein}
H.~Gebelein, ``Das statistische problem der korrelation als variations- und
  eigenwert-problem und sein zusammenhang mit der ausgleichungsrechnung,,''
  \emph{Zeitschrift f\''{u}r angew. Math. und Mech. 21}, pp. 364--379, 1941.

\bibitem{Reny}
A.~R\'{e}nyi, ``On measures of dependence,'' \emph{Acta Math. Hung.}, vol.~10,
  pp. 539--550, 1959.

\bibitem{Witsenhausen}
H.~Witsenhausen, ``On sequences of pairs of dependent random variables,''
  \emph{SIAM Journal on Applied Mathematics}, vol.~28, no.~1, pp. 100--113,
  Jan. 1975.

\bibitem{Anantharam}
V.~Anantharam, A.~Gohari, S.~Kamath, and C.~Nair, ``On maximal correlation,
  hypercontractivity, and the data processing inequality studied by {E}rkip and
  {C}over,'' \emph{https://arxiv.org/pdf/1304.6133.pdf}, Apr. 2013.

\bibitem{Golub}
G.~Golub, ``Some modified matrix eigenvalue problems,'' \emph{SIAM Review},
  vol.~15, no.~2, pp. 318--334, Apr. 1973.

\bibitem{Elgamal}
A.~E. Gamal and Y.-H. Kim, \emph{Network Information Theory}.\hskip 1em plus
  0.5em minus 0.4em\relax Cambridge University Press, 2012.

\end{thebibliography}
\bibliographystyle{IEEEtran}
\end{document}